%% file: ms.tex
\documentclass{mn2e}

\usepackage{graphics}
\usepackage{afterpage}

\newlength{\colwidth}
\newcommand{\cm}{{\rm cm}}

\newcommand{\kms}{{\rm km}\,{\rm s}^{-1}}
\newcommand{\K}{{\rm K}}
\newcommand{\pc}{{\rm pc}}
\newcommand{\kpc}{{\rm kpc}}
\newcommand{\Mpc}{{\rm Mpc}}

\newcommand{\Msun}{{{\rm M}_\odot}}
\newcommand{\yr}{{\rm yr}}

\newcommand{\ion}[2]{\hbox{#1\,{\sc #2}}}

\newcommand{\HI}{\ion{H}{I}}

\newcommand{\CII}{\ion{C}{II}}
\newcommand{\CIII}{\ion{C}{III}}
\newcommand{\CIV}{\ion{C}{IV}}

\newcommand{\NIV}{\ion{N}{IV}}
\newcommand{\NV}{\ion{N}{V}}
\newcommand{\OI}{\ion{O}{I}}
\newcommand{\OV}{\ion{O}{V}}
\newcommand{\OVI}{\ion{O}{VI}}
\newcommand{\SiII}{\ion{Si}{II}}

\newcommand{\SiIV}{\ion{Si}{IV}}
\newcommand{\FeII}{\ion{Fe}{II}}
\newcommand{\MgII}{\ion{Mg}{II}}
\newcommand{\AlII}{\ion{Al}{II}}
\newcommand{\CaII}{\ion{Ca}{II}}

\newcommand{\lya}{Ly$\alpha$} 
\newcommand{\lyb}{Ly$\beta$}

\newcommand{\ZC}{$[{\rm C}/{\rm H}]$}
\newcommand{\ZCgt}{$[{\rm C}/{\rm H}] > -1$}
\newcommand{\SiC}{$[{\rm Si}/{\rm C}]$}

\newcommand{\aap}{A\&A}
\newcommand{\apj}{ApJ}
\newcommand{\apjl}{ApJL}
\newcommand{\araa}{ARA\&A}
\newcommand{\mnras}{MNRAS}
\newcommand{\aj}{AJ}
\newcommand{\apjs}{ApJS}

\newcommand{\pasp}{PASP}

\newcommand{\procspie}{SPIE}

\def\lsim{~\rlap{$<$}{\lower 1.0ex\hbox{$\sim$}}}

\def\gsim{~\rlap{$>$}{\lower 1.0ex\hbox{$\sim$}}}

\title[metal-rich, intergalactic clouds]{A large population of
  metal-rich, compact, intergalactic \CIV\ absorbers --- 
  Evidence for poor small-scale metal mixing\thanks{ 
Based on public data obtained from the ESO archive of observations
from the UVES spectrograph at the VLT, Paranal, Chile.}}
\author[Schaye et al.]{Joop Schaye$^1$\thanks{E-mail:
  schaye@strw.leidenuniv.nl}, Robert F. Carswell$^2$, Tae-Sun Kim$^3$\\
$^{1}$ Leiden Observatory, Leiden University, P.O. Box 9513, 2300 RA Leiden,
  The Netherlands\\
$^2$ Institute of Astronomy, Madingley Road, Cambridge CB3 0HA, U.K.\\
$^3$ Astrophysikalisches Institut Potsdam, An der Sternwarte 16,
  D-14482 Potsdam, Germany.}  

\begin{document}

\pagerange{\pageref{firstpage}--\pageref{lastpage}} \pubyear{2006}

\maketitle

\label{firstpage}

\begin{abstract}
We carried out a survey for high-metallicity \CIV\ absorbers at redshift
$z\approx 2.3$ in 9 high-quality quasar spectra. Using
a novel analysis technique, based on detections of \CIV\ lines
and automatically determined upper limits on the column densities of
\HI, \CIII, \NV, and \OVI, we
find a large ($d{\cal N}/dz > 7$) population of photo-ionized, compact
($R\sim 10^2~\pc$), metal-rich ($Z\ga Z_\odot$)
\CIV\ clouds with moderate densities ($n_{\rm H}\sim
10^{-3.5}~\cm^{-3}$), properties that we show are robust 
with respect to uncertainties in the ionization model. 
In particular, local sources of ionizing radiation, overabundance of
oxygen, departures from ionization
equilibrium, and collisional ionization would all imply more compact
clouds. The clouds are too small to be self-gravitating and pressure
confinement is only consistent 
under special conditions. We argue that the clouds
are, in any case, likely to be short-lived and 
we demonstrate that this implies that the clouds could easily have
been responsible for the 
transport of all metals that end up in the intergalactic
medium (IGM). When metal-rich clouds reach pressure equilibrium
with the general, 
photo-ionized IGM, the heavy elements will still be concentrated in small
high-metallicity patches, but they will look like ordinary,
low-metallicity absorbers. We conclude that intergalactic metals are
poorly mixed on small scales and that nearly all of the IGM, and thus
the universe, may therefore be of primordial composition.
\end{abstract}


\section{Introduction}
There is strong observational evidence that correlated supernova
explosions drive gas flows into the halos of galaxies and into the
intergalactic medium (IGM) through galactic fountain activity and
massive superwinds (e.g., Veilleux et al.\ 2005). If the energy
powering the outflows originates in 
supernovae, one would expect the outflowing gas to be highly enriched
(e.g., Mac Low \& Ferrara 1999). Indeed, X-ray observations of a
starbursting dwarf galaxy indicate that nearly all the metals injected
by the starburst are contained in a superwind (Martin, Kobulnicky, \&
Heckman 2002). 

The outflowing gas will sweep up shells of gas, which may fragment due
to hydrodynamical and thermal instabilities and mix with the hot
wind. Eventually the wind cools and the outflow decelerates. The
shell fragments expand as the (ram) pressure drops, unless they
become self-gravitating. The metals carried by the outflow will either
rain back onto the galaxy or get mixed into the IGM, although this
mixing could remain incomplete (e.g., Dedikov \& Shchekinov 2004).

Careful studies of high-quality quasar absorption spectra have
recently produced a wealth of information regarding the distribution of heavy
elements in the high-redshift IGM. The abundance of
carbon is found to be far below solar (\ZC$\approx -3$ at ten times
the mean density and $z=3$; Schaye et al.\ 2003, hereafter S03; Simcoe
et al.\ 2004) 
but with a large scatter ($\sigma([{\rm C}/{\rm H}]) \approx 0.7$,
S03; Simcoe et al.\ 2004), to  
increase with density (S03), to be lower than that of
oxygen (Telfer et 
al.\ 2002) and silicon (Aguirre et al.\ 2004), and to depend on the
proximity to galaxies (e.g., Adelberger et al.\ 2003, 2005; Pieri et al.\
2006). It is often not mentioned that these measurements
implicitly smooth over the scales associated with typical \HI\ \lya\
absorbers, $R\la 10^2~\kpc$ depending on the density (see
Schaye \& Aguirre 2005 and S03 for discussions). On smaller
scales the distribution of metals is essentially unknown. 

If the heavy elements that end up in the IGM were initially
concentrated in gas clouds with very high metallicities, then what
about the intermediate phase? Where are the 
high-metallicity, intergalactic gas clouds? Observing this phase is
very important because it will give us valuable information regarding
the physics of 
galactic winds, a key ingredient of theories of galaxy
formation, and the enrichment of the IGM.

There are various strategies to look for high-metallicity gas flows
from galaxies. Perhaps the most obvious strategy is to look for
blueshifted absorption in the spectra of starburst galaxies. Such
studies have provided unambiguous evidence for outflows containing
heavy elements (e.g., Heckman et al.\ 2000; Pettini et al.\ 1998), but the
faintness of the galaxies makes it very difficult to obtain spectra
with sufficient resolution to resolve the substructure of the
lines, let alone measure the corresponding metallicities. Another problem
is that it is unclear how close to the 
galaxy the absorbing material is and therefore whether it can be
considered extragalactic. 

Some of these problems can be circumvented by
searching for metal-line absorption along sight lines to bright background
quasars which pass close to galaxies known to be driving winds (e.g.,
Norman et al.\ 1996, Chen et al.\ 2001;
Adelberger et al.\ 2003; 
Simcoe et al.\ 2006; Stocke et al.\ 2006). The drawback here is that 
only a very small fraction of observed galaxies are close to a
background source that is sufficiently bright to do high-resolution
spectroscopy. Ultraviolet metal line \emph{emission}, which could
remain observable above the background out to $\sim 10^2~\kpc$ from
galaxies (Furlanetto et al.\ 2004), is another promising tool,
particularly because it
would provide spatial information that cannot be obtained from single
line-of-sight absorption spectra.

Highly enriched gas clouds are regularly found in surveys for quasar
absorption lines, for a wide range of redshifts and \HI\ column
densities (e.g., Carswell, Schaye, \& Kim 2002; Tripp et al.\ 2002,
2006; Rigby et al.\ 2002; Charlton et 
al.\ 2003; Bergeron \& Herbert-Fort 2005; Aracil et al.\ 2006; Simcoe
et al.\ 2006; P\'eroux et al.\ 2006; Prochaska et al.\ 2006; Keeney et
al.\ 2006).  In their
survey of \OVI\ absorption at $z\sim 2.5$, Carswell et al.\ (2002)
found one unusual system that showed absorption by \HI, \CIV, \NV, and 
\OVI. The system is unusual in that the \HI\ absorption was very weak
compared to the metal lines, and that it showed clear \NV\ absorption.
If the system is photoionized, then it must be very compact ($\sim
10^2~\pc$). Carswell et al.\ (2002) were able to show that regardless
of the ionization mechanism, the metallicity must be high ($\ga$
solar) and the system cannot be self-gravitating, unless the gas
fraction is negligible. Similar clouds have later been found in the
survey for \OVI\ absorption by Bergeron \& Herbert-Fort (2005),
whereas the survey for strong \OVI\ absorption by Simcoe et al.\
(2006) turned up higher density analogs.  

Here we describe a survey for high-metallicity \CIV\
clouds in 9 high-quality quasar spectra at $z\approx 2.3$. We
introduce a novel analysis 
technique, combining measurements of \CIV\ column densities with
automatically determined upper limits on the column densities of \HI, \CIII,
\NV, \OVI, and \SiIV. Our reliance on blindly measured upper limits
for all transitions other than \CIV\ makes our analysis robust to
uncertainties due to blending, contamination, noise, and the presence
of multiple phases. We find a
large number of high-metallicity ($Z>0.1 Z_\odot$) clouds and
demonstrate that they are photo-ionized, moderately overdense,
compact, and not gravitationally confined.
In the second part of this paper we investigate the implications of
the existence of this population, and other metal-rich populations
found by others. We show that the clouds are probably short-lived, which
implies that they could easily have been responsible for the transport
of all the metals that end up in the IGM. Finally, we argue that
it is likely that most intergalactic metals reside in small patches of
high-metallicity gas, but that we generally cannot tell this
observationally once the clouds have reached pressure equilibrium with
their environments. This would mean that nearly 
all the gas in the universe is metal-free and that the generally
accepted low IGM metallicities are only appropriate when averaged over
large scales.

This paper is organized as follows. We begin with a brief description
of the observations. Section \ref{sec:selection} describes our
analysis technique and the selection of the metal-rich sample. Results
that are independent of the ionization model are described in
\S\ref{sec:properties}, while \S\ref{sec:ionmodeling} presents and
discusses the method, results, and uncertainties of the ionization model. In
\S\ref{sec:discussion} we compare our results with other
observations and we investigate the implications for the nature and
origin of the clouds, for the distribution of heavy elements, and for the 
interpretation of quasar absorption line studies. Finally, we provide
a concise summary in \S\ref{sec:conclusions}. Given the length of the
paper, we recommend reading the summary first. 

All abundances are given by number relative to hydrogen, and solar
abundances are taken to be the default values in CLOUDY96\footnote{See 
  \texttt{http://www.nublado.org/}}: $({\rm
  C}/{\rm H})_\odot = -3.61$, $({\rm N}/{\rm H})_\odot =
-4.07$, $({\rm O}/{\rm H})_\odot = -3.31$, and $({\rm Si}/{\rm
  H})_\odot = -4.46$.

\section{Observations}
\label{sec:observations}

We analyzed a sample of 9 high-quality ($6.6~\kms$ velocity resolution
(FWHM), signal-to-noise ratio $> 40$) absorption spectra of $2.1 \le z
\le 3.3$ quasars taken with the UV-Visual Echelle Spectrograph (UVES;
D'Odorico et al.\ 2000) on the Very Large Telescope (VLT). The quasars
and their emission redshifts are listed in columns 1 and 2 of
Table~\ref{tbl:oblist}. The data reduction procedures were the same as
those used by Kim et al. (2002).

We searched
for all \CIV\ systems between $z_{\rm min} \le z \le z_{\rm max}$
irrespective of the presence of the associated \HI. The minimum
redshift $z_{\rm min} \equiv {\rm max}(2.0,(1+z_{\rm
  em})1215.67/1548.20 - 
1)$ was chosen to ensure that the corresponding \OVI\ lines fall in
the observed wavelength range and that \CIV\ falls redwards of the
quasar's \lya\ emission line (to prevent contamination by \HI).
To avoid proximity effects, a high redshift cutoff was imposed
at\footnote{For HE0151$-$4326 a 
larger region was excluded to avoid broad, associated absorption
lines.}
$z_{\rm max} \equiv  z_{\rm em} - (1+z_{\rm 
  em})\Delta v/c$, where $\Delta v = 
\max(4000,8\,\Mpc\,H(z)/h)~\kms $ and $H(z)$ is the Hubble
parameter at 
redshift $z$ extrapolated from its present value ($H_0 \equiv
100h~\kms\,\Mpc^{-1}$) assuming $(\Omega_m,\Omega_\Lambda) =
(0.3,0.7)$. The search ranges and some information
on the detected high-metallicity \CIV\ systems are given in
Table~\ref{tbl:oblist}. 

\begin{table*}
\begin{minipage}{160mm}
\centering
\caption{
\label{tbl:oblist}
Observed quasars}
\footnote{Cols.\ (1) and (2) contain the quasar name and
  redshift. Cols.\ (3) and (4) contain the minimum and maximum 
  redshifts searched. Col.\ (5) shows the number
  of \CIV\ systems with at least one component with \ZCgt,
  while col.\ (6)  gives the redshifts
  of these absorption systems. Col.\ (7) and (8) shows, for each
  system, the total number of absorption line components with \ZCgt\
  and their maximum velocity separation. Finally, cols.\ (9)
  and (10) show, for the same systems, the number of
  components and their maximum velocity separation when all detected \CIV\
  components are taken into account, irrespective of their inferred lower
  limits on \ZC.} 
\begin{tabular}{lccccccccc} 
\hline
& & & & & & \multicolumn{2}{c}{\ZCgt} & \multicolumn{2}{c}{all \CIV} \\
QSO &  $z_{\rm em}$ & $z_{\rm min}$ &
$z_{\rm max}$ & \# sys. & $z_{\rm abs}$ & \# comp. & $\left | \Delta
v\right | (\kms)$ & \# comp. & $\left | \Delta v\right | (\kms)$ \\
(1) & (2) & (3) & (4) & (5) & (6) & (7) & (8) & (9) & (10)\\
\hline
Q0122$-$380   & 2.181 & 2.00 & 2.14 & 1 & 2.063 & 1 & 0 & 1 & 0\\
Pks0237$-$233 & 2.225 & 2.00 & 2.18 & 1 & 2.042 & 2 & 27 & 5 & 121\\
HE1122$-$1648 & 2.400 & 2.00 & 2.35 & 1 & 2.030 & 1 & 0 & 1 & 0\\
HE2217$-$2818 & 2.406 & 2.00 & 2.36 & 0 & 0 \\
Q0329$-$385   & 2.423 & 2.00 & 2.38 & 2 & 2.076, 2.352 & 1, 2 & 0, 10 &
1, 2 & 0, 10\\
HE1347$-$2457 & 2.534 & 2.00 & 2.49 & 1 & 2.116 & 1 & 0 & 3 & 69\\
HE0151$-$4326 & 2.740 & 2.00 & 2.52 & 2 & 2.417, 2.468 & 2, 2 & 346, 54
& 4, 3 & 373, 54\\
HE2347$-$4342 & 2.900 & 2.07 & 2.85 & 2 & 2.120, 2.275 & 1, 7 & 0, 201 & 2,
9 & 14, 445\\
Pks2126$-$158 & 3.266 & 2.36 & 3.21 & 2 & 2.394, 2.679 & 6, 2 & 699, 12
& 12, 4 & 718, 54\\
\hline
\end{tabular}
\end{minipage}
\end{table*}

\section{Selection of high-metallicity \CIV\ absorbers}
\label{sec:selection}
We wish to select \CIV\ lines that arise in gas with relatively high
metallicity, $Z > 0.1 Z_\odot$, and we wish to do this in a way that is
robust with respect to problems due to line blending, noise, contamination,
and uncertainties in the ionization models.

Our search for highly enriched, highly ionized gas clouds comprises
the following main steps: (1) Identify all detectable \CIV\ absorption
lines which fall redwards of the quasar's \lya\ emission line and
which have $z_{\rm min} < z < z_{\rm max}$ ($z_{\rm min}$ is chosen to
ensure that \OVI\ falls within the observed wavelength range and
$z_{\rm max}$ is chosen to avoid the region close to the quasar); (2)
Decompose the \CIV\ systems into Voigt 
profile components with parameters $z$, $N$ (column density), and $b$
(line width); (3) For each component, determine conservative
upper limits on the associated column densities of \HI, \NV, \OVI, and
other transitions of interest that fall within the observed wavelength
range (e.g., \CIII\ and \SiIV). We choose to work only with
upper limits because it is in general impossible to measure reliable
column densities for transitions that fall bluewards of the quasar's
\lya\ emission line due to line blending (most important for \HI) and
the presence of contamination (important for \CIII, \NV, and \OVI);
(4) Assuming photoionization by a given radiation field, 
select those components which must have a carbon abundance greater than
10\% solar based on the measured lower limits on $N_{\rm
CIV}/N_{\rm HI}$ plus those on $N_{\rm CIV}/N_{\rm NV}$ and/or $N_{\rm
  CIV}/N_{\rm OVI}$ (the selection turns out to be nearly
independent of the latter two measurements). In the following we will
discuss the last three steps in more detail. 

All detected \CIV\ systems were decomposed into Voigt profile
components using the VPFIT program described by Webb (1987) and Rauch et
al.~(1992) and updated by R.F.~Carswell et al.\footnote{See
   http://www.ast.cam.ac.uk/$\sim$rfc/vpfit.html.}. The
decomposition of any absorption system into Voigt profiles is generally
not unique and may depend on the signal-to-noise ratio of the
spectrum. 
However, we stress that the two processes that most complicate the 
decomposition of \HI\ \lya\ profiles, line blending and contamination, are
much less important for \CIV\ because the latter is a doublet, because
we limit our search to the region redwards of the quasar's \lya\
emission line, and because \CIV\ lines are much narrower and rarer
than \lya\ lines.

Because of the severe line blending and contamination in the \lya\ and
\lyb\ forests, we do not attempt to measure the column densities
associated with transitions that fall bluewards of the quasar's \lya\
emission line. Instead, we only consider robust upper limits on the
column densities, a method which is
conservative for our purposes. Note that this approach also
circumvents problems due to both the possible multiphase nature of the 
absorbers and projection effects in velocity space (i.e., the redshift
coincidence of absorption lines that arise in separate spatial
locations). For example, if the absorption by \HI\ at the
redshift of the \CIV\ absorber arises in different gas than the \CIV\
absorption, then the \HI\ column corresponding to the gas responsible
for the \CIV\ absorption will be lower than the upper limit we measure
from the data.     

We wrote a program to
automatically compute upper limits on the associated 
column densities. The algorithm is fed a list containing the redshifts,
Doppler parameters ($b$-values) and $b$-value error estimates for the 
\CIV\ absorbers, determines the possible Doppler parameters for other 
elements assuming a range from pure turbulent to pure thermal broadening, and 
returns upper limits
based on the spectrum, continuum and noise array. The algorithm is
described in more detail in the Appendix.

We can write the abundance of carbon as
\begin{eqnarray}
[{\rm C}/{\rm H}] &=& 
\log \left ({n_{\rm C} \over n_{\rm CIV}} {n_{\rm CIV}
  \over n_{\rm HI}} {n_{\rm HI} \over n_{\rm H}} \right ) -
\log\left ({n_{\rm C} \over n_{\rm H}}\right )_\odot \\
&=& 
\log \left ({N_{\rm CIV} \over N_{\rm HI}}\right )
-\log \left ( {n_{\rm CIV}/n_{\rm C} \over n_{\rm HI}/n_{\rm H}}
   \right ) - \log\left ({n_{\rm C} \over
     n_{\rm H}}\right )_\odot
\label{eq:metallicity},
\end{eqnarray}
where the first term is an observable and the second term depends on
the ionization balance.

\begin{figure}
\resizebox{\colwidth}{!}{\includegraphics{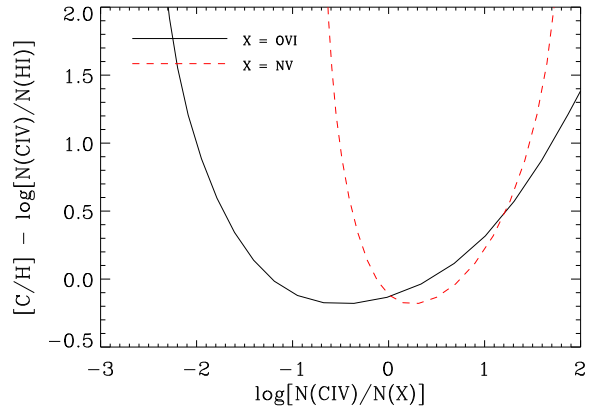}}
\caption{Metallicity as a function of the measured column density
  ratios. The CIV/HI column density ratio is assumed to be unity, the
  inferred metallicity is proportional to this ratio.}  
\label{fig:pzmet}
\end{figure}

Figure~\ref{fig:pzmet} shows how \ZC$-\log[N_{\rm CIV}/N_{\rm HI}]$
varies as a function of two observable proxies for the density,
$N_{\rm CIV}/N_{\rm OVI}$ (\emph{solid}) and $N_{\rm CIV}/N_{\rm NV}$
(\emph{dashed}). As is the case for 
all ionization models presented in this work, the ionization balance
was computed using the publicly available software package
CLOUDY (version 96; see Ferland et al.\
1998 and Ferland 2000 for details) and assuming ionization
equilibrium. The gas was further assumed to be optically thin, to have
solar relative abundances, to be at a temperature $T=10^4~\K$, and to be
exposed to the $z=2.3$ Haardt \& Madau (2001) model of the UV/X-ray
background from quasars and galaxies.  

The minima in the curves correspond to the maximum $n_{\rm
CIV}/n_{\rm HI}$ ratio. Clearly, using this value will
result in a robust lower limit on the metallicity. If the $x$-value
(\CIV/\NV\ or \CIV/\OVI) is known, then this extra 
information can be used to obtain tighter constraints on the
metallicity. In our case, we only have lower limits on the $x$-values,
which means we can improve the lower limit on the metallicity if the
observed $x$-value falls to the right of the minimum of the curve. It
turns out, however, that the vast majority of our data points fall to the
left of the minima.  

\begin{figure*} 
\resizebox{\colwidth}{!}{\includegraphics{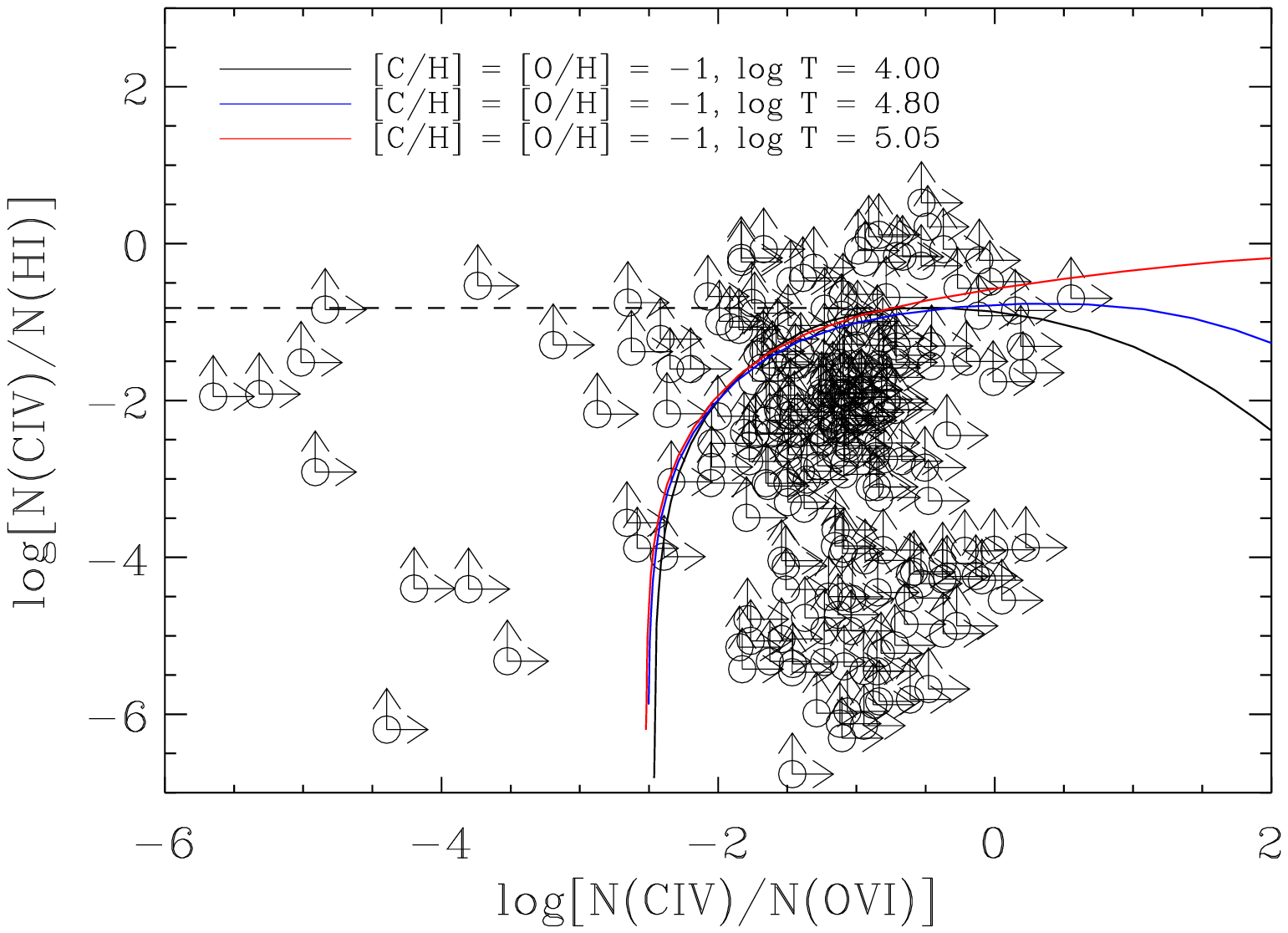}}
\resizebox{\colwidth}{!}{\includegraphics{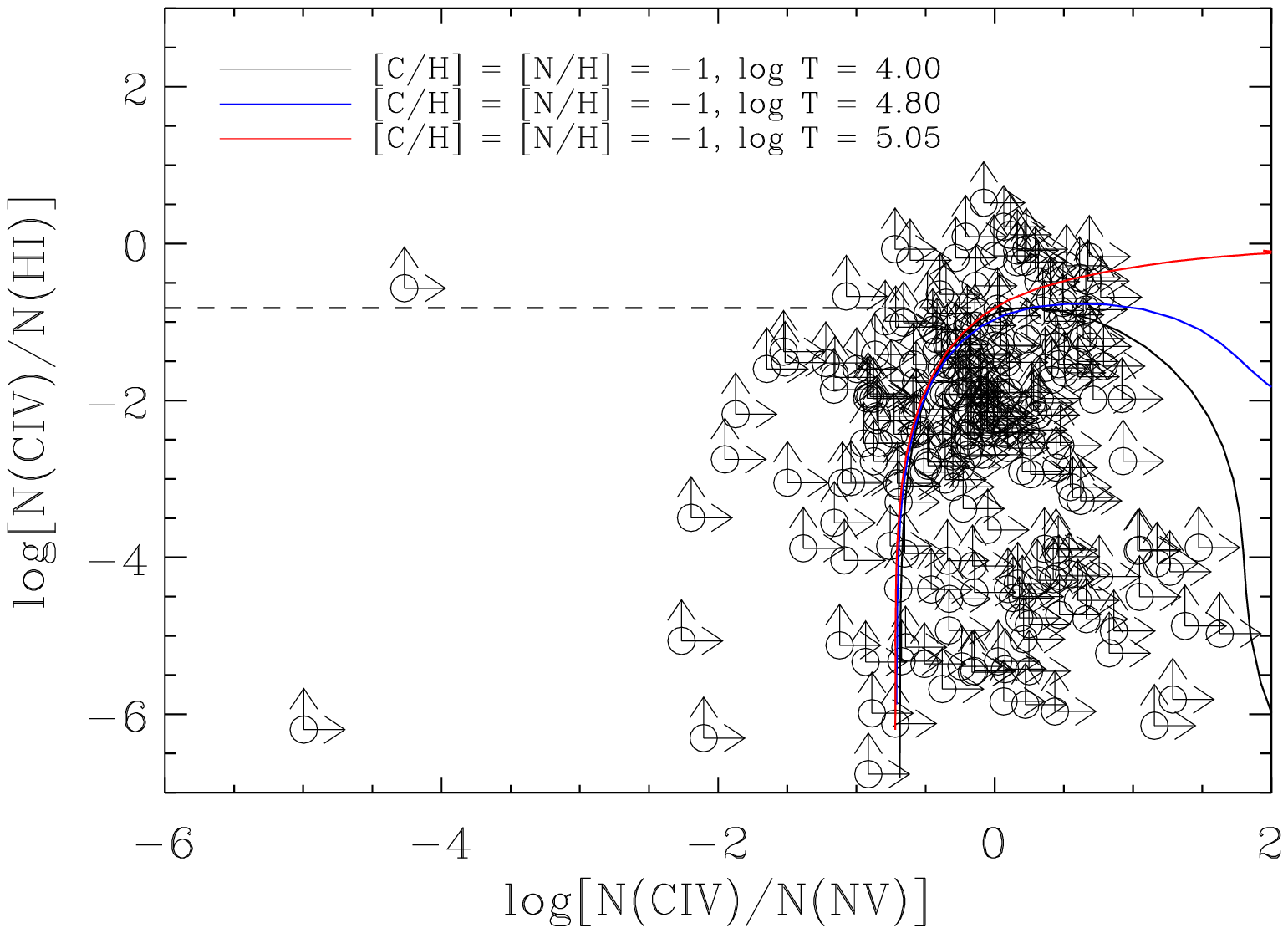}}
\caption{Data points indicate measured 1$\sigma$ lower limits on the
  \CIV/\HI\ column density ratio as a function of the
  1$\sigma$ lower limits on the column density
  ratios \CIV/\OVI\ (\emph{left}) and \CIV/\NV\ (\emph{right}). From
  bottom-to-top the solid curves show the predicted ratio for a
  metallicity of 10\% solar and temperatures of $\log[T (\K)] = 4.00$, 4.80,
  and 5.05, respectively. The latter value corresponds roughly to the
  maximum \CIV/\HI\ ratio. Curves exceed the dashed line only for temperature within
  a factor of about two from this value.
All points that are above at least one of the dashed curves
  (which coincide with the lowest solid curves to the right of their
  peaks) were selected as high-metallicity clouds.
\label{fig:selection}
}
\end{figure*}

Figure~\ref{fig:selection} shows the measured lower
limits\footnote{All limits on column density ratios involving \CIV\
take the error on $N_{\rm CIV}$ into account.} on $N_{\rm
CIV}/N_{\rm HI}$, for all 219 detected \CIV\ components, as a
function of the lower limits on $N_{\rm CIV}/N_{\rm OVI}$
(\emph{left}) and $N_{\rm CIV}/N_{\rm NV}$ (\emph{right}). The solid
curves indicate the predicted ratios for a metallicity of 10\% solar
and a fixed temperature of (\emph{from bottom-to-top}) $\log T = 4.0$,
4.8, and 5.05, respectively. Density varies along these curves.  

If our measurements of the \OVI\ and \NV\ column densities were detections
rather than upper limits, and if we would know the temperature of
the absorbers, then any point above the corresponding curve would have
a metallicity greater than 10\% solar. Since $N_{\rm CIV}/N_{\rm HI}$
is proportional to metallicity, the vertical offset between the points
and the curve can be used to determine the actual lower limit on the
metallicity. However, since we only have a
lower limit on the quantity plotted along the $x$-axes, we are allowed
to shift all the points to the right. For points to the left of the
maximum of the curve (which corresponds to the minimum of the curve
plotted in Fig.~\ref{fig:pzmet}), this would result in a
decreased lower limit on the 
metallicity. On the other hand, if a point lies above the maximum of
the curve, then shifting to the right will increase the lower limit on
the metallicity. Taking into account the fact that $x$-value (a proxy
for the density or ionization parameter) is only
constrained from below, we see that only the points that fall above the
dashed curve (which coincides with the $\log T=4.0$ solid curve beyond
its peak) are constrained to have metallicities greater than 10\%
solar. 

As we mentioned before, the peak in the curve corresponds to the
density (ionization 
parameter) for which $n_{\rm CIV}/n_{\rm HI}$ is maximum. For
temperatures around $10^4~\K$ collisional ionization is unimportant
and this ratio is insensitive to the temperature (see, e.g., fig.~3
of S03). From
figure~\ref{fig:selection} we can see that even for $\log T = 4.8$ the
maximum is nearly identical. For temperatures somewhat higher than
this, collisional ionization results in a slightly higher peak value for the
ratio $n_{\rm CIV}/n_{\rm HI}$. We find that the temperature has to
differ from $\log T = 5.05$ by less than about 0.2~dex for the peak to
exceed that obtained when photo-ionization dominates, and even then the
peak value is never boosted by more than a factor 5. Because of this
relatively small difference and because, as we shall show later,
such high temperatures are not favored by the data, we used the $\log
T = 4.0$ curve (dashed curve in Fig.~\ref{fig:selection}) to select
\CIV\ components with \ZCgt. 

For the selection of metal-rich clouds, the ratios $N({\rm
CIV})/N({\rm OVI})$ and $N({\rm CIV})/N({\rm NV})$ only provide
extra information for data points that fall above
the curve, but below and 
to the right of its peak. In our case this is true for only 2 of the
28 clouds in our high-metallicity sample\footnote{The two clouds are
  the ones at $z=2.119806$ in HE2347$-$4342 and at 
$z=2.393862$ in Pks2126$-$158}. Not coincidentally, these are
also the only two clouds for which only one of these two column
density ratios results in the selection of the cloud. Thus, our sample
would have been nearly identical if we had ignored our upper limits on
$N_{\rm OVI}$ and $N_{\rm NV}$.

The median lower limit on the metallicity for the sample of all 219
detected \CIV\ components is \ZC$> -2.5$. This value should not be
interpreted as a lower limit on the median carbon abundance of the
IGM, since we have ignored the many \HI\ absorbers for which \CIV\ was 
not detected.

Note that due to the conservative nature of our measurements and assumptions, our
sample of high-metallicity \CIV\ components may be highly
incomplete. That is, had we been able to measure accurate column densities for
\HI, \NV, and \OVI, then we might have found many more \CIV\
components with \ZCgt.  

\section{Direct observables}
\label{sec:properties}
Our search of 9 quasar spectra resulted in a sample of 12
\CIV\ systems which contain at least one component with \ZCgt. The
total number of Voigt 
profile components with \ZCgt, which we will refer to as ``high-metallicity
clouds'', is 28. 
Columns 5--7 of Table~\ref{tbl:oblist} list, for each quasar, the
number of high metallicity systems, their redshifts and their
number of high-metallicity \CIV\ components, respectively. 

We define
a system as a set of \CIV\ components which span a redshift range
$\Delta z < 0.01$. Of the 12 systems, 5 are consistent
with a single-cloud structure, but this number reduces to 3 if all
detected \CIV\ components belonging to these systems are taken into
account, rather than only those with inferred \ZCgt\ (the total number
of detected \CIV\ components is shown in col.~9). Column (8) of
Table~\ref{tbl:oblist} contains, for each system, the velocity
range spanned by the metal-rich clouds and column (10) shows the
corresponding value if all \CIV\ components are included.
The maximum velocity difference between the metal-rich (all) \CIV\
clouds is $699~\kms$ ($718~\kms$), while 8 (5) out of the 12 systems span
less than $30~\kms$.  

The low rate of incidence of high-metallicity clouds
ensures that the 
probability of a chance superposition on the velocity scales typical of the
absorption systems ($\Delta(v) \la 10^2~\kms$ which corresponds to
$\Delta(z) \la 10^{-3}$) is negligibly small. Hence, there is little
doubt that the components of a system are physically related.

The median and mean system redshifts are 2.28 and
2.25, respectively. The median redshift searched, i.e., the redshift
below which half of the redshift path is located, is 2.24. The total
redshift path searched is $\Delta z = 4.05$, which yields a rate of
incidence $d{\cal N}/dz$ of $3.0\pm 0.9$ for \ZCgt\ systems and 
$7\pm 1$ for the individual \ZCgt\ clouds\footnote{For
$(\Omega_m,\Omega_\Lambda)=(0.3,0.7)$ and $z=2.3$ these rates correspond to
number densities per unit absorption distance $d{\cal N}/dX$ of
$0.9\pm 0.3$ and $2.2\pm 0.4$ for systems and clouds, respectively.}. 
Because of the conservative
nature of our selection criteria, these rates are likely to be
underestimates. 

Figures
\ref{fig:0122_ulz2p062} -- \ref{fig:2126_ulz2p678} show the spectral regions of
interest for the 12 \ZCgt\ absorption systems and Table~\ref{tbl:vpfits}
lists the Voigt profile fits to the \CIV\ systems as well as the upper
on the column densities of other ions.

\begin{figure*}
\resizebox{0.33\textwidth}{!}{\includegraphics{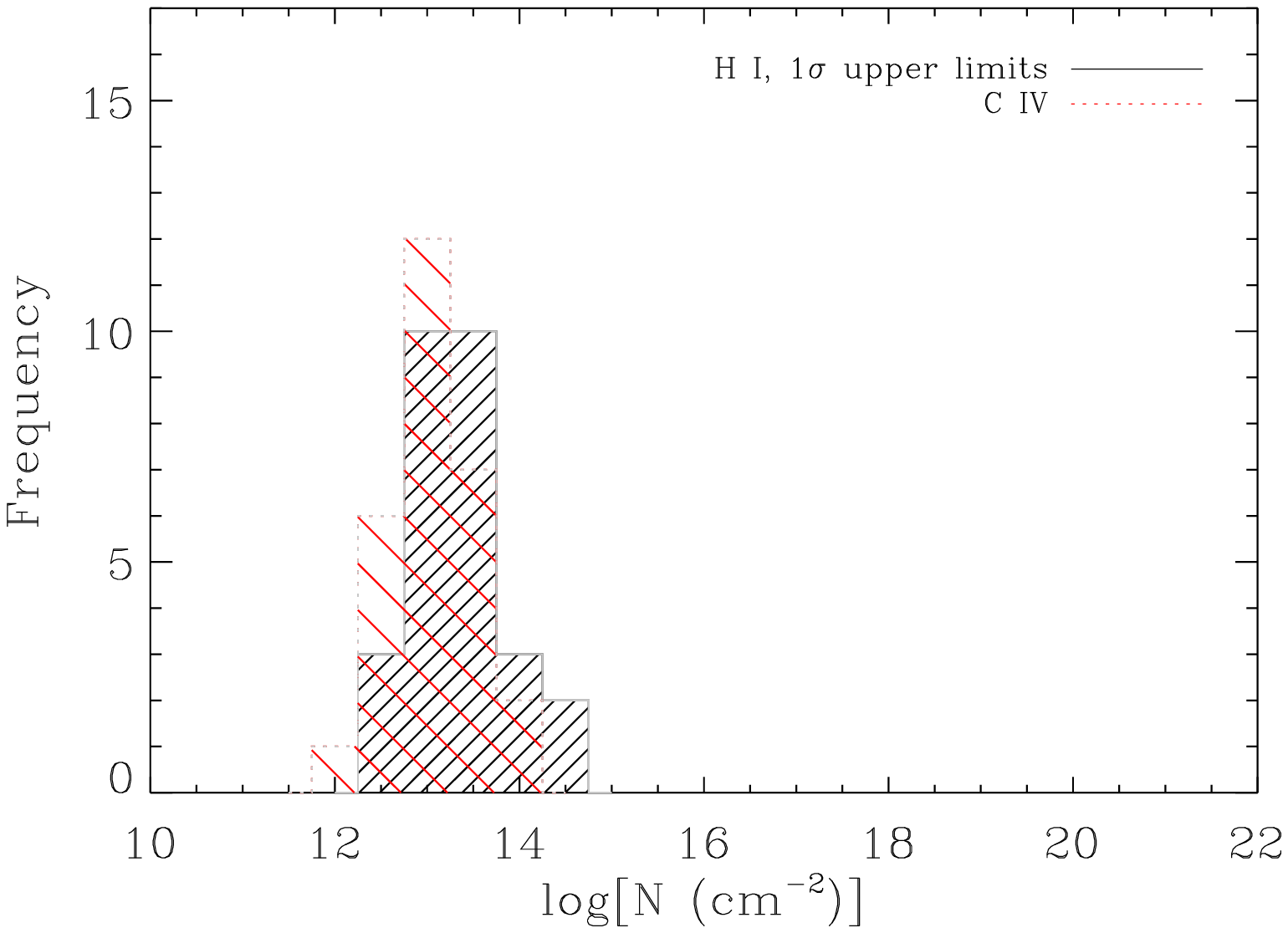}}
\resizebox{0.33\textwidth}{!}{\includegraphics{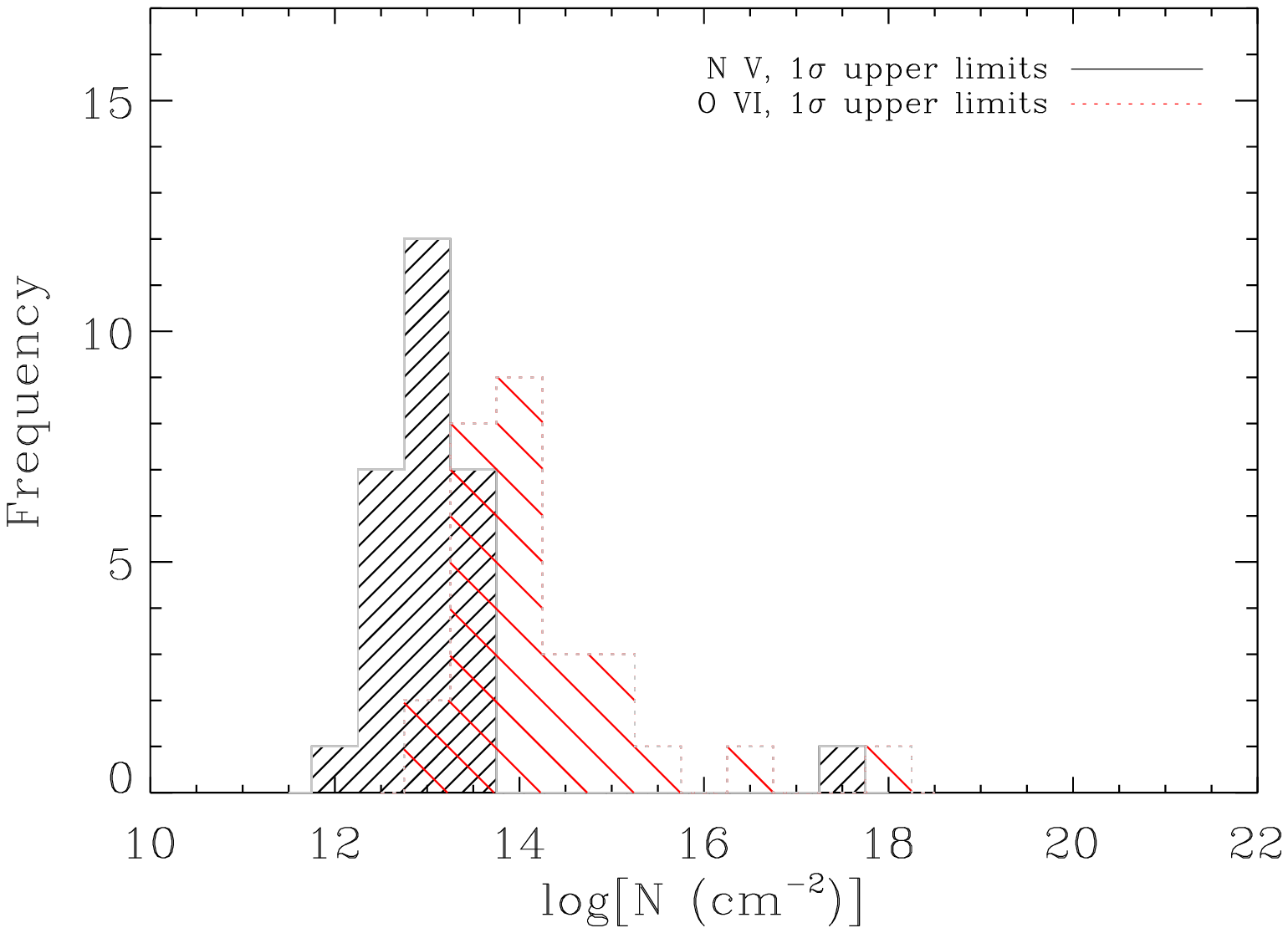}}
\resizebox{0.33\textwidth}{!}{\includegraphics{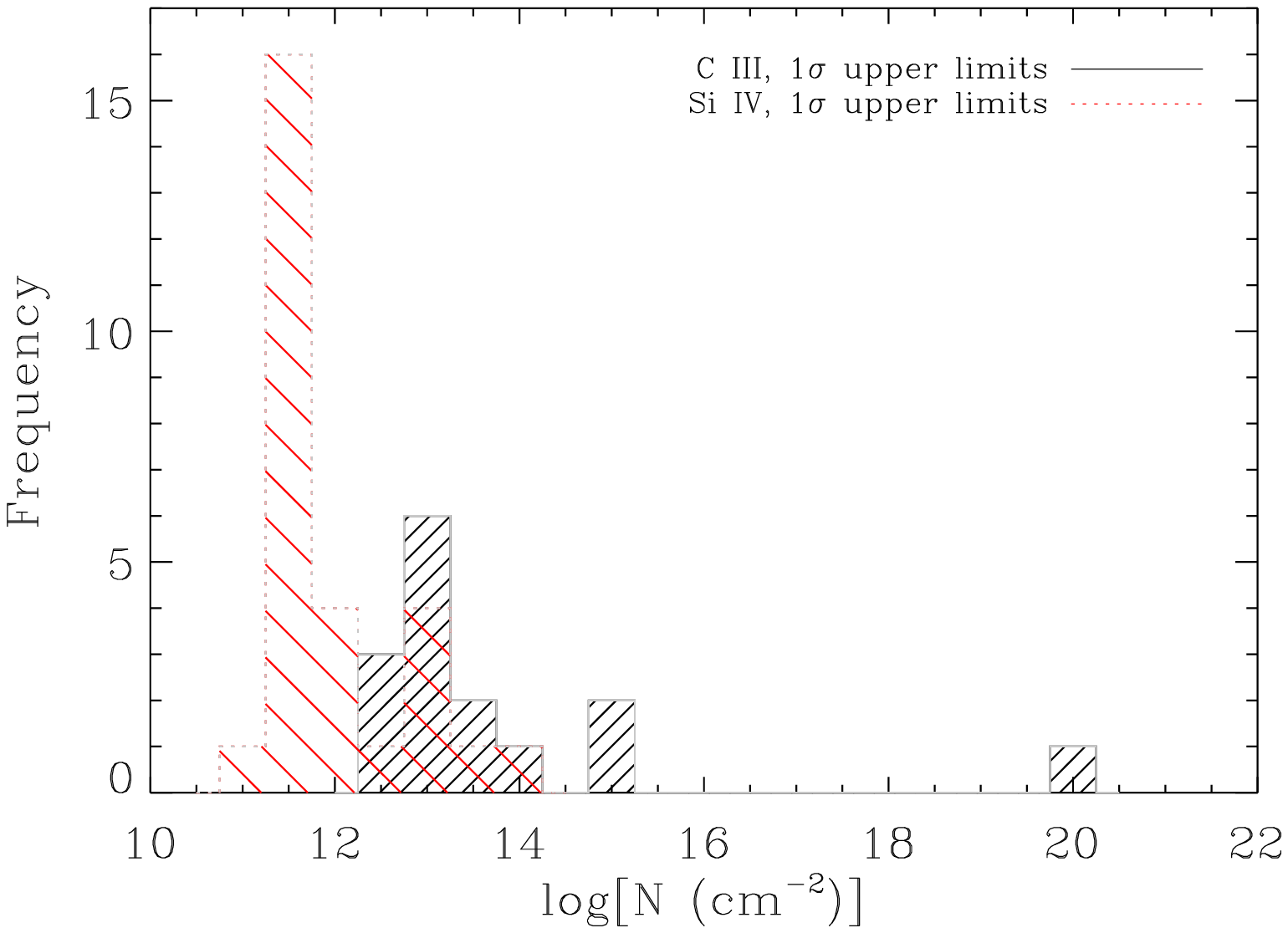}}
\caption{Histograms showing the column density distribution of HI
  (\emph{left, solid}), \CIV\ (\emph{left, dotted}), \NV\ (\emph{middle,
  solid}), \OVI\ (\emph{middle, dotted}), \CIII\ (\emph{right, solid}),
  and \SiIV\ (\emph{right, dotted}) for the high-metallicity clouds
 listed in Table~\protect\ref{tbl:vpfits}. Except for \CIV, all column
  densities are 1$\sigma$ upper limits.
\label{fig:histograms}}
\end{figure*}

Our spectral coverage enabled us to obtain $1\,\sigma$ upper limits
on the column densities of \HI, \NV, \OVI, and \SiIV\ for all the
clouds. For 15 of the 28 clouds we were also able to measure
upper limits on the \CIII\ column
densities. Figure~\ref{fig:histograms} shows the 
distributions of \HI\ (\emph{left, solid}), \CIV\ (\emph{left, dotted}),
\NV\ (\emph{middle, solid}), \OVI\ (\emph{middle, dotted}), \CIII\
(\emph{right, solid}), and \SiIV\ (\emph{right, dotted}) column
densities of the high-metallicity clouds. The clouds in our
sample span $\log [N_{\rm CIV}~(\cm^{-2})] = 11.8$ -- 13.9 (median
13.0) and the median upper limits on the column densities of the other ions
are 13.3 for \HI, 12.9 for \NV, 14.0 for \OVI, 12.7 for \CIII, and
11.7 for \SiIV. 

It is interesting to compare these values to the ones obtained for the
complete sample of \CIV\ absorbers. The median \CIV\ column for that
sample is 0.5 dex lower than for the metal-rich sample, while the
median upper limit on the column of \HI\ is 1.2 
dex higher. As expected from our selection criteria, the lower limits
on the \CIV/\HI\ ratio are much higher for the metal-rich
sample. Interestingly, the median upper limits on the \NV\ and \OVI\
column densities
are only lower by 0.3 dex for the complete sample, which suggests that
contamination and/or noise are important for these ions. For \CIII\
and \SiIV\ the situation is even worse: the median upper limits on the
column densities are lower for the metal-rich sample by 0.3 dex.

A worry for any quasar absorption line study, and particularly for
those focusing on rare systems, is that some of the absorbers may be
associated with 
(ejected by) the quasar rather than intervening. This, however,
appears not to be the case for the systems studied here. Ejected
systems tend to be very broad, show evidence
for partial coverage of the emission source, and have 
a preference for redshifts close to that of the quasar (e.g., Hamann
et al.\ 1997). The metal-rich clouds, on the other hand, 
are narrow (half of the \CIV\ lines have $b < 10~\kms$, see
Table~\ref{tbl:vpfits}, col.~11), the doublet ratios show no evidence for
partial coverage, and there is no evidence for an excess of absorbers
near the redshift of the quasar: the median and mean values of the
observable $(z_{\rm abs}-z_{\rm min})/(z_{\rm max}-z_{\rm min})$ are
0.26 and 0.38, respectively, for the metal-rich systems. The
Kolmogorov-Smirnov test gives a 
probability of 0.14 that the observable $(z_{\rm abs}-z_{\rm
min})/(z_{\rm max}-z_{\rm min})$ is drawn from a uniform distribution
between 0 and 1. Hence, there is a mild tendency for the absorbers to
reside in the low-redshift half of the redshift path searched. This
can be explained as a result of the lower contamination of \NV\ and
\HI, but is opposite to what one would expect for ejected systems.
Most of the absorbers are, in fact, at large separations from
the quasars: the median and mean velocity
separations are both $3.5\times 10^4~\kms$ and the velocity
difference is greater than $6.2\times 10^3~\kms$ in all cases. Thus, it is
very likely that the metal-rich absorbers are intervening gas clouds. 

Recently, Misawa et al.\ (2007) claimed that up to 17\% of narrow
absorption line \CIV\ systems that are blueshifted by at least
5,000~km/s may show evidence for partial covering. If 17\% of our
systems were intrinsic, even though none of them show evidence for
partial covering, our lower limit on the rate of incidence of
high-metallicity \CIV\ systems would have to be decreased by 17\%, but
this would not affect any of our conclusions.

As discussed in \S\ref{sec:observations}, regions close to the quasar
were excluded to avoid proximity effects. If we had not excluded the
regions between $z_{\rm em}$ and $z_{\rm max}$, then we would have
selected two more high-metallicity systems (both towards
HE2347$-$4342). The rate of incidence of \ZCgt\ systems in the
proximity zones is $d{\cal 
  N}/dz = 3 \pm 2$, which is consistent with the $3.0\pm 0.9$
found for the fiducial sample.  

Column 3 of Table~\ref{tbl:prob} lists for each metal-rich absorption
system the velocity separation from the nearest \HI\ \lya\ line with
$N_{\rm HI} > 10^{14.5}~\cm^{-2}$. For components in a complex the
system with the smallest separation from the nearest high column
density system has been chosen. The last column of the table gives the
probability that the velocity separation is smaller than observed if
the systems were positioned randomly between $z_{\rm min}$ and $z_{\rm
max}$. The data confirm the visual impression from figures
\ref{fig:0122_ulz2p062} -- \ref{fig:2126_ulz2p678}: 6 absorption
systems are probably associated with strong \HI\ lines, 1 is
possibly so ($z=2.352$ in Q$0329-385$, $z=2.275$ in HE$2347-4342$),
and another 4 are not.

\begin{table} 
\begin{minipage}{80mm}
\centering
\caption{Relation to high $N_{\rm HI}$
    absorbers
\label{tbl:prob}}
\footnote{Cols.\ (1)  
    and (2) contain the quasar name and the absorber redshift, 
respectively. Col.\ (3) gives the velocity 
 separation from the nearest \lya\ absorber with $N({\rm
   HI})>10^{14.5}~\cm^{-2}$ and col.\ (4) gives the probability that
 the separation is this small or smaller if the metal-rich systems were
 randomly distributed.}
\begin{tabular}{lcrc}
\hline
QSO & $z_{\rm abs}$ & $\Delta v (N_{\rm HI} > 10^{14.5})$ & 
$P(<\Delta v)$\\ 
& & ($\kms$)\\
(1) & (2) & (3) & (4)\\
\hline
Q0122$-$380   & 2.063 & 1910 & 0.564\\
Pks0237$-$233 & 2.042 &   59 & 0.019\\
HE1122$-$1648 & 2.030 &  300 & 0.140\\ 
Q0329$-$385   & 2.076 & 6785 & 0.800\\
Q0329$-$385   & 2.352 & 1021 & 0.191\\
HE1347$-$2457 & 2.116 &   89 & 0.038\\
HE0151$-$4326 & 2.417 &   75 & 0.049\\
HE0151$-$4326 & 2.468 & 1406 & 0.542\\
HE2347$-$4342 & 2.120 & 2208 & 0.800\\
HE2347$-$4342 & 2.275 &  485 & 0.250\\
Pks2126$-$158 & 2.394 &   21 & 0.067\\
Pks2126$-$158 & 2.679 &  121 & 0.123\\
\hline
\end{tabular}
\end{minipage}
\end{table}

In summary, the \ZCgt\ systems (clouds) have a rate of incidence
$d{\cal N}/dz > 3.0\pm 0.9$ ($7\pm 1$) at a median
redshift of $z_{\rm abs}=2.3$. They are a mixture of single cloud and
multiple clouds systems. Some reside in the vicinity of $N_{\rm HI} >
10^{14.5}~\cm^{-2}$-systems, but many do not. The main difference with
ordinary \CIV\ systems is that the metal-rich clouds have higher \CIV/\HI\
ratios (by selection). Their distribution along the line of sight,
their line widths, and their doublet ratios all suggest that they are
intervening systems, rather than ejected by the quasar.

\section{Ionization modeling}
\label{sec:ionmodeling}

In this section we will use ionization models to infer the physical
characteristics of the high-metallicity clouds listed in
Table~\ref{tbl:vpfits}. 

Assuming
photoionization and a fixed temperature, the column
density ratios of the different ions depend only on the ambient
radiation field, the gas density, and the relative 
abundances of the elements. In fact, for a fixed spectral shape of the
UV/X-ray radiation, the ionization balance depends only on the ratio
of the intensity to the gas density. It is therefore useful to
parametrize the intensity by the ``ionization parameter'', which
is defined as,
\begin{equation}
U \equiv {\Phi_{\rm H} \over n_{\rm H} c},
\end{equation}
where $\Phi_{\rm H}$ is the flux of hydrogen ionizing photons
(i.e., photons per unit area and time), $n_{\rm H}$ is the total
hydrogen number density, and $c$ is the speed of light. For our fiducial
radiation field, which we will take to be the 
Haardt \& Madau (2001) model of the $z=2.3$ UV/X-ray background from
quasars and galaxies, the ionization parameter is related to the gas
density by
\begin{equation}
\log U \approx -4.70 - \log [n_{\rm H} (\cm^{-3})].
\label{eq:ionpar}
\end{equation}
Although we will generally quote densities rather than
ionization parameters, it is important to keep in mind that only the
ionization parameter is measurable from the observations presented
here.

\begin{figure*}
\resizebox{\colwidth}{!}{\includegraphics{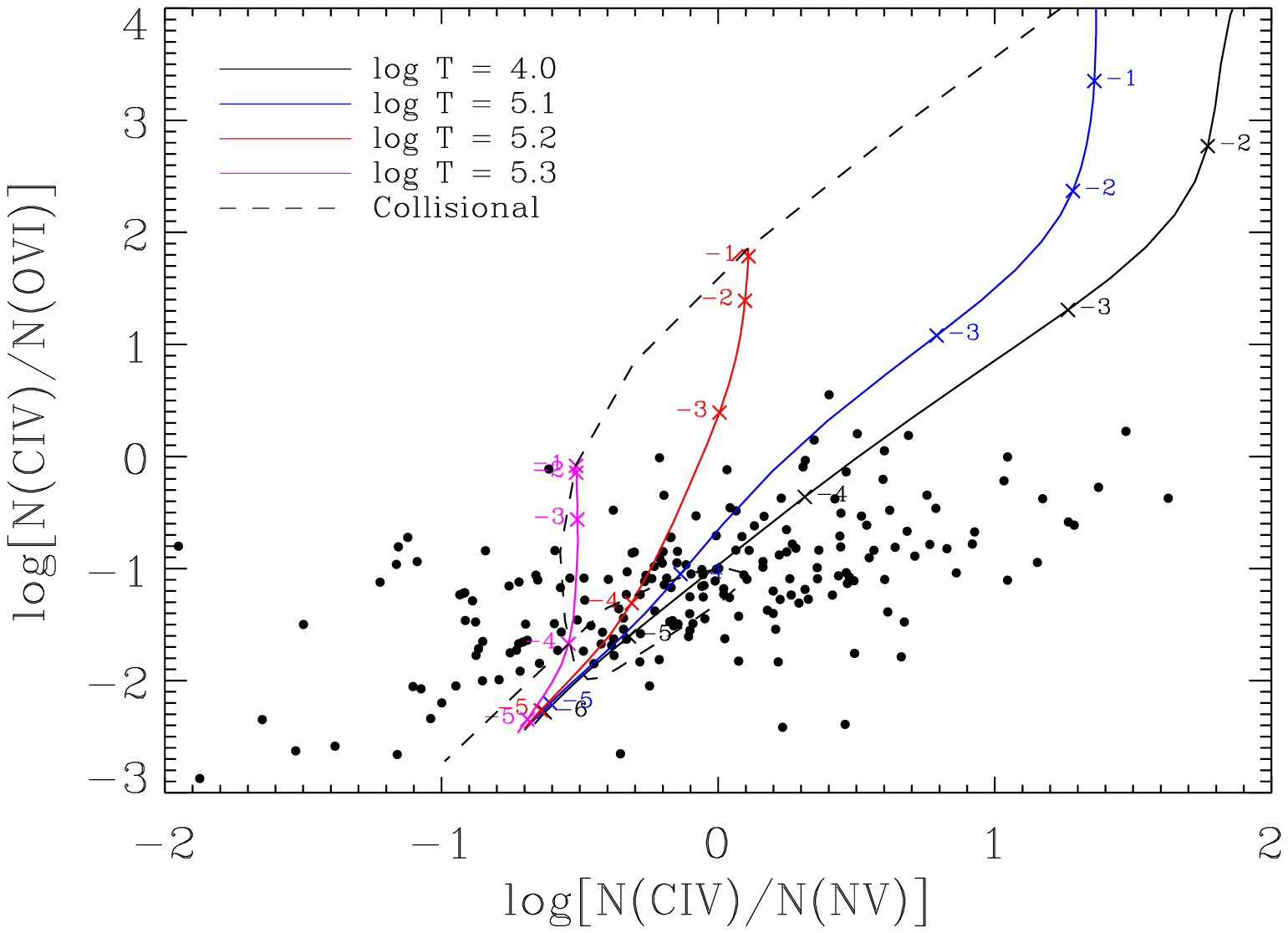}}
\resizebox{\colwidth}{!}{\includegraphics{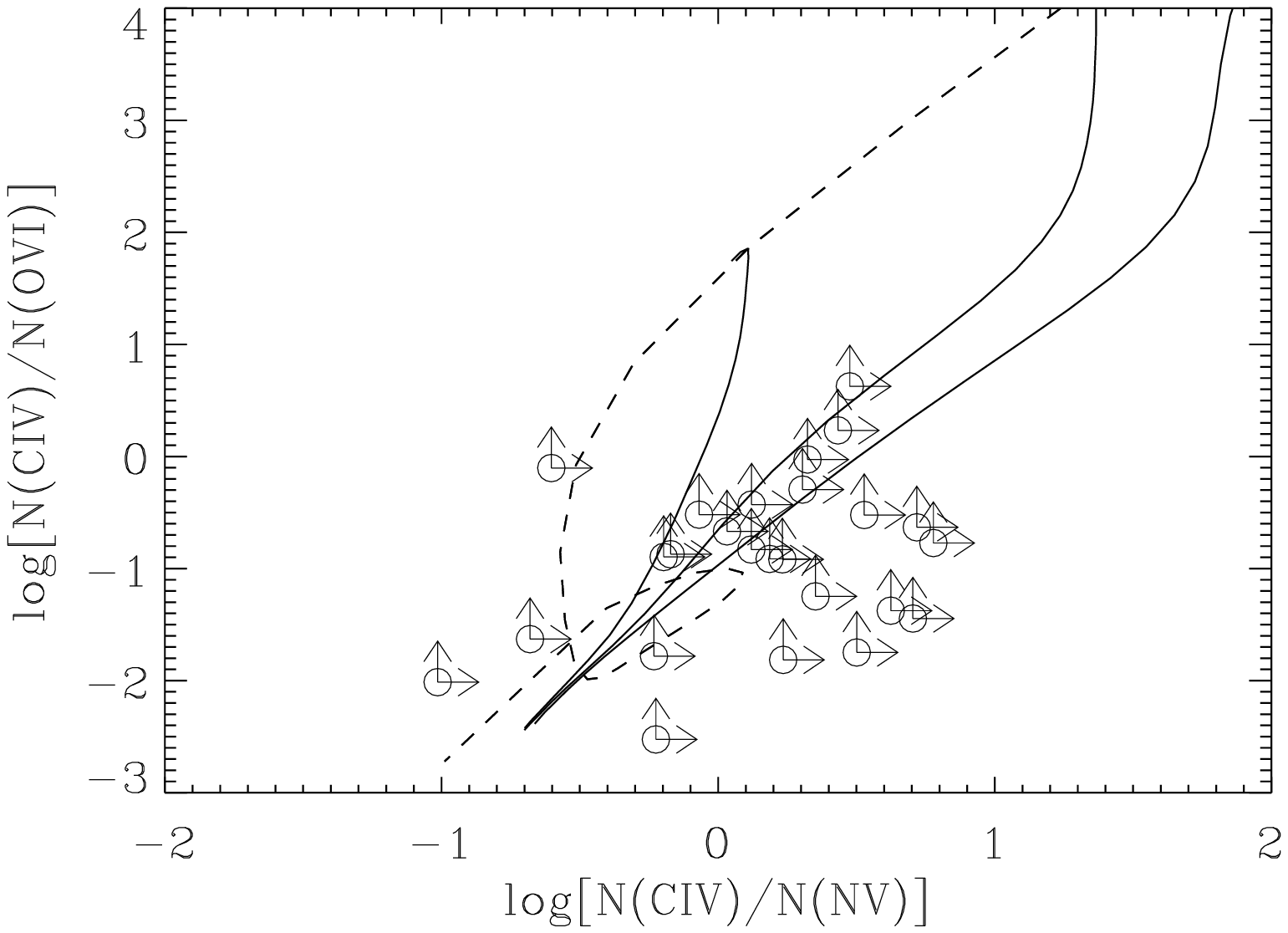}}
\caption{Measured $1\sigma$ lower limit on the column density ratio
$N({\rm CIV})/N({\rm OVI})$ versus the $1\sigma$ lower limit on the
  ratio $N({\rm CIV})/N({\rm NV})$ for all \CIV\ absorbers
  (\emph{left}) and for the metal-rich components only
  (\emph{right}). 
The curves indicate the predicted ratios 
assuming solar abundances and the $z=2.3$ Haardt \& Madau (2001) model
of the UV/X-ray background from quasars and galaxies.  Density varies
along the solid curves (values of $\log[n_{\rm H} (\cm^{-3})]$ are
indicated), but 
the temperature is kept fixed at 
(\emph{right-to-left}) $\log[T (\K)] = 4.0$,
5.1, 5.2, and 5.3, respectively. Temperature varies along the dashed
curve,
which assumes pure collisional ionization equilibrium (and is therefore
independent of the density). Parts of curves for which $n({\rm
  CIV})/n({\rm C}) < 10^{-6}$ are not plotted. Three data points are
missing from the 
plot on the right: one to the left and two below the plotted range,
many more are missing from the panel on the left (compare with
Fig.~\protect\ref{fig:selection}). In the left panel the arrows have
been suppressed for clarity.}  
\label{fig:colratios}
\end{figure*}

Since the \OVI\ fraction peaks at a higher ionization parameter (i.e.,
a lower density for a fixed radiation field) than
the \NV\ fraction, which in turn peaks at a higher ionization
parameter than the 
\CIV\ fraction, we can use the two independent ratios that can be
formed from these three column densities as indicators of the degree to which
the systems are ionized. For example, if the relative abundances of the
elements are approximately constant, more highly ionized clouds should
have lower \CIV/\OVI\ and \CIV/\NV. 

The solid curves in Figure~\ref{fig:colratios} indicate the expected
column density ratios for our fiducial radiation background, solar
abundance ratios, and temperatures of, from right-to-left, $\log T =
4.0$, 5.1, 5.2, and 5.3. For temperatures below $10^5~\K$ the results
are very close to those for $10^4~\K$. The dashed curve shows the
results for pure collisional ionization equilibrium. Density varies
along the solid curves (some values are indicated), while temperature
varies along the dashed curve.

The left panel shows all detected \CIV\ components while the right
panel shows only the high-metallicity sample (in the left panel arrows
have been suppressed for clarity). The points in the left
panel roughly follow the trend $\log[N_{\rm CIV}/N_{\rm OVI}]
\propto \log[N_{\rm CIV}/N_{\rm NV}]$, although with a large
scatter, as would be expected if the measured upper limits on $N_{\rm
OVI}$ and $N_{\rm NV}$ are dominated by contamination and are therefore 
independent of $N_{\rm CIV}$. This suggests that the
measurements are conservative lower limits, which may be significantly
lower than the true values. 

There are no data points with lower limits
on $N_{\rm CIV}/N_{\rm OVI}$ much greater than unity, which is no
surprise. Our high-metallicity clouds typically have
$N_{\rm CIV}\la 10^{13}$ which means that such points would need to
have $N_{\rm OVI} \ll 10^{13}$ which is much lower than the typical
contamination in the \OVI\ region. We would also not
expect this regime to be populated by \CIV\ absorbers because it
corresponds to densities beyond the peak in the \CIV\ fraction. 

If the absorbers are photo-ionized, then we can obtain a lower limit
on the density by shifting the points either upwards (which would
imply \OVI\ is contaminated but \NV\ is not) or to the right (implying
\NV\ is contaminated but \OVI\ is not),
until they fall on top of the rightmost solid curve which corresponds
to $\log T = 4.0$. For the metal-rich clouds the median, required
shift in the column density ratio is a modest 0.42 dex. However, if
both $N_{\rm OVI}$ and $N_{\rm NV}$ are contaminated, then the
required shifts will be greater. In fact, even in the absence of
contamination this should be the case because the presence of noise
implies that our $1\sigma$ upper limits on the \NV\ and
\OVI\ column densities typically exceed the true values.
The resulting lower limit on the density (upper limit on the
ionization parameter) is very robust, because taking a 
temperature high enough for collisional ionization to become
important yields higher densities (to compensate for the enhanced
ionization).  

\subsection{Method}
\label{sec:method}

We use the following procedure to constrain the physical
parameters of the absorbers. 

First, we make several assumptions: (1)
the absorbers are exposed to the $z=2.3$ Haardt
\& Madau (2001) model for the meta-galactic radiation field; (2) the
relative abundances of the heavy elements are solar; (3) the gas is in
ionization equilibrium; (4) the gas temperature is $\log[T (\K)] = 4.0$. In
\S\ref{sec:uncertainties} we will discuss both the validity of these
assumptions and the effects of relaxing them. 

\begin{figure}
\resizebox{\colwidth}{!}{\includegraphics{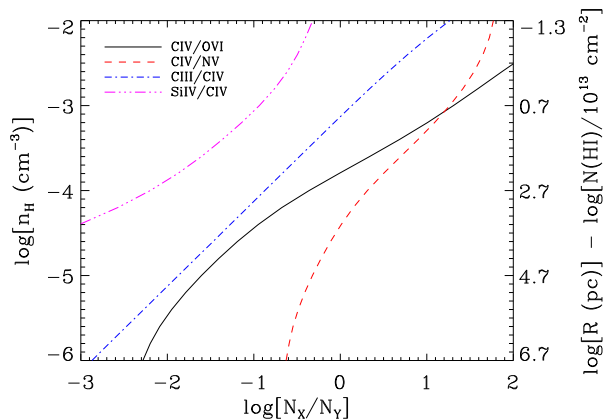}}
\caption{Hydrogen number density as a function of column
density ratio for CIV/OVI (\emph{solid}), CIV/NV (\emph{dashed}), CIII/CIV
(\emph{dot-dashed}), and SiIV/CIV (\emph{dot-dot-dot-dashed}),
respectively. The $y$-axis on the right-hand side 
indicates the corresponding cloud radius for $N_{\rm HI} =
10^{13}~\cm^{-2}$ (the inferred radius is proportional to the measured
HI column density).}
\label{fig:pdenssize}
\end{figure}

Second, we use our measurements of $N_{\rm CIV}$ and our upper limits
on $N_{\rm NV}$ and $N_{\rm OVI}$ to obtain lower limits on
two column density ratios (\CIV/\NV\ and \CIV/\OVI), which provide us
with two lower limits on the density. The greater of these two lower
limits is then our best estimate of the lower limit on the gas
density. Figure~\ref{fig:pdenssize} shows how the inferred density
varies as a function of these and other column density ratios.

Third, we obtain an upper limit on the size of the absorbers. The
absorber radius is given by
\begin{equation}
R = {1 \over 2}{N_{\rm HI} \over n_{\rm HI}}.
\label{eq:size}
\end{equation}
Since the neutral hydrogen fraction increases monotonically with
density, our lower limit on the density gives us also a lower limit on
$n_{\rm HI}$, the density of neutral hydrogen. Together with our
measured upper limit on $N_{\rm HI}$, we thus obtain an upper limit on
$R$. Figure~\ref{fig:pdenssize} (right $y$-axis) shows how the
inferred radius varies as a function of the column density ratios.

Fourth, we obtain an upper limit on the abundance of silicon relative
to carbon. Analogously to equation (\ref{eq:metallicity}), we can write
this relative abundance as
\begin{eqnarray}
[{\rm Si}/{\rm C}] &=&
\log \left ({N_{\rm SiIV} \over N_{\rm CIV}}\right ) 
-\log \left ( n_{\rm SiIV}/n_{\rm Si} \over {n_{\rm CIV}/n_{\rm C}
  } \right ) - \log\left ({n_{\rm Si} \over
     n_{\rm C}}\right )_\odot \nonumber \\
&=& \log \left ({N_{\rm SiIV} \over N_{\rm CIV}}\right ) - 
\log \left ( {N_{\rm SiIV} \over N_{\rm CIV}}\right )_{\odot,n_{\rm H}}
\label{eq:SiCrel},
\end{eqnarray}
where the last term denotes the predicted column density ratio $N_{\rm
SiIV}/N_{\rm CIV}$ for solar abundances and the density, $n_{\rm H}$,
appropriate for the particular absorber. In other words, if we know the
gas density, then \SiC\ is just the difference between the measured
value of $\log(N_{\rm SiIV}/N_{\rm CIV})$ and the value predicted for
solar abundances. Since we have measured an upper limit on $N_{\rm
SiIV}$, we can obtain an upper limit on \SiC\ provided we have a lower
limit on $(N_{\rm SiIV}/N_{\rm CIV})_{\odot,n_{\rm H}}$ [or,
equivalently, an upper limit on $(n_{\rm CIV}/n_{\rm C})/(n_{\rm
 SiIV}/n_{\rm Si})$]. As the dot-dot-dot-dashed curve in
figure~\ref{fig:pdenssize} shows, our lower limit on 
the density gives us precisely this. 

Finally, for those absorbers for which \CIII\ falls in the wavelength
range covered by our spectra, we perform a consistency check using our
upper limit on $N_{\rm CIII}$. Since \CIII/\CIV\ is a monotonically
increasing function of the density (see Fig.~\ref{fig:pdenssize},
\emph{dot-dashed} curve), our upper limit on $N_{\rm CIII}/N_{\rm
  CIV}$ translates into an upper limit on the density which
does not rely on any assumption about relative abundances. For each
absorber we check whether the upper limit derived from \CIII\
is greater than the lower limit derived from \NV\ and \OVI. Naturally,
the actual values for the upper limits are also of great interest.

\subsection{Results}
\label{sec:results}

Table~\ref{tbl:properties} gives the results of the ionization models
for all 28 high-metallicity clouds.

\begin{figure}
\resizebox{\colwidth}{!}{\includegraphics{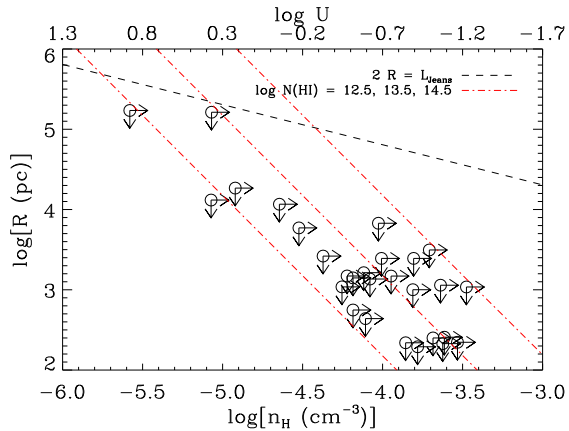}}
\caption{Inferred $1\sigma$ upper limits on the radius as a
  function of the inferred $1\sigma$ lower limits on the gas density
  (bottom $x$-axis) and ionization parameter (top 
  $x$-axis) for  all absorbers listed in
  Table~\protect\ref{tbl:vpfits}. The median upper limit on the
  radius is $\log[R (\pc)] < 3.2$ and the median lower limit on the gas
  density is $\log[n_{\rm H} (\cm^{-3})] =
  -4.0$. Note that the points cannot be shifted onto the dashed line,
  which indicates the radius-density relation for self-gravitating gas
  clouds, because the inferred radius decreases faster with increasing
  density than the Jeans scale (see text for details).}
\label{fig:dens-size}
\end{figure}

Figure~\ref{fig:dens-size} shows the upper limits on the size as a
function of the lower limits on the density (upper limits on the
ionization parameter, 
upper $x$-axis). The median upper limit on the radius is $\log[R (\pc)]
=3.2$ and the median lower limit on the gas density is $\log[ n_{\rm
  H} (\cm^{-3})] = -4.0$. Had we only made use of \CIV/\NV\
(\CIV/\OVI), then the median lower limit on the density would have been
-4.2 (-4.4). For comparison, the results for 
the complete sample of \CIV\ absorbers are $\log R < 5.3$ and $\log
n_{\rm H} > -4.4$. Thus, while
the upper limits on the density are similar for ordinary and
metal-rich \CIV\ absorbers, our upper limits to the sizes are two
orders of magnitude smaller for the high-metallicity clouds. 

There appears to be an anti-correlation between the upper limits on
the size and the lower limits on the density. The dot-dashed lines, which
indicate contours of 
$\log [N_{\rm HI} (\cm^{-2})] = 12.5$, 13.5, and 14.5, respectively, show that this
anti-correlation is directly related to the relatively small range in
neutral hydrogen columns covered by our data.

The absence of points in the bottom-left half of the diagram could be
due to a selection effect: such points correspond to neutral hydrogen columns
less than $10^{12.5}~\cm^{-3}$ or a central optical depth of less than
0.07 (for a $b$-value of $35~\kms$ which is the line width
corresponding to $b_{\rm CIV}=10~\kms$ for thermal broadening). The
severe blending of \lya\ lines and the presence of noise in the
data generally do not allow us to put such tight constraints on the
\HI\ column densities. The absence of points in the upper-right part of the
diagram is a bit more difficult to explain. This region corresponds to
$\log N_{\rm HI} > 14.5$. To pass our selection criterion,
such points would need to have very high \CIV\ column densities, which
are rare. However, one selection effect that may play a role here, is
that such \HI\ lines are strongly saturated which often results in
weak constraints on $N_{\rm HI}$ and hence on the metallicity.

As we already discussed, the upper limits on $N_{\rm NV}$ and $N_{\rm
OVI}$, from which the lower limit on the density is derived, are
likely dominated by contamination. Hence, the scatter in the density
may not be real. Since $R\propto n_{\rm HI}^{-1} \propto n_{\rm
  H}^{-2}$ (assuming a highly ionized plasma in photoionization
equilibrium), systematic errors in 
$n_{\rm H}$ would give rise to a spurious $R\propto 
n_{\rm H}^{-2}$ anti-correlation parallel to the contours of constant $N_{\rm
  HI}$, which is close to the observed scaling.

The dashed curve in Figure~\ref{fig:dens-size} shows the Jeans scale 
(for a purely gaseous cloud),
which is the expected size for a self-gravitating cloud and which is
also similar to the size of typical \lya\ forest absorbers (Schaye
2001), 
\begin{equation}
L_J \sim 6.5 \times 10^4~\pc ~\left ({n_{\rm
  H} \over 10^{-4}~\cm^{-3}}\right )^{-1/2} \left ({T \over
  10^4~\K}\right )^{1/2} f_g^{1/2},
\label{eq:LJ}
\end{equation}
where $f_g$ is the fraction of the mass in gas. All of our metal-rich
clouds are constrained to be smaller than the Jeans scale and many by
a large factor. Note that this conclusion cannot be avoided by
shifting the data points to the right, because an increase in the
density would result in an even larger decrease in the size: $R\propto
n_{\rm HI}^{-1}$ which scales with $n_{\rm H}^{-2}$ if the gas is
highly ionized and with $n_{\rm H}^{-1}$ if it is neutral. This dependence
is steeper than that for the Jeans length, which scales as $L_J\propto
n_{\rm H}^{-1/2}$. Thus, increasing the density makes the clouds less
gravitationally bound. 

If the clouds were self-gravitating, then their masses
would need to be strongly dominated by stars and/or dark
matter. Setting $2R=L_J$ and solving for the gas fraction, we obtain a
median upper limit on the gas fraction of $\log f_g < -3.2$. If we do
the same for the complete sample of \CIV\ absorbers, we obtain $\log
f_g < 0.7$ which is consistent with self-gravity for ordinary
gas fractions. Hence, the metal-rich clouds appear to differ from the
general population in that they are either not gravitationally
confined, or have negligible gas fractions.

For 15 of our high-metallicity clouds our spectra cover the \CIII\
region, enabling us to measure an upper limit on \CIII/\CIV\ and thus
on the density. Reassuringly, for all clouds the upper limits are
consistent with the lower limits shown in figure~\ref{fig:dens-size},
which were determined independently from the \CIV/\NV\ and \CIV/\OVI\
ratios. If we repeat the analysis for the full sample of 179 \CIV\
components for
which we can measure upper limits on $N_{\rm CIII}$, we again find
that every upper limit is greater than the corresponding lower
limit on the density (the minimum difference is in that case
0.02~dex). This gives us confidence in the robustness of the
constraints obtained from the ionization models.

The median upper limit is $\log n_{\rm H} = -3.0$, exactly one
order of magnitude above the median lower limit. The minimum difference
between the upper and lower limits is 0.5~dex. This result suggests
that $\log n_{\rm H} \approx -3.5$ is a reasonable estimate for the
typical gas density. Because $R\propto n_{\rm H}^{-2}$, this would mean that
the upper limit on the size should typically be reduced by an order of
magnitude, which would give a median value of $\log R < 2.2$. Note
that this is still an upper limit, insofar as it is based on an
upper limit on $N_{\rm HI}$. Finally, the median upper limit on the gas
fraction would be reduced to $\log f_g < -4.5$.

Assuming the clouds are roughly spherical, we can estimate their
gas masses. Typical values are
\begin{eqnarray}
M_g &\sim & {4\pi \over 3} {n_{\rm H} m_{\rm H} R^3\over  X}, \\
&\sim & 4\times 10^1~\Msun \left ( {n_{\rm H} \over
  10^{-3.5}~\cm^{-3}} \right ) \left ({R \over 10^2~\pc}\right
)^3,\nonumber  
\label{eq:mass}
\end{eqnarray}
where we assumed a hydrogen mass fraction $X=0.75$. We emphasize that
the mass is much more poorly constrained that the size, because
$M_g\propto R^3$.  

\begin{figure}
\resizebox{\colwidth}{!}{\includegraphics{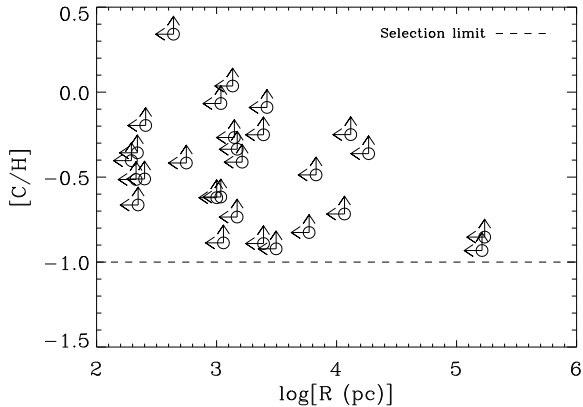}}
\caption{Inferred $1\sigma$ lower limits on the metallicity as a
  function of the inferred $1\sigma$ upper limits on the radius for
  all absorbers listed in Table~\protect\ref{tbl:vpfits}. The median
  lower limit on the metallicity is $[{\rm C}/{\rm H}] > -0.42$ and the
  median upper limit on the radius is $\log[R (\pc)] < 3.2$.} 
\label{fig:size-zmet}
\end{figure}

In Figure~\ref{fig:size-zmet} we plot the metallicity as a function of
the cloud size. By selection, the  
clouds have relatively high metallicities: the median lower limit
is $Z > 0.4 Z_\odot$, which is
much greater than the metallicities typical of the low-density IGM.
The absence of large, metal-rich clouds has the same origin, whatever
it is, as the absence of points in the upper, right part of
figure~\ref{fig:dens-size}. Because the inferred limits on the
metallicity and size scale as $Z\propto N_{\rm HI}^{-1} \propto
R^{-1}$ [see equations (\ref{eq:metallicity}) and (\ref{eq:size})], an
anti-correlation between $Z$ and $R$ is not unexpected. 

\begin{figure}
\resizebox{\colwidth}{!}{\includegraphics{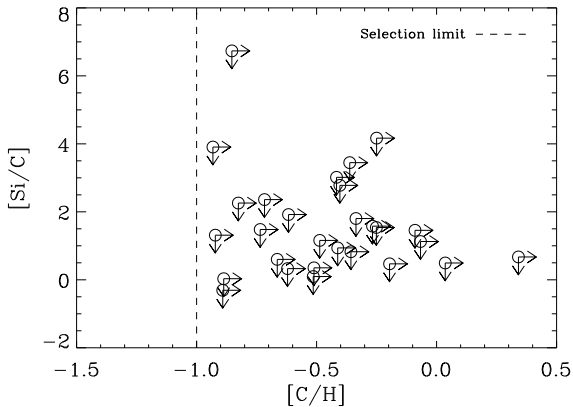}}
\caption{Inferred $1\sigma$ upper limits on the abundance of silicon
  relative to carbon as a 
  function of the inferred $1\sigma$ lower limits on the metallicity
  for all systems listed in 
  Table~\protect\ref{tbl:vpfits}. The median upper limit on the
  relative abundance of silicon is $[{\rm Si}/{\rm C}] < 1.5$ 
  and the median lower
  limit on the metallicity is $[{\rm C}/{\rm H}] > -0.42$.}
\label{fig:zmet-zsirel}
\end{figure}

Figure~\ref{fig:zmet-zsirel} shows that for all but one cloud the
abundance of silicon relative to that of carbon and oxygen is
consistent with the solar value level. The median upper limit is not
particularly interesting: \SiC$<1.5$ which reflects the fact that
we do not expect to detect \SiIV\ in these clouds (they are too
small and too low-density). 

We conclude that our metal-rich ($Z\ga Z_\odot$) clouds are typically
moderately overdense, 
$n_{\rm H} \sim 10^{-3.5}~\cm^{-3}$, which is about 50 times greater
than the mean density at $z=2.25$, and compact, $R\sim
10^2~\pc$. They are too small relative to their densities for them to be
gravitationally confined, unless their stars and/or dark matter
outweigh the gas by more than four orders of magnitude. 

\begin{figure}
\resizebox{\colwidth}{!}{\includegraphics{0122_ulz2p062dxv.ps}}
\caption{The system at $z=2.062560$ in
Q$0122-380$ showing (from bottom-to-top) the \HI\ \lya, \HI\ \lyb,
\CIII, \CIV, \NV, 
\OVI, and \SiIV\ regions on a velocity scale centered on the
redshift given at the bottom of the figure. The profiles corresponding
to the upper limits on the column densities are
shown as dashed lines. An integer offset has been
applied to separate the different transitions. Both lines of the \CIV\ 
doublet are clearly present, and there are blended features which may
contain the \NV\ and \OVI\ lines. \lya\ is weak, but
present.
\label{fig:0122_ulz2p062}}
\end{figure}

\begin{figure}
\resizebox{\colwidth}{!}{\includegraphics{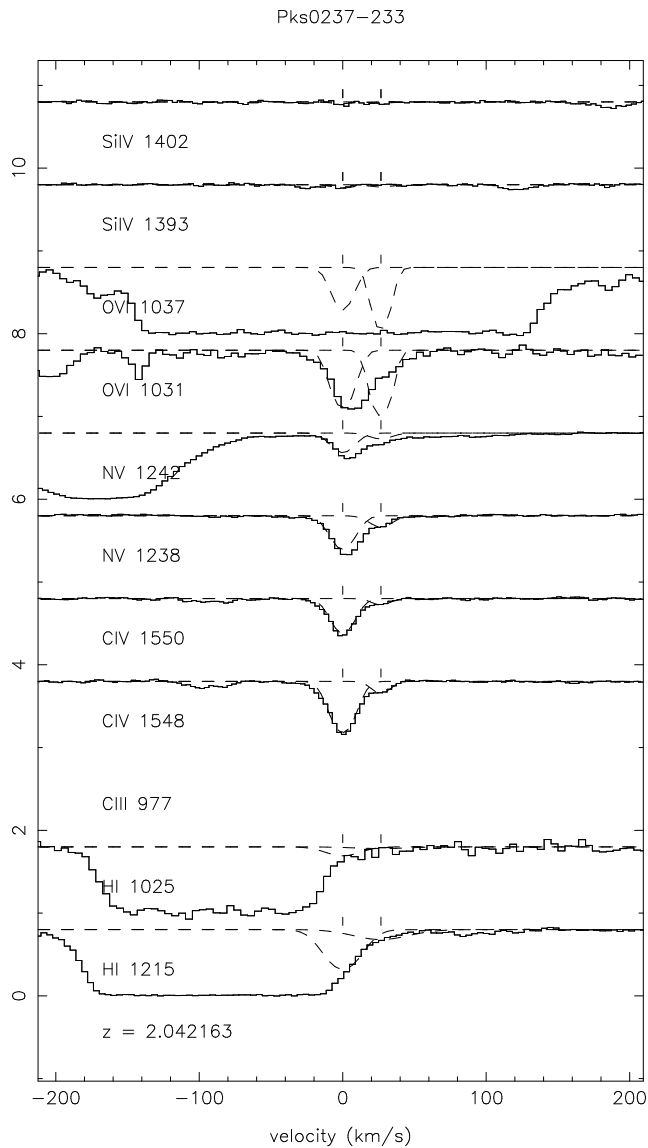}}
\caption{The $z=2.042163$ and 2.042433 components in
  Pks$0237-333$. \CIV\ and \NV\ are clearly present, the weaker
\OVI\ component is strongly contaminated. \lya\ falls in the wing of a
  strong \HI\ system.
\label{fig:0237_ulz2p042}}
\end{figure}

\begin{figure}
\resizebox{\colwidth}{!}{\includegraphics{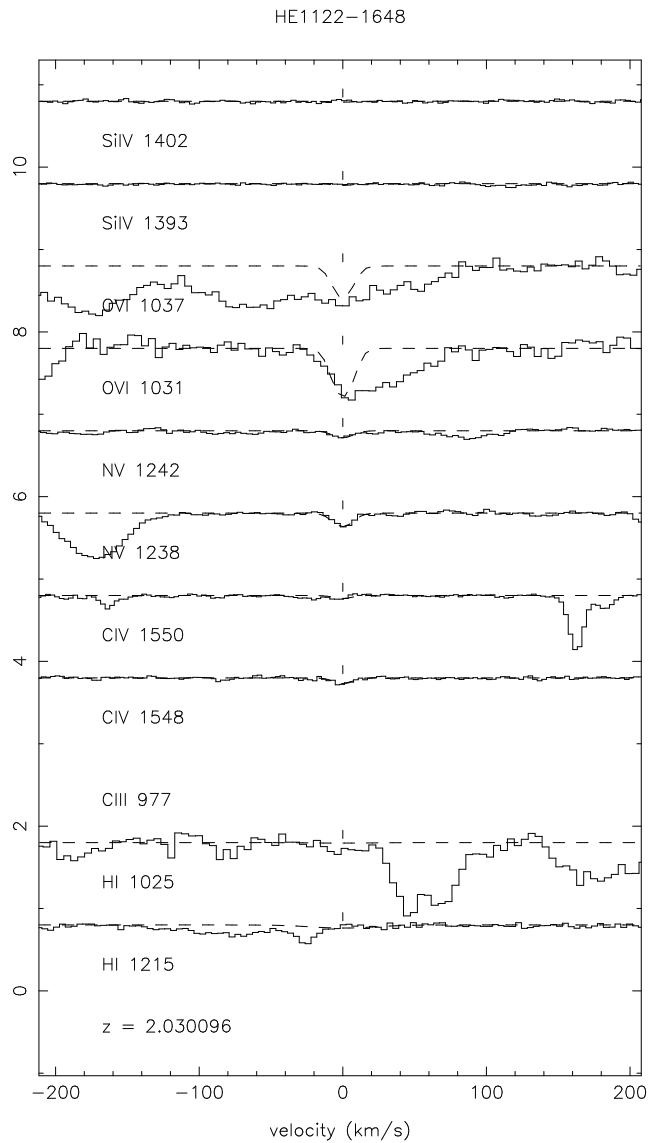}}
\caption{The system at $z=2.030096$ in HE1122$-$1648. This system was
  described and analyzed by Carswell et al.\ 
(2002). \CIV\ and \NV\ are clearly present, \OVI\ is
strongly contaminated, and \HI\ is very weak.
\label{fig:1122_ulz2p030}}
\end{figure}

\begin{figure}
\resizebox{\colwidth}{!}{\includegraphics{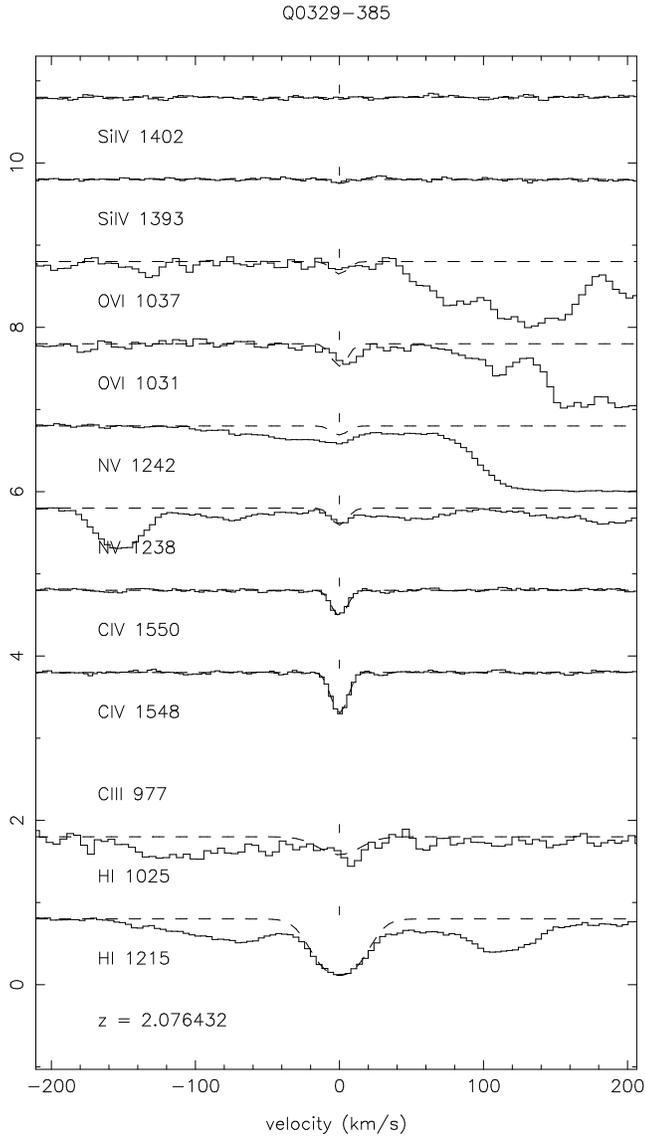}}
\caption{The $z=2.076432$ system in Q$0329-385$. \CIV\ has corresponding
unsaturated \lya\ and probably weak \NV. The \OVI\ doublet is noisy,
but not strongly contaminated.
\label{fig:0329_ulz2p076}}
\end{figure}

\begin{figure}
\resizebox{\colwidth}{!}{\includegraphics{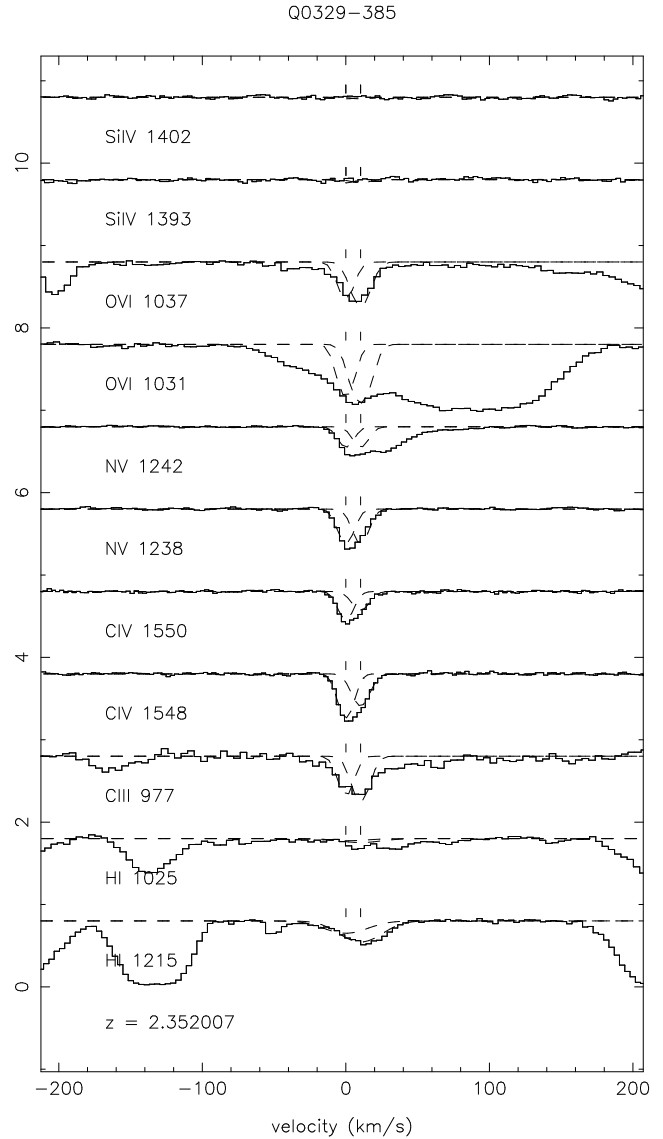}}
\caption{The $z=2.352007$ and 2.352122 components in
Q$0329-385$. The systems show clear \CIV\ and \NV. \OVI\ 1031 is
contaminated. \lya\ is also clearly distinguishable, 
but with the apparent dominant
component offset in velocity from the heavy element lines.
\label{fig:0329_ulz2p352}}
\end{figure}

\begin{figure}
\resizebox{\colwidth}{!}{\includegraphics{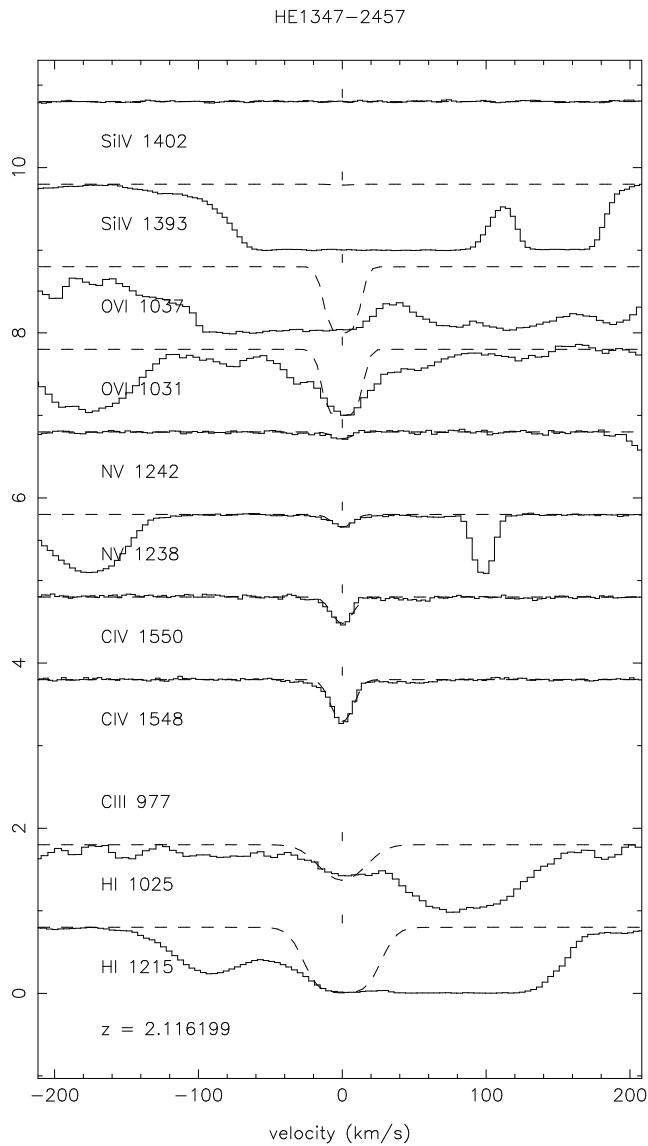}}
\caption{The system at $z=2.116199$ in HE1347$-$2457. \CIV\ and \NV\ are
  present, but \OVI\ is strongly contaminated.  
At $\log N_{\rm HI} < 14.1$, this is one of the strongest \lya\
lines in our sample. There is a stronger \lya\ line
at $+81~\kms$ for which no associated heavy elements are detected.
\label{fig:1347_ulz2p116}}
\end{figure}

\begin{figure}
\resizebox{\colwidth}{!}{\includegraphics{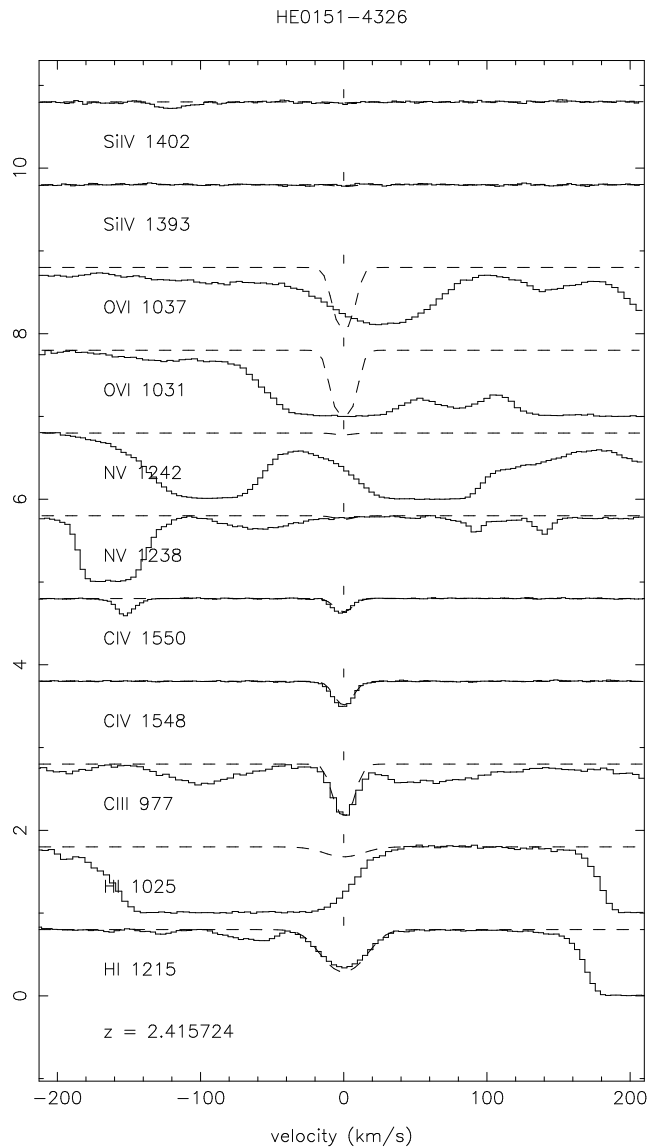}}
\caption{The $z=2.415724$ component of the system at $z=2.417$ in
  HE0151$-$4326. The other component of this system is shown in
  Fig.~\protect\ref{fig:0151_ulz2p419}. \HI, \CIII\ and both \CIV\
  lines are clearly seen, there is a tight upper limit on \NV, but
  \OVI\ is strongly contaminated.
\label{fig:0151_ulz2p415}}
\end{figure}

\begin{figure}
\resizebox{\colwidth}{!}{\includegraphics{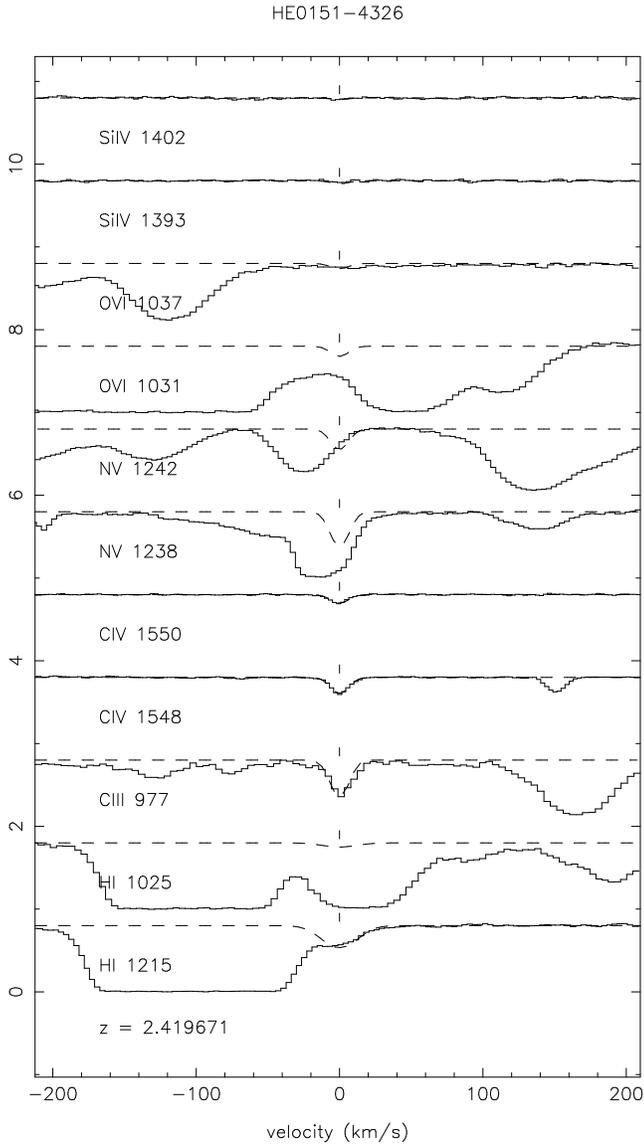}}
\caption{The $z=2.419671$ component of the system at $z=2.417$ in
  HE0151$-$4326. The other component of this system is shown in
  Fig.~\protect\ref{fig:0151_ulz2p415}. \HI\ falls in the wing of a
  strong system. \CIII\ and both \CIV\
  lines are clearly seen. \NV\ is strongly contaminated, but there is
  a tight upper limit on \OVI.
\label{fig:0151_ulz2p419}}
\end{figure}

\begin{figure}
\resizebox{\colwidth}{!}{\includegraphics{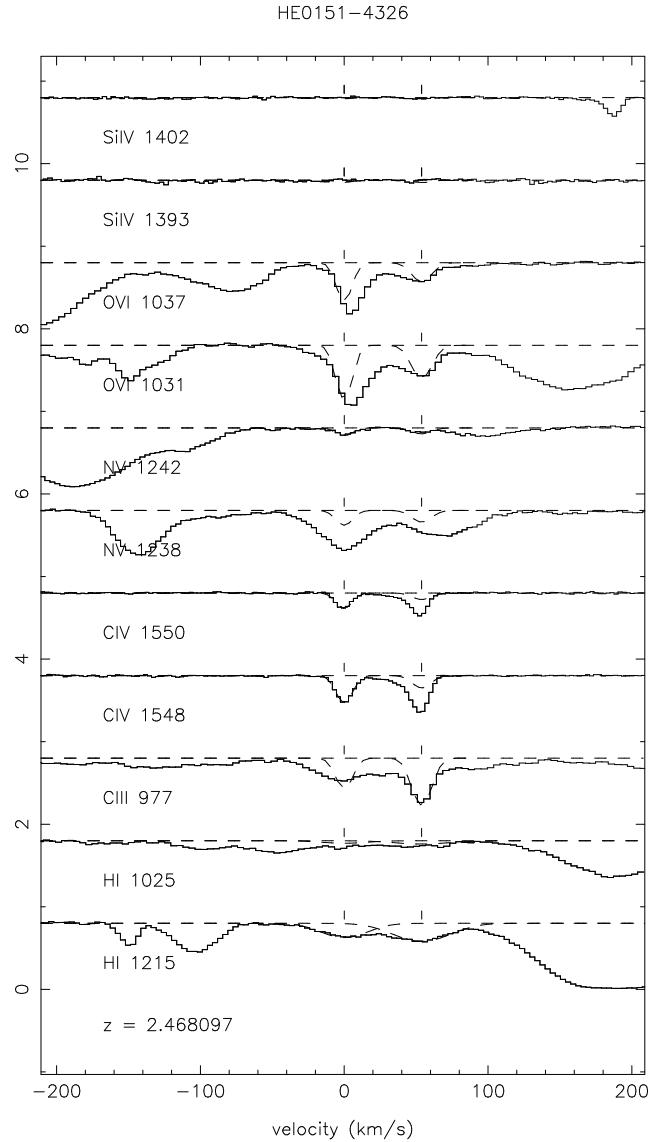}}
\caption{The $z=2.468097$ and 2.468720 components in Q$0151-4326$. 
For \CIV\ and \OVI\ it is clear that these components are blended with
others, which however, did not pass the selection criteria. Both \NV\
and \OVI\ appear to allow measurements of upper limits that are
relatively free from contamination by \HI. The \OVI\
lines are significantly broader than the
corresponding \CIV\ lines.
\label{fig:0151_ulz2p468}}
\end{figure}

\begin{figure}
\resizebox{\colwidth}{!}{\includegraphics{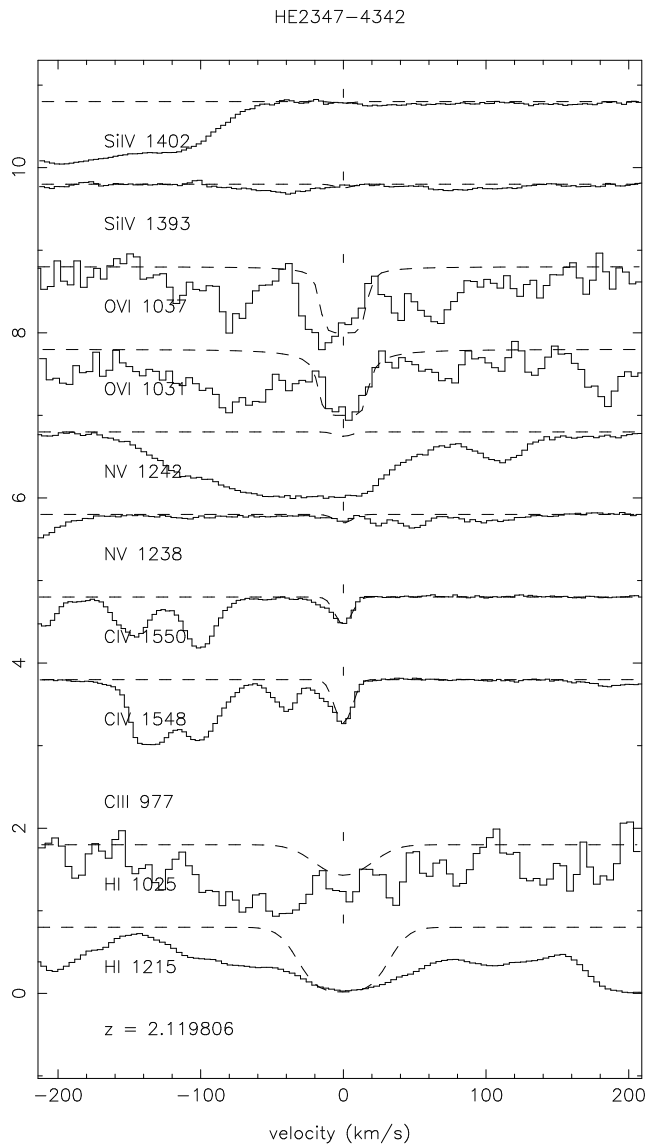}}
\caption{The $z=2.119806$ system at
  in HE2347$-$4342. Most of the absorption just to the left of the
  \CIV\ system is in fact \NV\ at $z=2.8968-2.8984$, in the quasar proximity
  zone. The effective continuum in the \HI\ \lyb\
  and \OVI\ regions was reduced to account for Lyman limit absorption. 
\label{fig:2347_ulz2p119}}
\end{figure}

\begin{figure}
\resizebox{\colwidth}{!}{\includegraphics{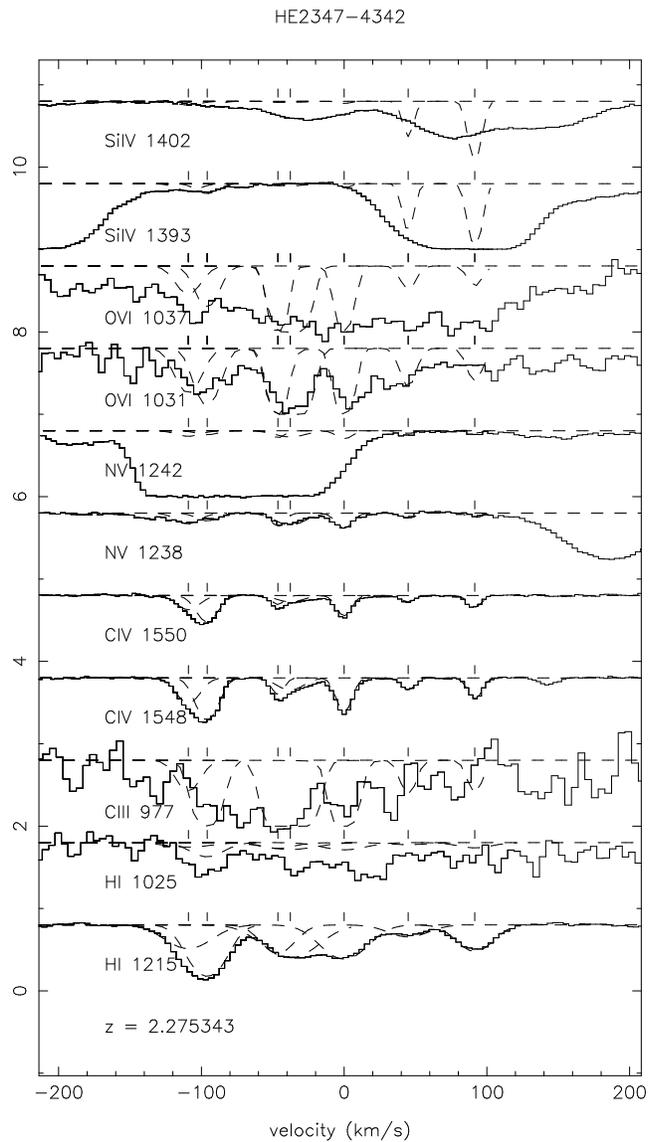}}
\caption{The $z=2.274151$, 2.274295, 2.274836, 2.274931, 2.275343,
  2.275833, and 2.276343 components from the system at 
  $z=2.275$ in HE2347$-$4342. 
 The effective continuum in the \HI\ \lyb, \CIII,
  and \OVI\ regions was reduced to account for Lyman limit
  absorption. Complex absorption by \HI, \CIV, and \NV\ is clearly 
visible.
\label{fig:2347_ulz2p274}}
\end{figure}

\begin{figure}
\resizebox{\colwidth}{!}{\includegraphics{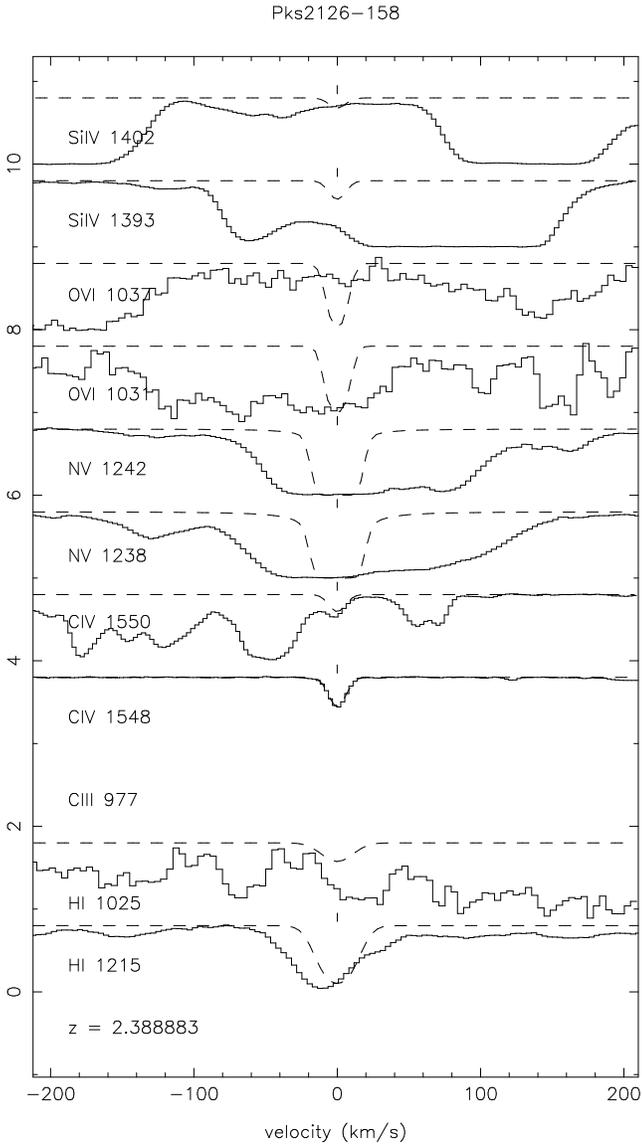}}
\caption{The $z=2.388883$ component of the $z=2.394$ system in
  Pks2126$-$158. The other components of this system are shown in
  Fig.~\protect\ref{fig:2126_ulz2p395}. \HI\ \lyb, \NV, \OVI, and
  \SiIV\ all suffer from strong contamination.
\label{fig:2126_ulz2p388}}
\end{figure}

\begin{figure}
\resizebox{\colwidth}{!}{\includegraphics{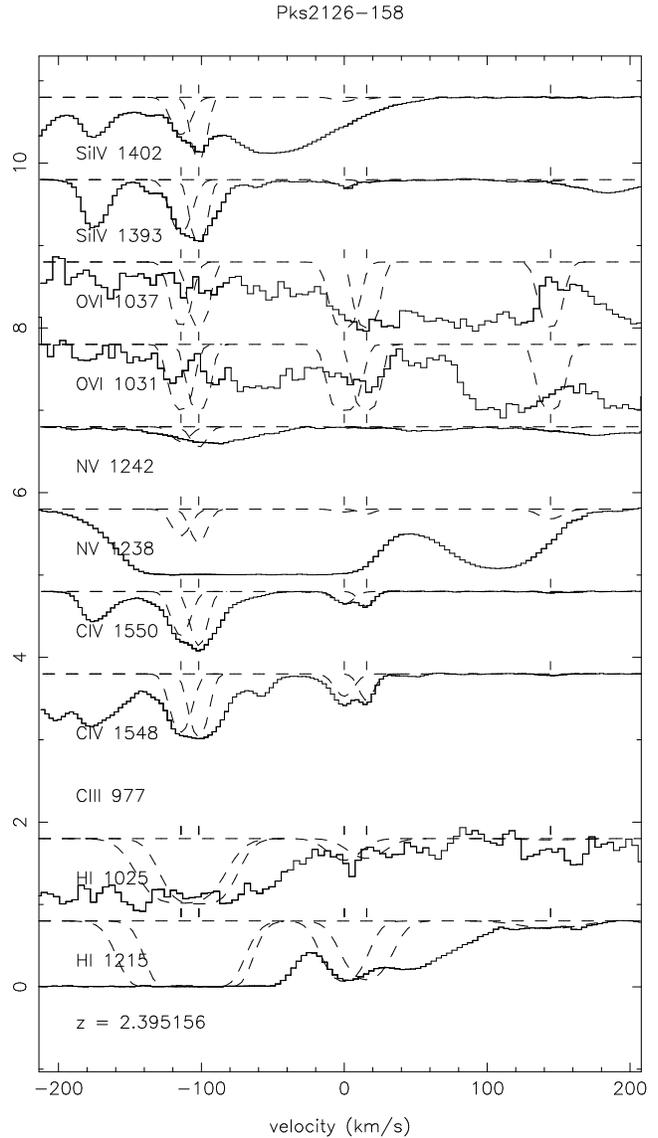}}
\caption{The $z=2.393862$, 2.394003, 2.395156, 2.395333, and 2.396791
  components 
  of the $z=2.394$ system in 
  Pks2126$-$158. The other component of this system is shown in
  Fig.~\protect\ref{fig:2126_ulz2p388}. \HI\ \lyb\ and \OVI\ are very
  noisy. \NV\ 1242 provides tight upper limits.
\label{fig:2126_ulz2p395}}
\end{figure}

\begin{figure}
\resizebox{\colwidth}{!}{\includegraphics{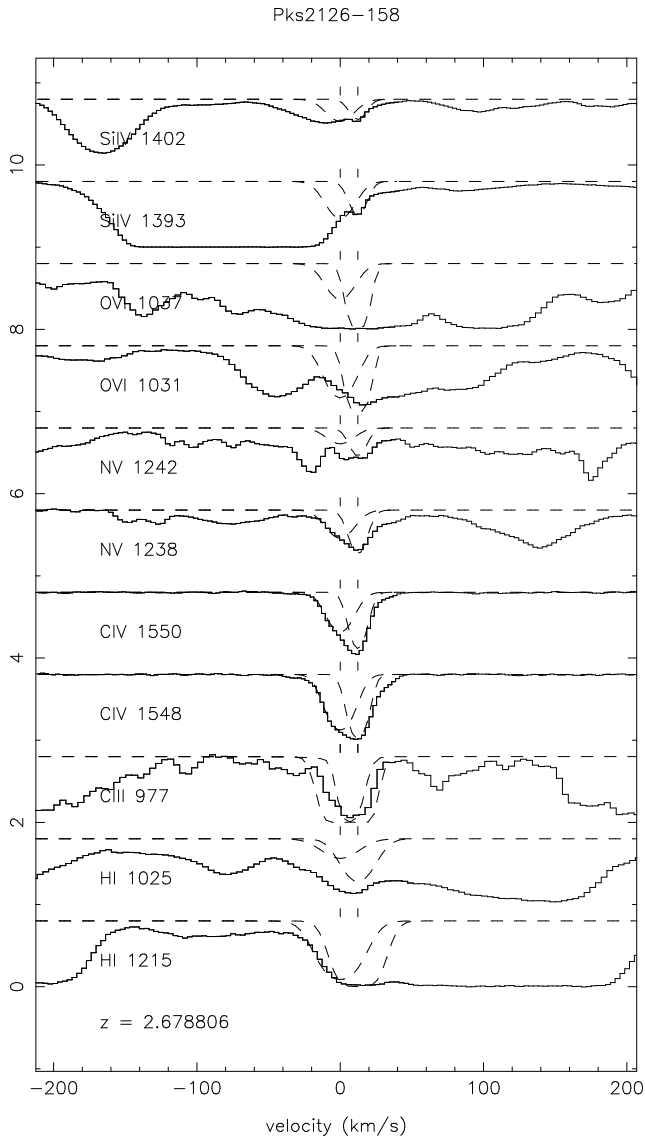}}
\caption{The $z=2.678806$ and 2.678957 components of the $z=2.679$
  system in Pks2126$-$158. Strong \HI\ and
  very strong \CIV\ are clearly present. \NV\ is probably also
  present. \OVI\ is strongly contaminated.
 \label{fig:2126_ulz2p678}}
\end{figure}

\subsection{Uncertainties}
\label{sec:uncertainties}

In this section we will investigate how robust our conclusions are
with respect to the assumptions on which our ionization models
rely. We will show
that our qualitative conclusions, namely that the clouds are
compact, not self-gravitating, and have high metallicities, are
robust. In particular, we will show that these conclusions would
be strengthened if: the radiation field were harder and/or more
intense, oxygen were overabundant relative to carbon, the
heavy elements were not yet in ionization 
equilibrium (which we will argue may be true), and if the temperature
were high enough for 
collisional ionization to be 
important (we will, however, argue that this is unlikely to be the case).

The main caveat appears to be the possibility of a much softer
radiation field, which would result in larger clouds and, for a subset
of the absorbers, lower metallicities. However, in practice a
significantly softer radiation field would require the presence of
local, soft sources (galaxies) and the resulting increase in the
hydrogen ionization rate would decrease the inferred cloud
sizes. Moreover, we shall argue that the rates of incidence of the
metal-rich absorbers is too high for local sources to dominate the
radiation field.

\subsubsection{The radiation field}
We assumed that the clouds are exposed to the integrated UV/X-ray
radiation from quasars and galaxies, as computed by Haardt
\& Madau (2001) for $z=2.3$. Given that the metal-rich clouds are
unlikely to be gravitationally confined and that they have high
metallicities, it would perhaps not be surprising if they resided in the
environments of galaxies. In that case the local radiation could be
both more intense and softer than the background (which includes a
substantial contribution from quasars). 

Schaye (2006, see also Miralda-Escud\'e 2005), showed that under the
assumption that the absorbers reside in the halos of the sources that
dominate the ionizing background, the ratio of the mean flux from local
sources to the flux from the background is given by
\begin{equation}
{\bar{F} \over F_{\rm bg}} = \left ({d{\cal N} \over dz}\right )_{\rm
  crit} \left ({d{\cal N} \over dz}\right )^{-1},
\end{equation}
where the critical rate of incidence is given by
\begin{equation}
\left ({d{\cal N} \over dz}\right )_{\rm crit} \approx 
0.6 \,{\left < f_{{\rm esc},N}\right > f_{\rm cov} \over f_{\rm esc}},
\end{equation}
where $f_{{\rm esc},N}$ and $f_{\rm esc}$ are the fractions of the
locally emitted radiation that is able to escape to the absorber and
out of the halo of the galaxy, respectively, $f_{\rm cov}$ is the covering
factor, and we assumed that the rate of incidence of Lyman limit
systems per unit redshift is 1.3 (P\'eroux et al.\ 2005) and that the \HI\
column density distribution has the form $d{\cal N}/dz \propto N_{\rm
  HI}^{-1.5}$. Since the \HI\ columns of our absorbers are much
smaller than those typical of galaxies, we expect $f_{{\rm esc},N}
\approx f_{\rm esc}$. The covering factor is unlikely to be much smaller than
unity given that more than half of our systems contain multiple
high-metallicity clouds. Plugging in our system rate of incidence
($d{\cal N}/dz = 3$), which is really a lower limit given our
conservative approach, we find that ${\bar{F} / F_{\rm bg}}\la
0.2$. Thus, in general we do not expect local sources to dominate the ionizing
radiation field to which our absorbers are exposed.

We will nevertheless examine the consequences of
changes in the intensity and spectral hardness, because we cannot be
sure that the Haardt \& Madau (2001) model is an accurate
description of the ionizing background.
As discussed in \S\ref{sec:ionmodeling}, from observed column
density ratios such as \CIV/\OVI, we can, under certain assumptions, 
measure the ionization parameter. Assuming that we know the intensity
$I$ of the ionizing radiation field, we can translate this into a gas
density. From equations (\ref{eq:ionpar}) and (\ref{eq:size}) it can
be seen that if the intensity $I$ of the 
radiation were higher, the clouds would be denser and smaller:  
$n_{\rm H} \propto I$ and $R\propto I^{-1}$ (for a fixed ionization
parameter the neutral hydrogen fraction is independent of the
intensity, therefore $R\propto n_{\rm HI}^{-1}\propto n_{\rm
  H}^{-1}\propto I^{-1}$). Both the metallicity and the relative abundance of
silicon would remain unchanged, as can be seen from
figure~\ref{fig:pzmet} and equation (\ref{eq:SiCrel}) (column
density ratios depend on the ionization parameter rather than the
density). 

Changing the spectral hardness of the radiation will affect the
inferred properties in the following way. Since the ionization
potentials of \OV\ and \NIV\ are higher than that of \CIII\ (114 and
77 versus 48~eV), softer
spectra yield higher \CIV/\OVI\ and \CIV/\NV\ ratios. Thus, if the
true radiation field is softer than the Haardt \& Madau (2001) model,
we will have 
overestimated the lower limit on the density, underestimated the size,
and overestimated 
the relative abundance of silicon (see Fig.~\ref{fig:pdenssize}). A
softer radiation field would, however, decrease 
the \CIII/\CIV\ ratio, which would give us higher upper limits on the
density. For 26 of the 28 clouds the constraint on the
metallicity would remain unchanged, because for these we assumed
$n_{\rm CIV}/n_{\rm HI}$ to have its maximum possible value. The
remaining 2 clouds could drop out of our sample because the curves in
figure~\ref{fig:selection} would shift to the right. Conversely, if
the radiation field were harder, we would infer higher densities,
smaller sizes, lower silicon abundances and, in some cases, higher
metallicities (which could also lead to an increased sample size). Note
that a much harder radiation field would lead to inconsistencies
between the lower limits on the density inferred from \CIV/\OVI\ and
\CIV/\NV\ and the upper limits derived from \CIII/\CIV.

If, contrary to the prediction of Schaye (2006), radiation from local
galaxies were important for our absorbers, then the radiation would 
have a softer spectral shape, but it would also be more intense than
in our fiducial model. This means that the effects on the density and
the size would be difficult to predict.

\subsubsection{Relative abundances}

We assumed that the relative abundances of carbon, nitrogen, and
oxygen are the same as for the sun. However, at high redshift oxygen
is often found to be overabundant relative to carbon (e.g., Telfer et
al.\ 2002), while nitrogen is often underabundant, at least relative
to oxygen (e.g., Bergeron et al.\ 2002). On the other hand, since
nitrogen and carbon are both produced by intermediate stars, we may
expect $[{\rm N}/{\rm C}]$ to be close to solar. Furthermore, it
is likely that the relative abundances of the elements are a stronger
function of metallicity than of time, in which case we would expect
the relative abundances to be close to solar, because the metallicity
of the clouds is typically of order solar. 

Nevertheless, let us examine the consequences of extreme variations in
the assumed relative abundances. We find that increasing the abundance
of oxygen relative to that of carbon by a factor of 10, leads
to the addition of three clouds to our sample (but no new systems),
while decreasing the N/C ratio by a factor of 10 removes one cloud
(and one system) from the sample. Increasing O/C (decreasing N/C) by
an order of magnitude increases (decreases) the median lower limit on
the density by 0.4 (0.3) dex, leaves the 
upper limits on the density nearly unchanged, decreases (increases) the
median upper 
limit on the size by 0.8 (0.4) dex, increases (decreases) the median
lower limit on the metallicity by 0.5 (0.4) dex, and decreases
(increases) the median upper limit on the relative abundance of
silicon by 0.7 (0.7) dex. Thus, increasing O/C has the opposite effect
of decreasing N/C, but the results tend to be more sensitive to O/C.

We conclude that even extreme changes in the assumed relative
abundances would not change our conclusions. In fact, the most
natural modification (enhanced oxygen relative to carbon), would make
the metal-rich clouds even more dense, compact, and metal-rich.

\subsubsection{Ionization equilibrium}
\label{sec:equil}

We assumed the gas to be in ionization equilibrium. However, if the
clouds were created by gas cooling from high temperatures, this
assumption could temporarily break down because for $T\sim 10^5 -
10^6~\K$ the cooling time can be much shorter than the relevant
photoionization (and recombination) timescales of the metals (some of
these timescales exceed $10^8~\yr$). Thus, if the clouds were 
created hot, $T\gg 
10^4~\K$, and have lifetimes $t\la 10^8~\yr$, then they may be much
more highly ionized than they would be in photoionization 
equilibrium. The ionization balance of the metals would then be
similar to that of collisionally ionized gas, even though the clouds
would typically have temperatures of $T\sim 10^4~\K$.
However, since for hydrogen the photoionization timescale 
is only $\sim 10^4~\yr$, ionization equilibrium is likely to be a very
good approximation for this element. Below we will investigate how the
properties of the clouds would change if the heavy elements had not
yet reached photoionization equilibrium.

As discussed in \S\ref{sec:ionmodeling}, the column density ratios
\CIV/\OVI\ and \CIV/\NV\ 
decrease with the degree to which the gas is ionized, which means they
increase with density in photoionization equilibrium.  By
assuming photoionization equilibrium, we may thus have underestimated
the gas density. The cloud size was computed using $R = N_{\rm
HI}/2n_{\rm HI}$ which, making use of the fact that hydrogen is likely
to be close to photoionization
equilibrium (i.e., $n_{\rm HI}/n_{\rm H}\propto n_{\rm H}$), implies
$R\propto n_{\rm H}^{-2}$. Hence, we may have overestimated
the cloud size.

The effects on the abundances are somewhat more difficult to
predict. From 
equation (\ref{eq:metallicity}) we can see that $10^{[{\rm C}/{\rm H}]} \propto
(n_{\rm HI}/n_{\rm H})(n_{\rm CIV}/n_{\rm C})^{-1} \propto n_{\rm H} 
(n_{\rm CIV}/n_{\rm C})^{-1}$, where number densities are computed assuming
solar abundances and we assumed that hydrogen is in photoionization
equilibrium (i.e., $n_{\rm HI}\propto n_{\rm H}^2$). Given that
\CIV/\OVI\ and \CIV/\NV\ are good ``ionization meters'', 
our estimate of the ratio \CIV/C is probably roughly right despite the
fact that we assumed ionization equilibrium. However, the equilibrium
assumption will have caused us to underestimate the gas density and
thus the metallicity. 

If we assume that the equilibrium
model that best fits the measured column density ratios \CIV/\OVI\ and
\CIV/\NV\ predicts approximately the right 
\SiIV/\CIV\ ratio, then it follows from equation (\ref{eq:SiCrel}) that
the inferred 
relative abundance of silicon is unaffected by the assumption
of ionization equilibrium. 

In short, if the clouds were created hot and are short-lived, then the
heavy elements might not yet be in photoionization equilibrium. In
that case our conclusions would again be strengthened: we would have
underestimated the limits on the density, substantially 
overestimated the upper limit on the size, probably underestimated the
lower limit on the metallicity, and
probably inferred roughly the right upper limit on $[{\rm Si}/{\rm C}]$.

\subsubsection{Temperature}
\label{sec:temperature}
We chose to keep the temperature fixed
rather than assume thermal equilibrium because the cooling rate 
depends on the metallicity and on the dynamical state of the clouds,
which we do not know a priori. We assumed that the clouds have a
temperature $T=10^4~\K$, which is typical of a photo-ionized plasma
and means that collisional ionization is unimportant. 

Figure~\ref{fig:colratios} demonstrates that
the metal column density ratios are insensitive to the temperature as long as
$T\sim 10^4~\K$. For $T\ga 10^5~\K$, collisional ionization becomes
increasingly important and the gas gets more highly ionized. The
dashed curves in figure~\ref{fig:colratios} indicate the predicted
column density ratios for the artificial case of pure collisional
ionization equilibrium (temperature is varied along these curves). 

From Figure~\ref{fig:colratios} it can clearly be seen that the data
do not favor temperatures high enough for collisional ionization to be
important. Bringing the high-temperature models into agreement with
the data requires much larger shifts of the data points. In fact, most limits
are incompatible with the $\log T = 5.2$ and 5.3 models. For the
(unrealistic) case
of pure collisional ionization, which corresponds to the high density
limit in the presence of an ionizing background, all but 4 points can be
shifted onto the dashed curve, but this would imply that for most
absorbers the upper limit on $N_{\rm OVI}$ exceeds the true value by
more than 2 orders of magnitude and that most have nearly identical
temperatures. Thus, it appears the data favor photo-ionization over
collisional processes. 

This conclusion is further strengthened by the
measured \CIV\ line widths. For collisional ionization to be effective,
temperatures of $T \ga 10^5~\K$ are required, corresponding to line
widths of
\begin{equation}
b = 40.5 ~\kms\ \left ({m_{\rm H} \over m}\right )^{1/2}\left ({T \over
  10^5~\K}\right )^{1/2}.
\end{equation}
As can be seen from Table~\ref{tbl:vpfits} (col.\ 11), more than half
of the \CIV\ lines are sufficiently narrow to rule out such high 
temperatures. 

In short, several independent pieces of evidence
suggest that most of the metal-rich clouds are at temperatures too low
for collisional ionization to be important.

Nevertheless, let us examine the effect of a higher temperature. 
If the gas were hot enough for collisional ionization to be important,
then the gas would be more highly ionized. As discussed in
\S\ref{sec:equil}, this would imply that we have underestimated the
limits on the densities. As discussed in \S\ref{sec:selection}, the
metallicities would generally remain unaffected, although the lower
limits would be slightly reduced if $\log T$ were within
about 0.2 dex of 5.05. For $\log T = (4.85, 5.05, 5.25)$ we find
(10,6,10) metal-rich systems and (24, 7, 18) metal-rich clouds. The
median lower limit on 
the density is increased by (0.2,0.2,0.5) dex, the median upper limit
on the size is increased by (0.3,0.4,0.3) dex, and the median lower
limit on the metallicity is decreased by (0.2, 0.4, 0.2) dex. If we
increase the temperature further, then we find many more 
high-metallicity systems and clouds, and we require much denser and
compacter clouds.  

To summarize, if the gas has a temperature high enough for the metals
to be collisionally ionized ($T\ga 10^5~\K$), which we argued is
unlikely, then our conclusions are generally strengthened: we find more
metal-rich clouds and infer higher densities and smaller sizes. Only
for temperatures very close to $\log T = 5.05$ do we find somewhat
fewer metal-rich clouds and infer slightly weaker constraints on the
sizes (but stronger constraints on the densities).

\section{Discussion}
\label{sec:discussion}

In the previous sections we showed that at $z\approx 2.3$ there exists
a population of 
highly ionized, metal-rich gas clouds that give rise to \CIV\
absorption in the spectra of quasars. These clouds often come in
groups, but some appear isolated. Some are located close to strong
\HI\ absorbers, but others are not. Groups of high-metallicity clouds
have a rate of incidence $d{\cal 
  N}/dz > 3.0\pm 0.9$, while the rate for individual clouds is $d{\cal
  N}/dz > 7\pm 1$. Their densities are typically $n_{\rm H}\sim
10^{-3.5}~\cm^{-3}$, they have sizes $R\sim 10^2~\pc$, and
metallicities $Z\ga Z_\odot$.

What could the high-metallicity clouds be? Intergalactic supernovae
remnants, interplanetary planetary nebulae, (fragments of) shells
driven by
galactic superwinds, dark matter minihalos, high-velocity clouds,
tidally stripped 
gas? We note that many of the
possible origins have been discussed in related contexts, such as weak \MgII\
systems (e.g., Rigby et al.\ 2002; Milutinovic et al.\ 2006 and
references therein) and strong \OVI\ absorbers (e.g., Simcoe et al.\
2006; Bergeron \& Herbert-Fort 2005), and that some of the
arguments used in those studies can also 
be applied to the high-metallicity \CIV\ clouds. In the rest of this
section we will discuss some of these scenarios in passing, but we
will focus on questions that can be addressed without knowledge of the
origin of the clouds.

\subsection{Cosmological significance}

It is interesting to investigate the cosmological significance of the
high-metallicity clouds. In this section we derive some generic
results that are independent of the interpretation of the nature of the
population. Our starting 
point will be the observed rates of incidence and the typical physical 
properties derived from the ionization model. In each case we will
show how the results scale with these and other parameters. 

Let $n$ be the \emph{comoving} number density of clouds and let each cloud
provide a \emph{proper} cross section $\pi R^2$. The total rate of
incidence is then given by 
\begin{equation}
{d{\cal N} \over dz} = n\pi R^2 {c \over H(z)} (1+z)^2.
\label{eq:dN/dz}
\end{equation}
Using $(\Omega_m,\Omega_\Lambda,h) = (0.3,0.7,0.7)$ and
$z=2.3$ we find,
\begin{equation}
n \approx 1.6 \times 10^4~\Mpc^{-3} \left ({d{\cal N}/dz \over 7}\right )
\left ({R \over 10^2~\pc}\right )^{-2}.
\label{eq:ncomoving}
\end{equation}
Comparing this with the number of bright Lyman-Break Galaxies
(LBG) at $z=3$, $n_g\sim 10^{-2} h^3\Mpc^{-3}$ down to $0.1L_\ast$ (Steidel et
al.\ 1999), it is clear that the population of clouds must be
enormous: the high-metallicity clouds outnumber bright galaxies by
about a million to one. 

Conversely, if we assume that the clouds reside in the
environments of a population of objects with comoving number density
$n_g$, we can estimate the proper radius, $r_h$, of the 
halo containing high-metallicity clouds around each such object,
\begin{equation}
r_h \approx  8.3\times 10^4~\pc~\left( {d{\cal N}/dz \over 3}\right )^{1/2}
  \left ({n_g \over 
  10^{-2}~\Mpc^{-3}}\right )^{-1/2} f_{\rm cov}^{-1/2},
\label{eq:rphys}
\end{equation}
where $f_{\rm cov}$ is the covering factor of high-metallicity clouds
  in the halo and $d{\cal N}/dz=3$
is the rate of incidence of metal-rich systems (as opposed to clouds). Given
that there are typically multiple high-metallicity clouds per
  system, we expect $f_{\rm cov}$ to be of order unity. If the clouds are
related to galactic winds, we would expect galaxies fainter than
$0.1L_\ast$ to contribute too, since winds may escape more easily
from small galaxies. Therefore, the halos containing metal-rich clouds
could typically be smaller than $10^2~\kpc$. Interestingly,
simulations predict that at $z\sim 3$, superwinds driven by
starbursting galaxies manage to propagate to scales $\la 10^2~\kpc$
(Aguirre et al.\ 2001). 

The product of the comoving number density and the cloud mass gives us
the cosmic gas mass density of the high-metallicity clouds. Expressed in
units of the critical density this becomes
\begin{eqnarray}
\Omega &=& {n M_g \over \rho_{\rm crit}}, \\
&\approx & 5.2\times 10^{-6}\left( {d{\cal N}/dz \over 7}\right )\left
({R \over 
  10^2~\pc}\right ) \left (  
{n_{\rm H} \over 10^{-3.5}~\cm^{-3}}\right ).\nonumber 
\end{eqnarray}
Note that if the absorption were to arise in shells centered on galaxies
rather than in spherical clouds, the total mass contained in the
population would be about the same (the mass of a shell exceeds that
of a spherical cloud by a factor of about $(r/R)^2$, where $r$ is the
radius of the shell, but the number
density of shells is down by the same factor relative to that of
spherical clouds).
Comparing the cosmic gas density of the high-metallicity clouds to the
cosmic baryon density $\Omega_b \approx 0.044$ 
(e.g., Spergel et al.\ 2003),
we see that the population of metal-rich clouds contains a negligible
fraction of the baryons. 

The high-metallicity clouds account for a cosmic metallicity of
\begin{eqnarray}
Z_{\rm cosmic} &=& {\Omega Z \over \Omega_b}, \\
&\approx & 1.2 \times 10^{-4}~ Z \left(
{d{\cal N}/dz \over 7}\right )\left ({R \over   10^2~\pc}\right )\\
&& \times \left ( {n_{\rm H} \over 10^{-3.5}~\cm^{-3}}\right ),\nonumber
\label{eq:zcosmic}
\end{eqnarray}
where $Z$ is the cloud metallicity. 
Comparing this estimate to the total carbon abundance in the diffuse
($\delta < 10^2$) IGM inferred from \CIV\ absorbers, $[{\rm C}/{\rm
    H}] = -2.8$  (S03, for their fiducial UV
background model), we find 
that the clouds contain of order 10\% of the intergalactic
metals associated with detectable \CIV\ absorption.

In \S\ref{sec:cossign2} we will show that the cosmological
significance of the clouds would be drastically increased if the
clouds are short-lived, as we will argue is likely to be the case.

\subsection{Cloud confinement}

There are in general two ways to confine a gas cloud: gravity and
external pressure. We already showed that gravitational confinement of
the high-metallicity clouds is only possible if the gas mass is
negligible. Below we will explore this possibility, as well as
confinement by external pressure. We will show that gravitational
confinement by dark matter minihalos with mass $M\sim 10^6 -
10^7~\Msun$ would be  possible, except that there probably are not
enough minihalos to 
account for the large number of high-metallicity clouds. We will also
show that confinement by external pressure is generally only possible if the
confining medium has an overdensity of a few or less and a temperature
greater than $10^5~\K$. 

\subsubsection{Self-gravity}

In \S\ref{sec:results} we found that for the high-metallicity
clouds to be self-gravitating, the median gas mass fraction would have to
be smaller than $\log f_g < -3.2$. This estimate was based on the
lower limits on the densities derived from the lower limits on the
column density ratios 
\CIV/\NV\ and \CIV/\OVI, and the upper limits on the cloud sizes
obtained by combining the upper limits on $N_{\rm HI}$ and the lower
limits on the densities. However, the observed
upper limits on $N_{\rm CIII}/N_{\rm CIV}$ suggested that the
densities are typically about 0.5~dex greater than our lower limits,
which in turn reduces the upper limits on the sizes by
1~dex. From equation (\ref{eq:LJ}) we can see that the gas fraction
required for gravitational confinement then becomes
\begin{equation}
f_g \sim 7.7\times 10^{-6} \left ({n_{\rm H} \over
  10^{-3.5}~\cm^{-3}}\right ) \left ({R \over 10^2~\pc}\right )^2
  T_4^{-1},
\end{equation}
where $T_4\equiv T/(10^4\,\K)$. 
Using equation (\ref{eq:mass}) this gives a total mass of
\begin{equation}
M \sim 5.7\times 10^6 ~\Msun \left ({R \over 10^2~\pc}\right ) T_4.
\end{equation}
Note that the required total mass is much less uncertain than
the gas mass, which scales as $M_g\propto n_{\rm H}R^3$.

If we identify the total mass with the virial mass of a halo
collapsing at redshift $z_c$, then we find a virial radius of
\begin{equation}
r_{\rm vir} \sim 1.2 \times 10^2~\pc 
\left ({R \over 10^2~\pc}\right )^{1/3} T_4^{1/3}
\left ({1+z_c \over 11}\right )^{-1},
\end{equation}
which agrees remarkably well with the characteristic size we inferred
from the ionization modeling ($R\sim 10^2~\pc$) if the halo collapsed
at high redshift. Such a halo would have a circular velocity of
\begin{eqnarray}
v_c & \equiv & \left ( {G M \over r_{\rm vir}}\right )^{1/2} \\
& \sim & 15 ~\kms \left ({R \over 10^2~\pc}\right )^{1/3} T_4^{-1/6}
\left ({1+z_c \over 11}\right )^{1/2}, \nonumber
\end{eqnarray}
and a virial temperature of
\begin{eqnarray}
T_{\rm vir} & \equiv & {\mu m_{\rm H} v_c^2 \over 3 k} \\
& \sim & 5.0 \times 10^3~\K 
\left ({R \over 10^2~\pc}\right )^{2/3} T_4^{-1/3}
\left ({1+z_c \over 11}\right ). \nonumber
\end{eqnarray}
Since the virial temperature is smaller than $10^4~\K$, stars could
only have formed in such a minihalo if it collapsed before
reionization. 

The comoving number 
density of minihalos is $n(M > 10^6~\Msun) \sim 10^{2.6} ~\Mpc^{-3}$
at $z=10$, increasing to about $10^3$ at $z=6$ (e.g., Reed et al.\
2007). This is significantly lower than the
comoving number density of high-metallicity clouds, which we estimated
to be $2\times 10^4 ~\Mpc^{-3}$ [see eq.~(\ref{eq:ncomoving})]. The
discrepancy is actually worse, since many of the minihalos that
collapsed before reionization will have merged with galaxies by
$z=2.3$ and it is only the remainder that we should consider here. We
therefore do not think that minihalos from reionization can account
for a significant fraction of the high-metallicity clouds.

\subsubsection{External pressure}

We have two useful constraints on a possible confining medium: it must have
about the same pressure as the clouds and it must give rise to
\HI\ absorption that is consistent with our upper limit on the \HI\
column density. 

The high-metallicity clouds have only a modest pressure, 
\begin{equation}
{P \over k} = nT \sim 10^{0.5}~\cm^{-3}\,\K \left ({n_{\rm H} \over
  10^{-3.5}~\cm^{-3}}\right ) T_4.
\end{equation}
This pressure is relatively well constrained because much higher
densities are ruled out by our upper limits on the \CIII/\CIV\ ratio
and much higher temperatures are inconsistent with the line widths as
well as the metal column ratios (see \S\ref{sec:temperature}). If the
confining medium has a higher density, then this would imply a
temperature smaller than $10^4~\K$. Such low temperatures are only
expected for clouds that are optically thick to ionizing radiation,
which is clearly ruled out by our constraints on $N_{\rm HI}$.

A higher temperature, on the other hand, would imply an overdensity
lower than 50, which means that the medium would have a
density significantly lower than is typical of virialized
objects. This is, however, certainly not impossible. The confining gas
could for example have been heated by accretion shocks onto
large-scale filaments or by shocks associated with galactic winds. The
medium must, however, be sufficiently hot that collisional ionization
suppresses its neutral hydrogen absorption to the very low levels we
observe.

\begin{figure*}
\resizebox{\colwidth}{!}{\includegraphics{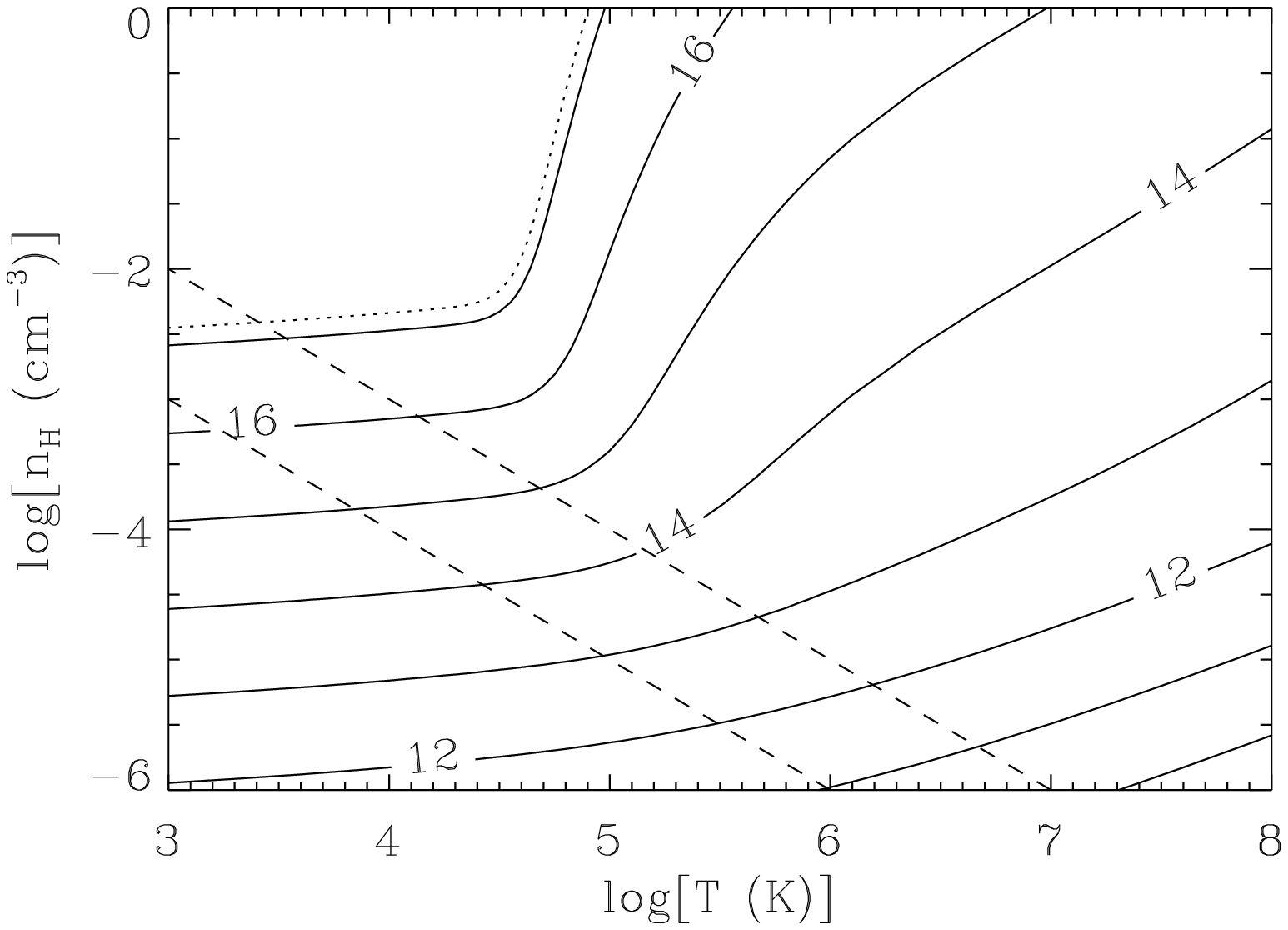}}
\resizebox{\colwidth}{!}{\includegraphics{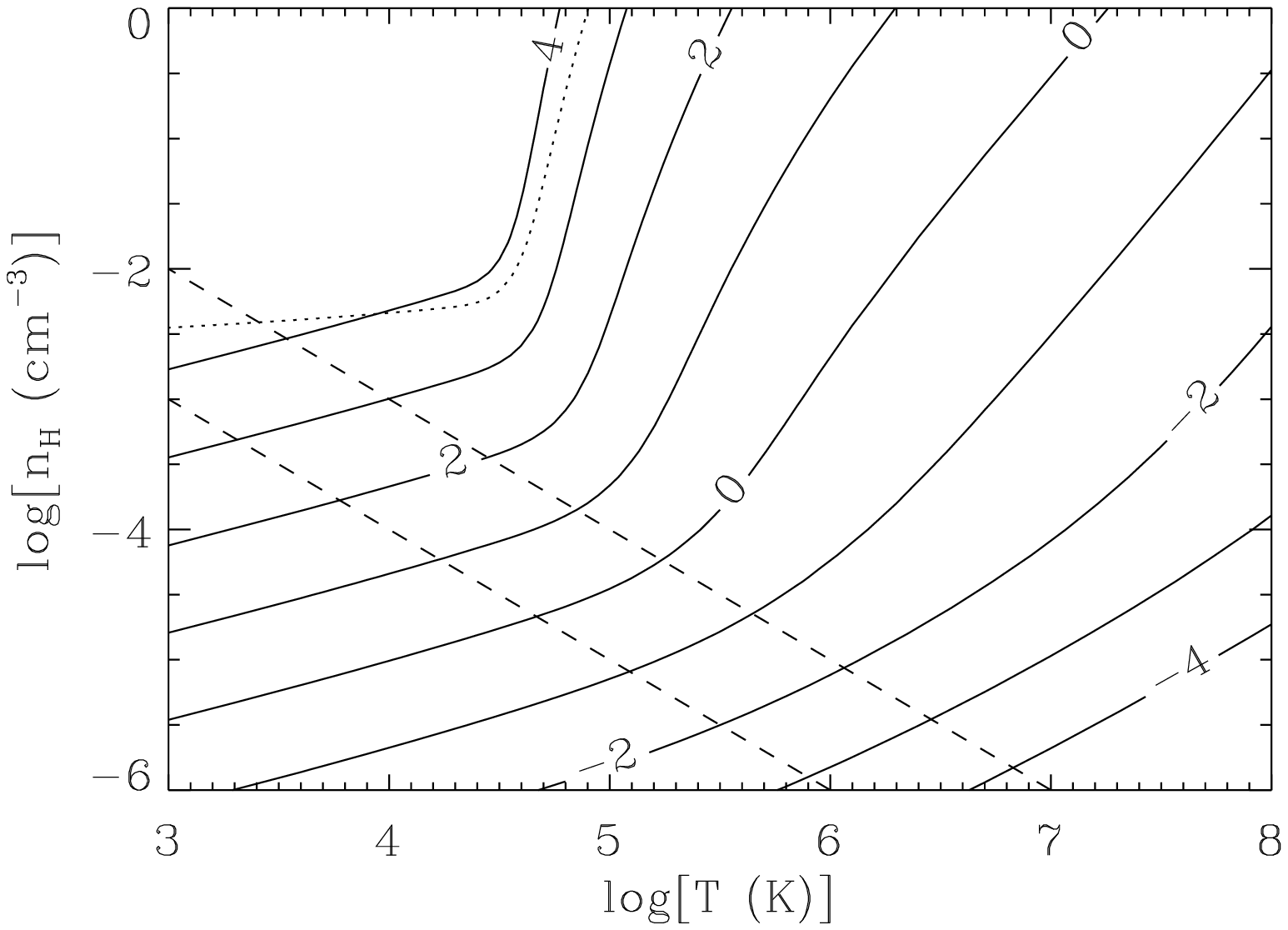}}
\caption{Typical \HI\ column densities and \lya\ optical depths
  expected for self-gravitating clouds as a function of the cloud
  density and temperature. The solid curves are contours of constant
  neutral hydrogen column density 
  (\emph{left}) and \HI\ \lya\ optical depth at the line center
  (\emph{right}). The dotted contours
  indicate the self-shielding limit, $\log[N_{\rm HI}~(\cm^{-2})] =
  17.2$, and the dashed contours correspond to pressures of $\log[P/k
  ~(\cm^{-3}\K)] = 0$ (\emph{bottom}) and 1 (\emph{top}). The cloud
  was exposed to the Haardt \& Madau model for the $z=2.3$ UV/X-ray
  background and assumed to be optically thin.}
\label{fig:contours}
\end{figure*}

Figure~\ref{fig:contours} shows as solid contours the typical neutral
hydrogen column density (\emph{left}) and \HI\ \lya\ optical depth
at the line center (\emph{right}) associated with a self-gravitating
($2R=L_J$) gas cloud as a function of its gas density and
temperature. The dashed lines are lines of constant pressure,
corresponding to $P/k = 
10^0$ and $10^1~\cm^{-3}\,\K$, respectively. The region enclosed
by the dashed 
lines roughly satisfies the pressure constraint. To be
consistent with the observed \HI\ absorption we typically
require\footnote{We are assuming 
  that the \HI\ absorption is dominated by the high-metallicity
  cloud. If it were dominated by a confining medium, then the
  high-metallicity clouds would have metallicities $Z\gg
  Z_\odot$ and sizes $R\ll 10^2~\pc$.} $N_{\rm HI} 
\ll 10^{13}~\cm^{-2}$ or, for $T \gg 10^5~\K$, $\tau_c \ll 10^{-1}$
(note that the column density constraint was inferred under the
assumption that $b_{\rm HI}=(m_{\rm C}/m_{\rm H})^{1/2} b_{\rm
 CIV}$). A confining medium with $n_{\rm H}\sim 10^{-5}~\cm^{-3}$ and
$T\sim 10^{5.5}~\K$ would satisfy all constraints. In general,
the density cannot be high compared with the cosmic mean and the
temperature must exceed $10^5~\K$. 

If the confining medium is not self-gravitating, as could for example
be the case if it were expanding, hot, wind fluid, then its size will
be smaller than in the hydrostatic equilibrium case, making it
possible to satisfy the absorption constraints for higher densities. 

\subsection{Lifetimes}

If the clouds are confined neither by self-gravity nor by external
pressure, as seems plausible from the above discussion, then they will
freely expand until they reach pressure equilibrium with their
environment. 

The free expansion timescale is just the ratio of the cloud size and
internal sound speed $c_s$, 
\begin{eqnarray}
t_{\rm exp} &\sim & {R \over c_s} 
= R\sqrt{{\mu m_{\rm H} \over \gamma k T}} \\
&\sim & 6.4\times 10^6~\yr \left ({R \over 10^2~\pc} \right
)T_4^{-1/2}, \nonumber  
\label{eq:t_exp}
\end{eqnarray}
where we used a mean particle weight $\mu=0.59$ and a ratio of
specific heats $\gamma = 5/3$. The expansion timescale is very small 
compared to the Hubble time, which means that the high-metallicity
clouds may be a transient phenomenon.

In fact, the clouds would likely be short-lived, even in the presence
of a confining medium. Unless it is gravitationally confined, a cloud
of size $R$ and density $\rho$ that is moving with velocity $v$
relative to a medium with density 
$\rho_{\rm m}$, will be destroyed through Kelvin-Helmholtz and
Rayleigh-Taylor 
instabilities on a timescale (e.g., Jones, Kang, \& Tregillis 1994) 
\begin{equation}
t_{\rm inst} \sim \left ({\rho \over \rho_{\rm m}} \right )^{1/2} {R
 \over v} ,
\end{equation}
which we can write as
\begin{equation}
t_{\rm inst} \sim t_{\rm exp}\left ({\rho \over \rho_{\rm m}} \right
)^{1/2}{c_s \over v} 
\sim t_{\rm exp}\left ({P \over P_{\rm m}}\right )^{1/2} {c_{s,m}
  \over v}, 
\end{equation}
where $c_s$ is the internal sound speed of the cloud and $c_{s,m}$ is
the sound speed in the confining medium. If the cloud is falling into
or rotating within a hot, hydrostatic gaseous halo, then we would
expect $v\sim c_{s,{\rm m}}$ in which case the cloud will be 
destroyed in an expansion timescale if the medium has a pressure
similar to that of the cloud. In general, $t_{\rm inst} \gg t_{\rm
  exp}$ would require $v \ll c_{s,m}$ for pressure-confined
clouds. Since the cloud has an 
overdensity of only about 50 and the medium is unlikely to be underdense
by a large factor, we expect $c_{s,m}\la \sqrt{50} c_s$. Thus, $v \ll
c_{s,m}$ would imply velocities of only a few km/s, which seems
implausible. Note that the 
clouds would have to be born with such small velocities because, as we
will show below, deceleration from $v\sim c_s$ takes longer than
$t_{\rm inst}$. Clouds could perhaps form with such small relative
velocity if they formed from an expanding, hot wind
fluid. However, in that case the cloud lifetime would still be limited
by the free expansion timescale of the wind fluid, which is typically much
smaller than the Hubble time. We conclude that hydrodynamical 
instabilities would likely change the cloud on a timescale similar to the
expansion timescale.

Note that although two-dimensional magnetohydrodynamic simulations
have shown that magnetic fields inhibit the disruption of supersonic
clouds through fluid instabilities (Mac Low et al.\ 1994), 
three-dimensional simulations show that the interaction with a
magnetic field perpendicular to the direction of motion actually
accelerates the development of the Rayleigh-Taylor instability (Gregori
et al.\ 1999).

Unless the cloud is driven by ram-pressure, for example in the form of
a galactic wind, a moving cloud will decelerate on the timescale that
it sweeps up its own mass,
\begin{equation}
t_{\rm dec} \sim {\rho \over \rho_{\rm m}} {R \over v} \sim t_{\rm
  exp} {\rho \over \rho_{\rm m}} {c_s \over v}.
\end{equation}
For clouds that are denser than their environment this exceeds $t_{\rm
  inst}$, which means that deceleration due to conservation of
  momentum cannot save the cloud from being disrupted.  

Note that free expansion will not lead to temperatures much lower than
$10^4~\K$ because the photo-heating timescale becomes extremely short
at such low temperatures for highly ionized clouds\footnote{If the clouds
went through an optically thick phase, then they may have been
colder in the past.}. For example, for
$T=10^3~\K$ and $n_{\rm H}\sim 10^{-3.5}~\cm^{-3}$ the photo-heating
timescale is more than an order of magnitude smaller than the
timescale for adiabatic cooling resulting from free
expansion\footnote{Radiative cooling is extremely inefficient below
  $10^4~\K$ for a photo-ionized plasma.}.

In summary, we expect the clouds to be short-lived, with a
characteristic lifetime of order the free expansion timescale,
equation (\ref{eq:t_exp}).

\subsection{Implications of short lifetimes}

If the high-metallicity clouds are short-lived, as we argued is
likely, then this has important implications for their cosmological
significance as well as for scenarios accounting for their origin. 

\subsubsection{Cosmological significance}
\label{sec:cossign2}

If the high-metallicity clouds are short-lived, then the amount of metal
contained in the population at any given time is much smaller
than the total amount of metal that has passed through such a
phase. 
Assuming a steady population of freely expanding clouds, the total
cosmic metallicity that has gone through the phase corresponding to
the high-metallicity clouds is of order
\begin{eqnarray}
Z_{\rm cosmic} {t_H \over t} 
& \approx & 5.2\times 10^{-2}~Z \left ({ t \over 10^7~\yr}\right )^{-1}\left(
{d{\cal N}/dz \over 7}\right )\\
&& \times \left ({R \over   10^2~\pc}\right ) \left (   
{n_{\rm H} \over 10^{-3.5}~\cm^{-3}}\right ). \nonumber 
\label{eq:zcosmicovertime}
\end{eqnarray}
If the lifetime $t\propto R$, as in equation (\ref{eq:t_exp}),
then this estimate is independent of the cloud size and lifetime. 

Comparing this again to the metallicity of the diffuse IGM, $Z/Z_\odot
\sim 10^{-2.8} \sim 1.6\times 10^{-3}$ if carbon is used as a tracer
(S03), we see that if the clouds are short-lived, all of
the carbon contained in the IGM could in principle at one time have
resided in high-metallicity clouds. Note, however, that many of the high
metallicity clouds may end up inside galaxies and that the same gas
could be cycled through multiple generations of high-metallicity
clouds.

\subsubsection{The nature of the clouds}

Galactic winds are a plausible origin for the high-metallicity
clouds. Both observations and simulations show that the wind fluid is
highly clumpy (e.g., Veilleux et al.\ 2005). Hot gas sweeps up shells
which fragment due to thermal and hydrodynamical
instabilities. The hot gas pushes itself out in between the fragments,
sweeping up new shells in the process. Shell fragments are constantly
being formed and destroyed.\footnote{Models of shells driven by
winds have been invoked to explain other classes of 
absorbers. These models usually assume a hot wind fluid within a
spherical, unfragmented shell. We believe that such an idealized model
is inconsistent 
with both observations and simulations of galactic winds in a
realistic context, which predict a clumpy fluid and anisotropic
outflows.} Hence, the lifetimes of individual clouds may be much
shorter than the duration of the wind phenomenon (which could, in fact,
exceed the Hubble time). It is easy to see that this would indeed be
necessary for the high-metallicity clouds.

Starbursting galaxies at high redshift are observed to generate outflows with
velocities of $v\sim 10^2 - 10^3~\kms$ (e.g., Pettini et al.\
1998). Hence, we would expect the typical age of a
metal-rich halo of size $r_h$ to be at least
\begin{eqnarray}
t &=& 9.8\times 10^7~\yr ~ \left ({r_h \over 10^4~\pc}\right )
\left ({v \over 10^2~\kms}\right )^{-1} \\
&\sim & 8.1\times 10^8~\yr ~ \left ( {d{\cal N}/dz \over 3}\right )^{1/2}
  f_{\rm cov}^{-1/2} \\
&& \times \left ({n_g \over  
  10^{-2}~\Mpc^{-3}}\right )^{-1/2} \left ({v \over
      10^2~\kms}\right )^{-1}, \nonumber 
\label{eq:t_wind}
\end{eqnarray}
where we used equation (\ref{eq:rphys}) in the last step.
Unless the galaxy number density $n_g \gg 10^{-2}~\Mpc^{-3}$ or the
(sustained) wind velocity $v \gg 10^2~\kms$, the travel time exceeds
the characteristic cloud lifetime $t_{\rm exp} \sim 10^7~\yr$
[equation (\ref{eq:t_exp})]. In the wind scenario this would suggest
that the clouds are continuously being destroyed and
formed. Comparing the wind propagation
timescale with the minimum age of a starburst, $\sim 
10^7~\yr$ (i.e., the lifetime of a $8~\Msun$ star), we see that if
high-metallicity clouds
arise in the winds of starbursting galaxies, the associated galaxies
may already have faded by the time
their winds have expanded into the regime of our high-metallicity
clouds. 

The high-metallicity clouds could also be part of individual,
intergalactic supernova remnants. For example, Shelton (1998) showed
using non-equilibrium models of supernova remnants expanding in the
lower Galactic halo that such objects can give rise to high ionization
lines such as the ones associated with our high-metallicity \CIV\
clouds. 

If each component is associated with a single, distinct supernova
remnant, then the population traces a comoving star formation rate
density of: 
\begin{eqnarray}
\dot{\rho}_\ast &=& 10^2 nM_{100}/t \\
&\sim & 1.6\times 10^{-1} ~\Msun\,\Mpc^{-3}\yr^{-1} ~ M_{100}\,
\left ({d{\cal N}/dz \over 7}\right ) \\
&& \times \left ({r \over 10^2~\pc}\right )^{-2} \left ({t \over
  10^7~\yr}\right )^{-1},\nonumber 
\end{eqnarray}
where $r$ and $t$ are the radius and age of the remnant, respectively,
and $M_{100}$ is the average mass in stars for which the initial mass
function predicts a single supernova. We used as default values the
ones appropriate for our high-metallicity clouds: $r=R$ and
$t=t_{\rm exp}$. Note that in that case $\dot{\rho}_\ast \propto
R^{-1}$. Comparing to 
the observed, global star formation density, $\dot{\rho}_\ast \approx 0.1
~\Msun\,\Mpc^{-3}\yr^{-1}$ (e.g., Fardal et al.\ 2006), we see that
for the high-metallicity clouds to be supernova
remnants, essentially all star formation should be intergalactic,
which of course is highly unlikely. 

If the high-metallicity clouds are intergalactic
planetary nebulae, then we can use the same calculation, provided we
adjust $M_{\rm 100}$, and interpret $\dot{\rho}_\ast$ as averaged over
the range of lifetimes of the progenitor stars. Since we now have
$M_{100} \ll 1$, this scenario can probably not be ruled out on this
basis.

\subsubsection{Evolution of individual clouds}
\label{sec:evolution}

\begin{figure*}
\resizebox{\colwidth}{!}{\includegraphics{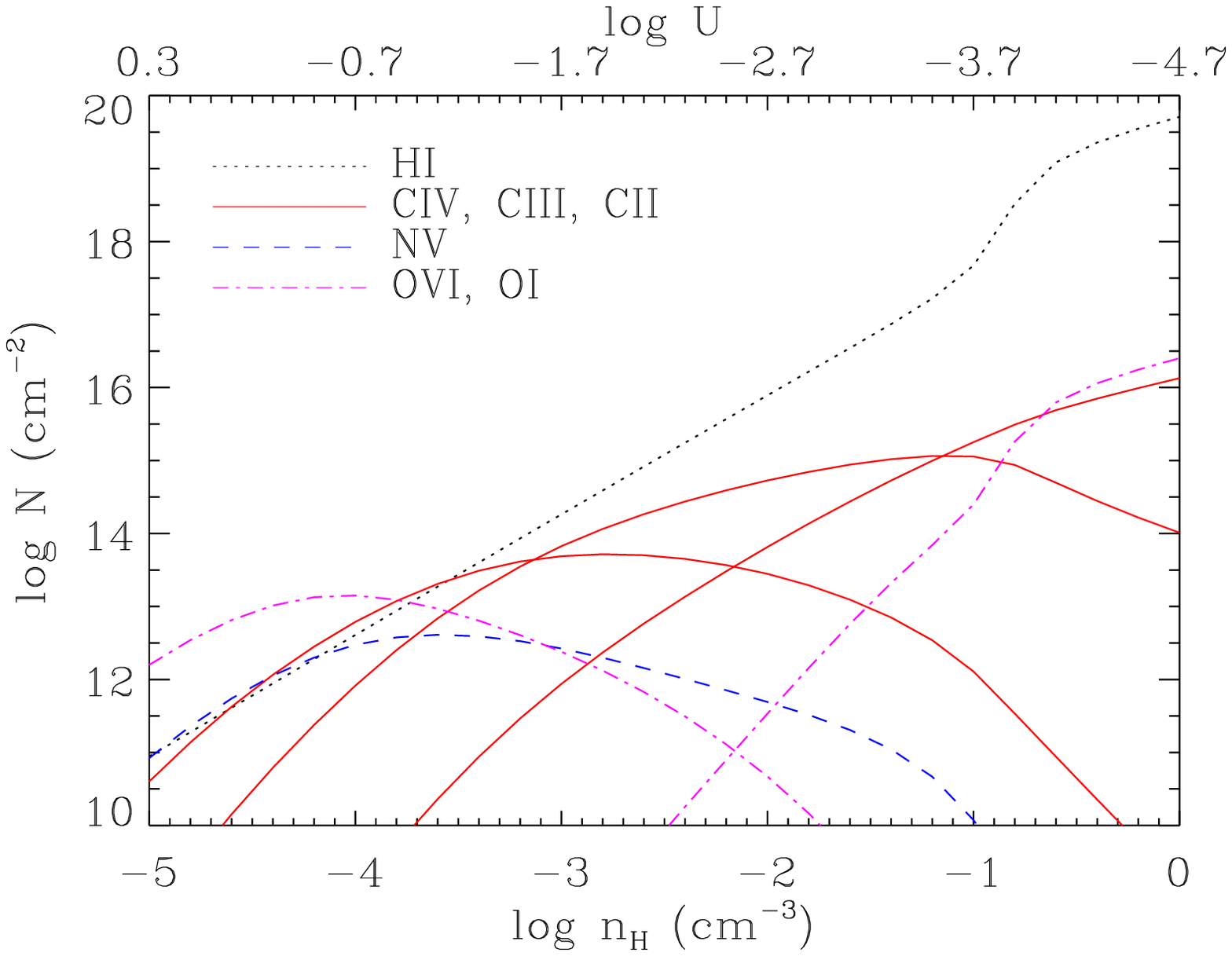}}
\resizebox{\colwidth}{!}{\includegraphics{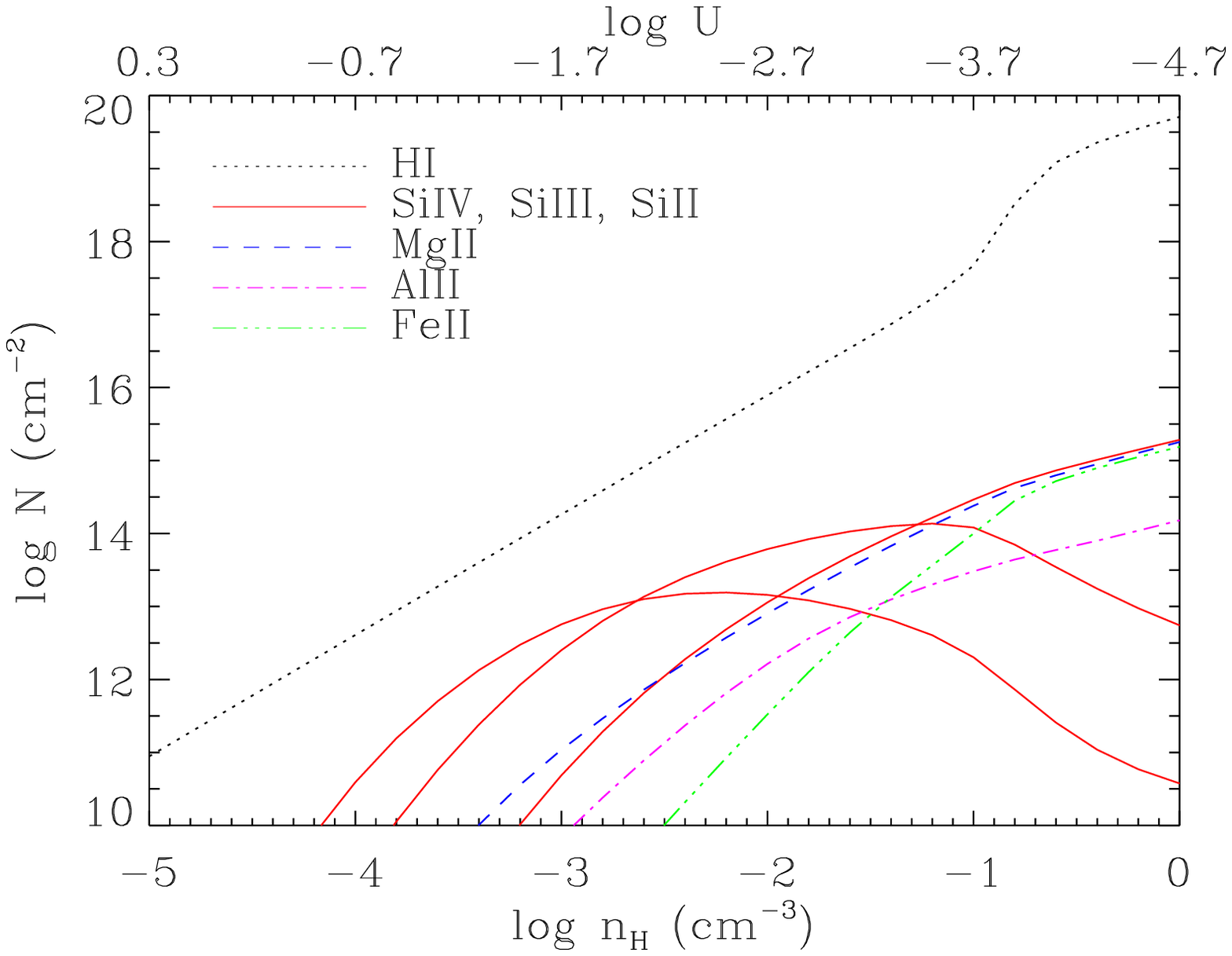}}
\caption{Column densities of various ions as a functions of density
(lower $x$-axis) and ionization parameter (upper $x$-axis) for a sight
  line through the center of a spherical $10^2~\Msun$ gas cloud. The
  cloud was assumed to have a constant density, a temperature of
  $10^4~\K$, and solar abundances. We used the Haardt \& Madau (2001)
  model for the integrated $z=2.3$ UV/X-ray background from galaxies
  and quasars.}
\label{fig:colevol}
\end{figure*}

If the clouds are short-lived, as we argued is likely, it is
interesting to ask what they look like at different points in their
lifetimes. Figure~\ref{fig:colevol} shows how the column densities of
a number of potentially observable ions vary as a function of
the density/ionization parameter. We assumed a path length
equivalent to that of a sightline passing through the center of a
spherical, constant density gas cloud with mass $M_g=10^2~\Msun$. The
cloud has solar abundances, a constant temperature of $10^4~\K$, and
is exposed 
to the model of the $z=2.3$ metagalactic radiation field of Haardt \&
Madau (2001) from galaxies and quasars. Provided self-shielding is
unimportant, which is true for $N_{\rm HI} < 10^{17}~\cm^{-2}$, all
curves can be shifted vertically in proportion with $M^{1/3}$ and in
proportion to the elemental abundances. 

Figure~\ref{fig:colevol} shows 
that unless the clouds are much more massive than $10^2~\Msun$ or much
more metal-rich than the sun, they would be undetectable for $n_{\rm H} \ll
10^{-4}~\cm^{-2}$ ($U \gg 10^{-1}$). Thus, if our clouds were to
expand by more than a factor of a few, we would no longer be able to
detect them. The \CIV\ line remains detectable
up to much higher densities, $n_{\rm H} \la 10^{-1}~\cm^{-3}$, but for
$n_{\rm H} \ga 10^{-3}~\cm^{-3}$ \NV\ and particularly \OVI\ would
typically be swamped by contamination and noise. For
high-density absorbers we predict $N_{\rm CIV} \ll N_{\rm HI}$, which
means that 
we would only be able to recognize them as high-metallicity clouds if
we knew they had high densities. Since we would not know this (because
our upper limits on the columns of \NV\ and \OVI\ would exceed the true
values), clouds with $n_{\rm H} \ga 10^{-3}~\cm^{-3}$ will not be
included in our sample. 

\subsection{Comparison with other observations}

\subsubsection{Weak \MgII\ absorbers}

Clouds with densities much higher than the clouds in our
sample can be detected in low-ionization lines
(see Fig.~\ref{fig:colevol}). The \MgII\ 
($\lambda\lambda$ 2796, 2803) doublet is a particularly 
effective tracer because it is strong, does not suffer from
contamination by Lyman series lines, and is easily identifiable. Its
disadvantage is that because of its large rest wavelength, studies in
the optical are only possible for $z\sim 1$, in which case space-based
UV-spectroscopy is required to measure hydrogen column densities
(without which metallicities and sizes cannot be determined) as well as to
measure most other metal lines (which are required to constrain the
ionization parameter).

Rigby et al.\ (2002) find that at $z\la 1$, single-cloud,
weak (equivalent width $< 0.3$~\AA) \MgII\ absorbers are metal-rich ($Z
> 0.1 Z_\odot$) and that at least the subset of systems for which \FeII\ is
detected must be compact ($R\sim 10~\pc$). Rigby et al.\ estimate that the
weak \MgII\ clouds with 
iron detections have densities $n_{\rm H} \sim 10^{-1}~\cm^{-3}$,
assuming that they are irradiated by the integrated UV radiation from
quasars as modeled for $z=1$ by Haardt \& Madau (1996). This gives a
typical cloud mass of order $10~\Msun$, which is similar to that of
our high-metallicity sample and is thus suggestive of a connection
between the two types of clouds.  

Rigby et al.\ (2002) estimate a rate of incidence of $d{\cal N}/dz\approx 0.25$ 
for the subset of weak \MgII\ systems showing \FeII\ absorption. Assuming
$(\Omega_m,\Omega_\Lambda,h) = (0.3,0.7,0.7)$ and $z=1$ we find from
equation (\ref{eq:dN/dz}) a comoving number density of $n\sim 8\times
10^4\,\Mpc^{-3}~(R/10~\pc)^{-2}$. This is only about a factor 5
greater than our high-metallicity systems at $z=2.3$
[equation~(\ref{eq:ncomoving})], which is remarkably close, particularly if we
take into account the substantial uncertainties in both of these
estimates. One might
expect the number densities of the two types of absorbers to be
similar if each cloud contained both a low and a high-ionization
phase (e.g., in a core-halo structure). However, although one would
expect high density peaks giving rise to low-ionization lines to be
surrounded by more dilute, highly ionized gas, it 
would seem natural to also have a population of gas clouds with
central densities that are too low to give rise to low-ionization
lines. Moreover, even if every high-metallicity cloud contained a weak
\MgII\ core, there is no obvious reason why the two phases should
contain similar masses. 

The mass coincidence could be accounted for if the high-metallicity
clouds were created as little dense knots, for example through
hydrodynamic and/or thermal instabilities in a
high-pressure environment, which then expanded to become weak \MgII,
and later high-metallicity \CIV, clouds on timescales that are short
compared with the 
Hubble time (and thus with the difference in age between the $z=1$ and
$z=2.3$ universe). However, there is no reason why the lifetimes of
the \MgII\ and \CIV\ phases should be similar, as would need to be the
case for their number densities to agree.
On the other hand, given that
that the two number densities were determined for different redshifts and
that the rates of incidence and cloud sizes are highly uncertain, it
would be surprising if the number densities agreed 
even if the \CIV\ and weak \MgII\ and timescales were
identical.

\subsubsection{\OVI\ systems}

Simcoe et al.\ (2006) used ionization models and a range of low- and
high-ionization lines to determine the properties of strong \OVI\
systems within 100--200 kpc from starbursting galaxies at $z\approx
2.5$. They found typical densities $n_{\rm H}\ga 10^{-3}~\cm^{-3}$,
sizes $R\sim 
10^2-10^4~\pc$ and metallicities $Z\sim 10^{-2}-10^0~Z_\odot$. The
systems selected by Simcoe et al.\ tend to have very strongly
saturated \HI\ absorption which should mean that most of their sizes
are in fact upper limits and most of their metallicities lower
limits. Their selection criteria appear to result in a sample of clouds with a
range of physical properties, but their results clearly show that
there exist compact, metal-rich clouds in the vicinity of galaxies.

Bergeron \& Herbert-Fort (2005) also found a range of properties for
\OVI\ absorbers at this redshift, including a subset of metal-rich
clouds. Like Carswell et al.\ (2002) and the present study, they found
that the high-metallicity clouds cannot be self-gravitating. Bergeron \&
Herbert-Fort (2005) also showed that the majority of metal-rich \OVI\
clouds are within $500~\kms$ of strong \HI\ absorption ($\tau_{{\rm
    Ly}\alpha} > 4$) and that at least 
half of their \OVI\ lines are too narrow for collisional ionization to
be effective.

\subsubsection{\HI\ 21~cm emitters in the halo of the Galaxy}

If we increase the cloud density above that of the weak \MgII\
systems, the \MgII\ lines would become too strong be included in Rigby
et al.'s sample of ``weak'' lines. Figure~\ref{fig:colevol} shows that
the clouds would be Lyman limit systems (i.e., $N_{\rm HI} >
10^{17}~\cm^{-2}$) with strong \CII, \OI, \MgII, \AlII, \SiII, and
\FeII\ absorption. However, given their small cross sections, the rate
of incidence would be extremely small if the number density of such
systems were similar to that of the high-metallicity \CIV\ clouds and
one would require a very large sample of quasars to see just a few.  

If such a population
of compact, metal-rich Lyman limit absorbers were present in the halo
of our own galaxy, its 21~cm emission might be
detectable. Interestingly, Lockman (2002) recently discovered that as
much as half of the \HI\ in the inner Galaxy's halo is made up of tiny
clouds with $N_{\rm HI}\sim 10^{19}~\cm^{-2}$, $R\sim 10~\pc$, $n_{\rm
H}\sim 10^{-1} - 1~\cm^{-3}$, and $M\sim 10 -
10^2~\Msun$. Such clouds may correspond to high-velocity \CaII\ absorbers
(Richter et al.\ 2005). Unfortunately, metallicities are not available
for any of 
the clouds. Regardless of the metal content of the clouds, this
observation does show that galaxies are capable of forming large
numbers of compact gas clouds in the mass range of the weak \MgII\ and
high-metallicity \CIV\ clouds.

\subsubsection{Direct constraints on cloud sizes}
\label{sec:sizes}
Observations of gravitationally lensed quasars confirm the small sizes
of low-ionization clouds derived from ionization modeling.
Rauch, Sargent, \& Barlow (1999) found a low-ionization (\CII, \OI, and
\SiII) absorber with a size $\sim 10~\pc$ and solar metallicity.
Several studies have shown that while \MgII\ complexes are larger than
about 0.5~kpc, individual \MgII\ clouds cannot be traced over
distances $\gg 10^2~\pc$ (e.g., Petitjean et al.\ 2000; Rauch et al.\ 2002;
Churchill et al.\ 2004; Ellison et al.\ 2004).

Intriguingly, Hao et al.\ (2006) recently found that a $z=1.48$ system
with strong \MgII\ and \FeII\ absorption, observed towards a $z=4.05$
Gamma-Ray Burst (GRB), showed strong variability over timescales of
days. They found that this implies absorber sizes similar to the GRB
beam size, $\sim 10^{-2}~\pc$. Such small sizes had been predicted by
Frank et al.\ (2006) because they could explain the different
rates of incidence of strong \MgII\ absorption in sight lines
towards quasars and GRBs (Prochter et al.\ 2006). 

Tzanavaris \& Carswell (2003) used evidence for partial coverage of
$z\approx 3$ \CIV\ components in the spectrum of one quasar to infer a
lower limit 
on the size of\footnote{Tzanavaris \& Carswell (2003) overestimated
the sizes by a factor of 2 due to an error, see Ellison et al.\
(2004). We quote the correct results here.} 150~pc and a most likely
size $R\sim 1~\kpc$. The study of Rauch, Sargent, \& Barlow (2001),
who analyzed \CIV\ 
coincidences in three gravitationally lensed quasar pairs at $z\sim
2$--3, is most relevant for us. Their figures~8 and 9 show
that the fractional difference in the \CIV\ columns of \emph{single
components} (but not complexes\footnote{There are other studies of \CIV\
absorption in quasar pairs that looked at complexes (e.g., Ellison et
al.\ 2004), but we have not found any other studies that also looked at
differences between individual clouds. Clearly, only individual
components can be compared with the cloud sizes inferred from our
ionization models.}) 
increases sharply from $\la 10$\% for separations less than 30~pc to
20--90\% for separations above 50~pc. They find that if the redshift
of the lens of one of the quasars is 1.32 rather than 0.73, then the
transition would be shifted to about 100~pc. 

Thus, it appears that \CIV\ clouds (with $N_{\rm CIV} \ga
10^{13}~\cm^{-2}$) have characteristic sizes $R\sim 10^2 - 10^3~\pc$,
which agrees remarkably well with the median upper limit on the size we
inferred for our high-metallicity clouds ($R\la 10^3~\pc$), but is
slightly greater than our preferred size ($R\sim 10^2~\pc$). Note that
for 6 out of 28 clouds, our robust and conservative upper limits are
$R < 10^{2.5}~\pc$. On the other hand, the 5 clouds with the
weakest limits on the density all have upper limits greater than
$10^4~\pc$. If there is scatter in the sizes, which seems
natural, then our selection criteria for high-metallicity will
preferentially pick out small clouds. This is because the inferred
limits on the metallicity and size scale as $Z \propto N_{\rm HI}^{-1}
\propto R^{-1}$. 

If the clouds are expanding, then we predict higher
density clouds (i.e., clouds with lower ionization) to be smalller and
clouds with lower densities (which would be difficult to detect in
\CIV\ but may still be visible in \OVI) to be larger. Such a trend is
consistent with our data (see Fig.~\ref{fig:dens-size}), with sizes
inferred from quasar pairs, and with ionization models of other
populations.

\subsection{Implications for the mixing of metals}

We argued that our high-metallicity clouds are likely
short-lived and that they were probably born with high densities. What
will be their ultimate fate? Presumably, they will expand
until they reach pressure equilibrium with their environment. If that
environment is the diffuse IGM, then it will have the same
temperature as the clouds, $T \sim 10^4~\K$. Pressure equilibrium 
then implies that the densities would also be the same. At that point, the
term cloud would be somewhat of a misnomer, ``metal
concentration'' might better describe the situation. 

When the cloud is done expanding, its \HI\ column density will be much
smaller than that associated with the environment. To see this, note
that when we observe the clouds, their densities are about 1.5 dex
higher than the cosmic mean. They will thus typically have to increase
in size 
by 0.5 dex to reach pressure equilibrium with their environment, or
less if the latter is overdense. At that point they will still be much
smaller than typical, moderately overdense \HI\ clouds, which are
observed to have sizes $R\sim 10^2~\kpc$ (e.g., Bechtold et al.\
1994). Since their densities and temperatures are now the same as those of
the environment, so will their \HI\ densities. Thus, $N_{\rm HI} =
n_{\rm HI} R$ will be much greater for the environment than for the
clouds. 

Another way to see this goes as follows. Unless the
environment is underdense, the hydrodynamical timescales are short
enough that its density will fluctuate on the local Jeans scale
(Schaye 2001). Hence, the \HI\ column associated with the environment
will typically be $N_{\rm HI} \sim n_{\rm HI} L_J(n_{\rm H},T)$, which
is much greater than the \HI\ column associated with the
high-metallicity gas because we found that the clouds have $R\ll L_J$
(note that if $R 
< L_J$ initially, then expansion at fixed temperature will not change this
because [see equations (\ref{eq:LJ}) and (\ref{eq:mass})]
$R/L_{\rm J} \propto n_{\rm H}^{-1/3} / n_{\rm H}^{-1/2} \propto
n_{\rm H}^{1/6}$).  

If the \HI\ column associated with the environment
is much greater than that associated with the metals, then we would
no longer consider the cloud to be a high-metallicity
system. If the high-metallicity gas is at rest with respect to its
environment, then we would not be able to tell whether the
metals are mixed throughout the \HI\ cloud or not, except through
observations that can probe the sizes of the metal-containing gas directly,
such as those comparing quasar pairs. 

Thus, we can only recognize a cloud as having high-metallicity as long
as it is denser than its environment. The density at which a
metal-rich cloud will stop  
expanding, will vary according to the density of its
environment. After it has reached pressure equilibrium, it will look
like an ordinary, low-metallicity absorption system. This suggests
that the metals in ordinary systems may also be concentrated in patches
that are very small compared to the \HI\ coherence length. 

In fact, clouds in high-density, photo-ionized environments would look
like ordinary metal systems even while they are still expanding. From
our data we can only tell that the metallicity is high when their
\CIV/\HI\ fraction is close to maximum (see
\S\ref{sec:evolution}). Clouds that end up in higher density 
environments presumably also exist, but would at no point in
their evolution be included in our sample. 

As we discussed in \S\ref{sec:sizes}, direct observations of the sizes
of individual metal-line clouds, including \CIV, typically find very
small sizes. Significantly, these observations do not attempt to
select high-metallicity gas, except for the unavoidable condition that
the metal line be detectable. This suggests that intergalactic
metals generally reside in small patches of gas. The metallicity we
infer from absorption studies is then not determined by the abundances
of heavy elements on the size of the metal concentrations, but by the
metallicity smoothed over the size of the \HI\ absorber, which is well
known to be much greater. 

We are therefore inclined to conclude that the intergalactic metals
are transported from galaxies in the form of dense, high-metallicity
clouds. This is also supported by the large amount of
metals that may have passed through the high-metallicity cloud
phase [equation (\ref{eq:zcosmicovertime})]. Although the clouds
expand until they become part of the IGM, the metals remain
poorly mixed on scales greater than $10^3~\pc$ for very low
overdensities and on even smaller scales for higher densities. 

This scenario has some profound implications. When smoothed on
small scales ($\ll \kpc$), most of the IGM (which contains most of the
baryons in the universe) may be of primordial composition. A very
small amount of 
intergalactic gas is, however, metal-rich. Such pockets of metal-rich
material will cool more efficiently, which may change the physics of
galaxy formation. The number of
metal-line components per \HI\ absorber depends on  the
number of metal concentrations along the line of sight. The absence of
associated absorption by heavy elements, even in a spectrum with an infinite
signal-to-noise ratio, does not necessarily imply that the \HI\
absorber is metal-free.

\section{Summary and conclusions}
\label{sec:conclusions}

We performed a search for high-metallicity \CIV\ absorption in
9 high-quality spectra of $2.2 < z < 3.3$ quasars taken with the UVES
spectrograph at the VLT telescope. We used a novel approach to
robustly select high-metallicity absorbers. First all \CIV\ absorbers
were identified and decomposed into Voigt profiles using
VPFIT. We then used a new algorithm to automatically measure
conservative upper limits on the column densities of associated
transitions. The resulting lower limits on $N_{\rm CIV}/N_{\rm HI}$
together with the maximum predicted \CIV/\HI\ fraction, were used to
set lower limits on the metallicity and to select 26 \CIV\ components
with \ZCgt. Two additional clouds were 
determined to have \ZCgt\ after observed lower limits on the density,
obtained from the observed lower limits on $N_{\rm CIV}/N_{\rm NV}$
and $N_{\rm CIV}/N_{\rm OVI}$, were taken into account.
Due to our reliance on upper limits, both our sample selection and the
physical properties inferred from our ionization model are robust with
respect to line blending, noise, contamination, and the presence of
phases other than the one responsible for the \CIV\ absorption. 

In total we selected 28 high-metallicity components in 12
different systems, giving rates of incidence of $d{\cal N}/dz > 7\pm
1$ and $> 3.0\pm 0.9$ for components and systems, respectively. The
mean system redshift is 2.25 and the redshift below which half of the
redshift path searched is located is 2.24. Like
the general population of \CIV\ absorbers, some of the clouds appear
isolated, others are not. Some are close to strong \HI\ systems,
others are not. The high-metallicity clouds have median \CIV\ column
density $\log[N_{\rm CIV} (\cm^{-2})] = 13.0$. The upper 
limits on $N_{\rm NV}$ and $N_{\rm OVI}$ are similar to those for the general
population of \CIV\ clouds, but the \CIV/\HI\ ratio is (by selection)
much higher ($\log(N_{\rm CIV}/N_{\rm HI})>-0.8$). 

We find no evidence for clustering around the redshift
of the quasar\footnote{Regions closer than $4,000~\kms$ to
  the quasar were excluded.}. Combined with the narrow line widths
(typically $b_{\rm CIV} < 10~\kms$) and the absence of evidence for
partial coverage, this 
strongly suggests that the clouds are intervening rather than ejected.

Assuming that the clouds are irradiated by the metagalactic
radiation field from quasars and galaxies as modeled by Haardt \&
Madau (2001) for $z=2.3$ and that they have temperatures $T=10^4~\K$,
we derived  
lower limits on the density by comparing an ionization
model to the measured lower limits on the ratios $N_{\rm CIV}/N_{\rm
  NV}$ and $N_{\rm CIV}/N_{\rm OVI}$. Similarly, for the 15 clouds for
which our spectra covered the associated \CIII, we used the upper
limits on $N_{\rm CIII}/N_{\rm CIV}$ and $N_{\rm SiIV}/N_{\rm CIV}$ to
obtain upper limits on the density and the abundance of silicon
relative to carbon, respectively. 

The lower and upper limits on the density are consistent in all
cases. The median limits are \ZC$> -0.42$, $\log[n_{\rm
H}~(\cm^{-3})]> -4.0$, $\log[R~(\pc)] < 3.2$, and $[{\rm Si}/{\rm C}] <
1.5$. There is considerable scatter around the medians, with evidence
for anti-correlations between the limits on size and density as well
as between metallicity and size, but these trends may
be caused by selection effects. The median upper limit on the density
is $\log[n_{\rm 
H}~(\cm^{-3})]< -3.0$, which suggests a typical value $n_{\rm H}\sim
10^{-3.5}~\cm^{-2}$, about 50 times the mean density at
$z=2.25$. Using this density rather than the median lower limit 
yields $R\sim 10^2~\pc$ and a gas mass $M_g \sim 10^2~\Msun$.

The high rate of incidence suggest that local sources do not dominate
the ionizing radiation field. Both the distribution of the clouds in the
\CIV/\OVI--\CIV/\NV\ plane and the line widths suggest that collisional
ionization is unimportant. In any case, the conclusions that the
clouds are compact and metal-rich would 
be strengthened if (a) the clouds were exposed to a radiation field
that is harder and/or more intense than the Haardt \& Madau (2001)
model; (b) the gas was hot enough for collisional ionization to
be important; (c) oxygen were overabundant relative to carbon; (d) the
metals were more highly ionized than they would be in photo-ionization
equilibrium.

The high-metallicity clouds are far too small (relative to their
densities) to be self-gravitating. For the clouds to be
gravitationally confined, their gas fraction would need to be
completely negligible, with dark matter (or stellar) masses of order
$10^6-10^7~\Msun$. Dark matter minihalos collapsing at $z\sim 10$
would have virial radii similar to the cloud sizes and would loose
their gas content during reionization (or earlier due to feedback from
star formation), but the predicted halo number densities are too small
to account for a significant fraction of the high-metallicity clouds.

If the clouds are confined by an external medium, then this medium
needs to have the same (low) pressure and it must not give rise to
\HI\ absorption that is stronger than allowed by the observations. We
found that if the medium is self-gravitating, this implies a
temperature $T>10^5~\K$ and a density not much higher than the cosmic
mean. Higher densities are, however, possible if the confining medium
is expanding.

If the clouds are unconfined, then their natural lifetime is the
timescale for free expansion, which is only $\sim
10^7~\yr$. If there is a confining medium, then hydrodynamical
instabilities would still limit the lifetime of the clouds to similar
values, unless they move through the medium with velocities $\ll
10~\kms$. 

If the clouds are short-lived, they will have been denser in the
past. If we decrease 
their size by  factor of a few, they will look like weak \MgII\ clouds
(e.g., Rigby et al.\ 2002). If we again compress them by a factor of a
few, they will look like the \HI\ clouds discovered by Lockman (2002) in
the halo of our Galaxy. On the other hand, if we let them expand by a
factor of a few, they will become undetectable in \CIV.  

Given their small cross sections, the clouds must be extremely
numerous for us to see as many as we do. We estimate a comoving number
density of order $10^4~\Mpc^{-3}$ which exceeds that of observable
galaxies (down to $0.1~L_\ast$) by about a factor of $10^6$. If the
clouds were to reside in halos centered on such galaxies, then we
would require halo radii of order $10^2~\kpc$ (for a cloud covering
factor of order unity). If the clouds were formed close to the galaxy,
it would taken them $10^9~\yr$ to reach $10^2~\kpc$ if they traveled
at $10^2~\kms$, a typical speed for superwinds. Since this is much
greater than the characteristic lifetime of the clouds, for the clouds
to originate in winds, they would need to be 
continuously formed and destroyed. This is expected in a galactic
wind, where swept up shells fragment due to thermal and hydrodynamical
instabilities. Observations of local starburst galaxies indeed reveal
a clumpy medium embedded in an (initially) hot fluid.

There are other possibilities, besides galactic winds, for the origin
of the clouds. Most have already been discussed in the context of other
populations of metal-rich, compact clouds, such as weak \MgII\
and strong \OVI\ absorbers. At least one scenario, intergalactic
supernovae remnants, can be ruled out if the clouds are
short-lived because it would imply that most star formation would be
intergalactic. 

The high-metallicity clouds contain only a negligible fraction of
the baryons and about an order of magnitude less carbon than can be
traced by \CIV\ in the diffuse IGM. However, if they are short-lived,
then the amount of metal that 
has passed through this phase would be greater than the amount
contained in the clouds at a given time by a factor of order
$t_H/t$. Although gas can be cycled through clouds multiple times and
although many of the clouds may end up in galaxies, this does suggest that
all of the metals in the diffuse IGM may at one point have been
contained in compact, metal-rich clouds.  

When the clouds reach pressure equilibrium with the diffuse,
photo-ionized IGM, they will still have sizes that are more than an
order of magnitude smaller than those of the diffuse \HI\
clouds, which are known to be of order $10^2~\kpc$. At that
point the \HI\ column of the confining medium will completely wash out
the \HI\ associated with the metal concentration and the cloud will
look like an ordinary, low-metallicity absorber. In fact,
even high-metallicity \CIV\ clouds that are still expanding will look
like ordinary metal-line systems until their densities are low enough
for the \CIV/\HI\ fraction to be close to its maximum.

This suggests that all intergalactic metals may reside in small
patches of highly-enriched gas. This scenario is supported by
independent constraints, such as those obtained from observations of
metal-line anti-coincidences and partial covering in the images of
gravitationally lensed quasars. We are therefore inclined to conclude
that metals are 
transported into the IGM in the form of dense, high-metallicity
clouds. These clouds expand as they move out, until they end up at
rest and in pressure equilibrium with 
the IGM, but at that point they still fill only a very small fraction
of the volume.  

Thus, most
of the IGM, and therefore the baryons in the universe, may be of
primordial composition. This is not inconsistent with previous
work on the distribution of metals in the IGM, because those studies
implicitly smooth the gas on the scale of the \HI\ absorbers. It does
mean, however, that much of the scatter in the metallicity found in
studies of quasar absorption spectra may be due to scatter in the
number of metal-rich patches intersected by the line of sight, rather
than by scatter in the metallicity on the scale of the \HI\
absorbers. In particular,  
the absence of metal-line absorption associated
with an \HI\ line, even in a spectrum with an infinite
signal-to-noise ratio, does not necessarily mean the \HI\ absorber
contains no metals.

Poor small-scale mixing of metals not only requires a
re-interpretation of previous observational analyses, it may also
affect the physics of the IGM and the formation of galaxies. For
example, the cooling rates depend on local abundances, rather than
the the metallicity obtained after smoothing over $\sim 10^2~\kpc$.

\section*{Acknowledgments}

We are grateful to the ESO Archive, without their help this work would
not have been possible. We would also like to thank the referee, Rob
Simcoe, for a helpful report. JS gratefully acknowledges support from
Marie Curie Excellence Grant MEXT-CT-2004-014112.

\begin{table*}
\caption{High-metallicity components \label{tbl:vpfits}}
\begin{tabular}{lrrrrrrrrrrr}
\hline
\input{highmetalcomponents.combined.tex}
\hline
\end{tabular}
\end{table*}

\begin{table*}
\caption{High-metallicity component properties \label{tbl:properties}}
\begin{tabular}{lrrrrrrr}
\hline
\input{Highmetalcomponents.properties.combined.tex}
\hline
\end{tabular}
\end{table*}

\bsp

\appendix

\input{appendix1.tex}

\label{lastpage}

\end{document}

%% file: highmetalcomponents.combined.tex
QSO & $z$ & $\log N_{\rm HI}$ & $\log N_{\rm CIII}$ & $\log N_{\rm CIV}$ & Error & $\log N_{\rm NV}$ & $\log N_{\rm OVI}$ & $\log N_{\rm SiIII}$ & $\log N_{\rm SiIV}$ & $b_{\rm CIV}$ & Error \\
(1) & (2) & (3) & (4) & (5) & (6) & (7) & (8) & (9) & (10) & (11) & (12) \\
\hline
Q0122$-$380 & 2.062560 & $<13.050$ &           & 12.932 &  0.039 & $<12.900$ & $<13.600$ & $<12.150$ & $<12.100$ &  17.90 &   5.10 \\
HE0151$-$4326 & 2.415724 & $<13.400$ & $<13.100$ & 12.954 &  0.031 & $<12.250$ & $<14.400$ & $<11.050$ & $<11.400$ &  17.20 &   2.70 \\
HE0151$-$4326 & 2.419671 & $<12.950$ & $<12.800$ & 12.747 &  0.009 & $<13.350$ & $<12.850$ & $<11.150$ & $<11.600$ &   7.20 &   0.90 \\
HE0151$-$4326 & 2.468097 & $<12.850$ & $<12.650$ & 12.970 &  0.008 & $<12.850$ & $<13.800$ & $<11.050$ & $<11.400$ &  12.20 &   0.50 \\
HE0151$-$4326 & 2.468720 & $<13.050$ & $<13.000$ & 12.628 &  0.225 & $<12.800$ & $<13.500$ & $<11.100$ & $<11.500$ &  10.50 &   0.20 \\
PKS0237$-$233 & 2.042163 & $<13.300$ &           & 13.571 &  0.056 & $<13.450$ & $<14.000$ & $<12.100$ & $<11.700$ &  10.10 &   3.60 \\
PKS0237$-$233 & 2.042433 & $<12.750$ &           & 12.618 &  0.051 & $<12.850$ & $<14.400$ & $<12.350$ & $<11.550$ &  11.00 &   1.00 \\
Q0329$-$385 & 2.076432 & $<13.700$ &           & 13.223 &  0.008 & $<12.900$ & $<13.250$ & $<11.300$ & $<11.650$ &  18.10 &   1.60 \\
Q0329$-$385 & 2.352007 & $<12.750$ & $<12.800$ & 13.281 &  0.011 & $<13.350$ & $<13.800$ & $<11.850$ & $<11.550$ &  12.10 &   0.70 \\
Q0329$-$385 & 2.352122 & $<13.000$ & $<12.950$ & 13.103 &  0.014 & $<13.300$ & $<14.000$ & $<11.900$ & $<11.600$ &  41.70 &  27.00 \\
HE1122$-$1648 & 2.030096 & $<12.300$ &           & 12.270 &  0.041 & $<12.950$ & $<13.900$ & $<19.900$ & $<11.400$ &   6.30 &   1.70 \\
HE1347$-$2457 & 2.116199 & $<14.100$ &           & 13.351 &  0.008 & $<12.850$ & $<15.100$ & $<11.050$ & $<11.150$ &   7.70 &   0.30 \\
HE2347$-$4342 & 2.119806 & $<14.100$ &           & 13.286 &  0.027 & $<12.600$ & $<18.100$ & $<11.400$ & $<11.650$ &   7.60 &   0.20 \\
HE2347$-$4342 & 2.274151 & $<13.100$ & $<12.800$ & 13.082 &  0.070 & $<12.850$ & $<14.000$ & $<11.500$ & $<12.100$ &   7.60 &   0.20 \\
HE2347$-$4342 & 2.274295 & $<13.500$ & $<13.700$ & 13.368 &  0.035 & $<12.650$ & $<14.000$ & $<11.400$ & $<12.200$ &   5.40 &   0.10 \\
HE2347$-$4342 & 2.274836 & $<13.100$ & $<19.900$ & 12.607 &  0.045 & $<12.700$ & $<16.300$ & $<11.550$ & $<11.300$ &  25.10 &   3.10 \\
HE2347$-$4342 & 2.274931 & $<13.400$ & $<14.900$ & 12.775 &  0.128 & $<13.000$ & $<15.300$ & $<11.900$ & $<11.650$ &   9.50 &   1.00 \\
HE2347$-$4342 & 2.275343 & $<13.300$ & $<14.900$ & 13.085 &  0.018 & $<12.850$ & $<14.900$ & $<11.150$ & $<11.600$ &   9.40 &   2.90 \\
HE2347$-$4342 & 2.275833 & $<12.700$ & $<12.700$ & 12.486 &  0.024 & $<12.300$ & $<13.400$ & $<11.050$ & $<12.950$ &   6.60 &   0.80 \\
HE2347$-$4342 & 2.276343 & $<13.050$ & $<12.700$ & 12.778 &  0.010 & $<12.250$ & $<13.300$ & $<11.650$ & $<13.700$ &  14.40 &   0.50 \\
PKS2126$-$158 & 2.388883 & $<13.600$ &           & 13.035 &  0.004 & $<17.300$ & $<13.300$ & $<11.700$ & $<12.400$ &   9.80 &   0.20 \\
PKS2126$-$158 & 2.393862 & $<14.400$ &           & 13.633 &  0.086 & $<13.200$ & $<13.400$ & $<15.300$ & $<13.150$ &   5.70 &   2.40 \\
PKS2126$-$158 & 2.394003 & $<14.400$ &           & 13.776 &  0.075 & $<13.300$ & $<13.150$ & $<15.900$ & $<14.200$ &   6.60 &   0.10 \\
PKS2126$-$158 & 2.395156 & $<13.600$ &           & 12.928 &  0.013 & $<12.150$ & $<13.700$ & $<11.900$ & $<12.000$ &  17.10 &   3.60 \\
PKS2126$-$158 & 2.395333 & $<13.350$ &           & 12.924 &  0.011 & $<12.300$ & $<14.300$ & $<11.650$ & $<11.550$ &   8.10 &   0.40 \\
PKS2126$-$158 & 2.396791 & $<12.400$ &           & 11.786 &  0.060 & $<12.800$ & $<13.800$ & $<11.900$ & $<11.400$ &  24.60 &   1.30 \\
PKS2126$-$158 & 2.678806 & $<13.700$ & $<13.250$ & 13.705 &  0.077 & $<13.400$ & $<14.000$ & $<13.600$ & $<13.000$ &   9.90 &   0.60 \\
PKS2126$-$158 & 2.678957 & $<14.100$ & $<14.000$ & 13.852 &  0.060 & $<13.500$ & $<15.100$ & $<14.700$ & $<12.750$ &  26.00 &   4.50 \\

%% file: Highmetalcomponents.properties.combined.tex
QSO & $z$ & $\log n_{\rm H}$ & $\log n_{\rm H}$ & $\log R$ & \ZC & $\log f_g$ & $[{\rm Si}/{\rm C}]$ \\
 &  & ($\cm^{-3}$) & ($\cm^{-3}$) & (pc) &  &  &  \\
(1) & (2) & (3) & (4) & (5) & (6) & (7) & (8) \\
\hline
Q0122$-$380 & 2.062560 & $>-4.22$ &          & $< 3.17$ & $>-0.34$ & $<-3.49$ & $< 1.80$ \\
HE0151$-$4326 & 2.415724 & $>-3.62$ & $<-2.96$ & $< 2.33$ & $>-0.51$ & $<-4.58$ & $< 0.10$ \\
HE0151$-$4326 & 2.419671 & $>-3.86$ & $<-3.07$ & $< 2.34$ & $>-0.36$ & $<-4.79$ & $< 0.83$ \\
HE0151$-$4326 & 2.468097 & $>-4.25$ & $<-3.44$ & $< 3.04$ & $>-0.07$ & $<-3.80$ & $< 1.13$ \\
HE0151$-$4326 & 2.468720 & $>-4.52$ & $<-2.56$ & $< 3.77$ & $>-0.83$ & $<-2.60$ & $< 2.26$ \\
PKS0237$-$233 & 2.042163 & $>-4.08$ &          & $< 3.13$ & $> 0.04$ & $<-3.42$ & $< 0.50$ \\
PKS0237$-$233 & 2.042433 & $>-4.92$ &          & $< 4.27$ & $>-0.36$ & $<-2.00$ & $< 3.45$ \\
Q0329$-$385 & 2.076432 & $>-3.81$ &          & $< 3.00$ & $>-0.62$ & $<-3.42$ & $< 0.33$ \\
Q0329$-$385 & 2.352007 & $>-4.11$ & $<-3.60$ & $< 2.64$ & $> 0.34$ & $<-4.44$ & $< 0.67$ \\
Q0329$-$385 & 2.352122 & $>-4.37$ & $<-3.27$ & $< 3.42$ & $>-0.09$ & $<-3.15$ & $< 1.46$ \\
HE1122$-$1648 & 2.030096 & $>-5.07$ &          & $< 4.12$ & $>-0.25$ & $<-2.45$ & $< 4.17$ \\
HE1347$-$2457 & 2.116199 & $>-3.81$ &          & $< 3.39$ & $>-0.89$ & $<-2.64$ & $<-0.31$ \\
HE2347$-$4342 & 2.119806 & $>-3.64$ &          & $< 3.06$ & $>-0.89$ & $<-3.14$ & $< 0.03$ \\
HE2347$-$4342 & 2.274151 & $>-4.18$ & $<-3.34$ & $< 3.15$ & $>-0.27$ & $<-3.50$ & $< 1.58$ \\
HE2347$-$4342 & 2.274295 & $>-3.61$ & $<-2.78$ & $< 2.41$ & $>-0.20$ & $<-4.41$ & $< 0.47$ \\
HE2347$-$4342 & 2.274836 & $>-4.64$ & $< 1.67$ & $< 4.07$ & $>-0.72$ & $<-2.13$ & $< 2.36$ \\
HE2347$-$4342 & 2.274931 & $>-5.07$ & $<-1.33$ & $< 5.21$ & $>-0.93$ & $<-0.26$ & $< 3.91$ \\
HE2347$-$4342 & 2.275343 & $>-4.12$ & $<-1.60$ & $< 3.21$ & $>-0.41$ & $<-3.31$ & $< 0.94$ \\
HE2347$-$4342 & 2.275833 & $>-4.18$ & $<-2.90$ & $< 2.75$ & $>-0.42$ & $<-4.30$ & $< 3.02$ \\
HE2347$-$4342 & 2.276343 & $>-3.78$ & $<-3.20$ & $< 2.29$ & $>-0.40$ & $<-4.81$ & $< 2.78$ \\
PKS2126$-$158 & 2.388883 & $>-3.95$ &          & $< 3.17$ & $>-0.74$ & $<-3.22$ & $< 1.48$ \\
PKS2126$-$158 & 2.393862 & $>-3.71$ &          & $< 3.50$ & $>-0.92$ & $<-2.33$ & $< 1.31$ \\
PKS2126$-$158 & 2.394003 & $>-3.48$ &          & $< 3.04$ & $>-0.62$ & $<-3.02$ & $< 1.92$ \\
PKS2126$-$158 & 2.395156 & $>-3.53$ &          & $< 2.35$ & $>-0.66$ & $<-4.45$ & $< 0.60$ \\
PKS2126$-$158 & 2.395333 & $>-3.68$ &          & $< 2.40$ & $>-0.51$ & $<-4.50$ & $< 0.35$ \\
PKS2126$-$158 & 2.396791 & $>-5.58$ &          & $< 5.23$ & $>-0.85$ & $<-0.73$ & $< 6.74$ \\
PKS2126$-$158 & 2.678806 & $>-4.01$ & $<-3.51$ & $< 3.39$ & $>-0.25$ & $<-2.84$ & $< 1.54$ \\
PKS2126$-$158 & 2.678957 & $>-4.03$ & $<-2.93$ & $< 3.83$ & $>-0.49$ & $<-1.98$ & $< 1.16$ \\

%% file: appendix1.tex
\section{Column density upper limits}

If the redshift and Doppler parameter are determined for some
reference ion, for example \CIV, then for other ions within the same
region the redshift will be the same and one can estimate a range of
acceptable Doppler parameters $b$ under the assumption that there is a
turbulent component $b_{\rm turb}$ which is the same for all ions and
a thermal component $b_{\rm T}$ which is proportional to the square
root of the mass $m$ of the ion.
Then the two components add in quadrature to give the actual Doppler
parameter, so
\[ b^2= b_{\rm turb}^2+b_{\rm T}^2.\]
where
$b_{\rm T}=\sqrt{m\over m_{\rm ref}}~b_{\rm T,ref}$
where the subscript 'ref' is the value for the reference ion.
Possible extreme values for $b$ are then determined assuming that
the Doppler parameter for the reference ion is fully turbulent and
fully thermal. In the implementation used here this range was extended
by using the 1-$\sigma$ error estimates for the reference ion, so
reference Doppler parameters $b_{\rm ref}\pm\sigma$ were used and
the most extreme values of $b$ adopted to give the Doppler parameter 
ranges for each ion.

Using the redshift for the reference ion, and a sequence of Doppler
parameters from the minimum to maximum obtained in this way, a grid of
Voigt profiles convolved with the instrument profile was constructed
for the transitions of the test ion available in the observed
range. For Lyman lines these could, in principle, be several of the 
Lyman series, but for the application here generally only Ly$\alpha$ and 
Ly$\beta$ were used since they are covered for most of the sample. 
The line profiles were compared with the data, and a $\chi^2$ determined
for pixels where the Voigt profile was below the data value plus
1-$\sigma$, and within $2b$ of the line center for each transition. If
no pixels satisfied this criterion then the column density was
increased until some did.  The column density limit for each $b$-value
was taken as the highest value for which the $\chi^2$ value over this
range had a probability of occurring by chance of less than 0.16
(corresponding to a 1-$\sigma$ one-sided deviation). The final overall
limit adopted was the maximum of these over the range of Doppler
parameters. This yields a maximum possible column density
for the ion even in the presence of blends, since it is effectively only the
pixels where the trial fitted profile is too low which contribute to the
significance level.

The chance probability criterion for accepting or rejecting possible line
profiles is arbitrary, so there seemed little point
in iterating or interpolating to achieve high accuracy. We adopted a
column density step of $\Delta\log N=0.05$ for $\log N < 13.5$, and
double this for higher values. The quantities given in Table 3
reflect this choice.